\documentclass[twocolumn]{aastex701}

\usepackage{amsmath}
\usepackage{ragged2e}
\usepackage{booktabs, makecell, tabularx}
\usepackage{tabularx}
\usepackage{xspace} 
\usepackage{hyperref}
\usepackage{soul}

\defcitealias{Draine_2007}{DL07}
\defcitealias{Thomson_2011}{TC11}
\defcitealias{Sako_2018}{S18}
\defcitealias{Rubin_2025}{R25}
\defcitealias{DESICollaboration2025}{DESI-DR2}

\newcommand{\OmValueThree}{\ensuremath{0.352^{+0.026}_{-0.025}}\xspace}
\newcommand{\alphavalueThree}{\ensuremath{0.171\pm 0.009}\xspace}
\newcommand{\betaBvalueThree}{\ensuremath{2.16^{+0.23}_{-0.21}}\xspace}

\newcommand{\betaRhighvalueThree}{\ensuremath{3.26^{+0.15}_{-0.14}}\xspace}
\newcommand{\betaRlowvalueThree}{\ensuremath{4.30^{+0.28}_{-0.25}}\xspace}
\newcommand{\stepmassvalueThree}{\ensuremath{10.114^{+0.061}_{-0.040}}\xspace}
\newcommand{\deltazerovalueThree}{\ensuremath{0.037^{+0.017}_{-0.016}}\xspace}
\newcommand{\deltahvalueThree}{\ensuremath{0.82^{+0.13}_{-0.25}}\xspace}

\newcommand{\OmValueThreeOne}{\ensuremath{0.344^{+0.026}_{-0.025}}\xspace}
\newcommand{\alphavalueThreeOne}{\ensuremath{0.164 \pm 0.007}\xspace}
\newcommand{\betaBvalueThreeOne}{\ensuremath{2.20^{+0.23}_{-0.20}}\xspace}

\newcommand{\betaRhighvalueThreeOne}{\ensuremath{3.29^{+0.13}_{-0.12}}\xspace}
\newcommand{\betaRlowvalueThreeOne}{\ensuremath{4.48^{+0.20}_{-0.18}}\xspace}
\newcommand{\stepmassvalueThreeOne}{\ensuremath{10.115^{+0.044}_{-0.053}}\xspace}
\newcommand{\deltazerovalueThreeOne}{\ensuremath{0.037\pm 0.014}\xspace}
\newcommand{\deltahvalueThreeOne}{\ensuremath{0.73^{+0.18}_{-0.26}}\xspace}

\newcommand{\nTotThree}{2087}
\newcommand{\nMassThree}{995}
\newcommand{\nNoMassThree}{1092}
\newcommand{\fracWithMassThree}{47.7}
\newcommand{\fracNoMassThree}{52.3}

\newcommand{\nTotLow}{682}
\newcommand{\nTotMid}{1335}

\newcommand{\nLSTPlow}{559}
\newcommand{\fracLSTPlow}{82.0}
\newcommand{\nNoLSTPlow}{123}

\newcommand{\nLSTPmid}{968}

\newcommand{\fracLSTPmid}{72.5}

\newcommand{\fracNoMassThreeOne}{10.7}

\newcommand{\woDESIuThreeOne}{\ensuremath{w_{0} = -0.719\pm 0.084}}
\newcommand{\waDESIuThreeOne}{\ensuremath{w_{a} = -0.95^{+0.29}_{-0.26}}}

\begin{document}

\title{Union3.1: Reducing Systematics in Supernova Cosmology with Self-consistent Measurements of Host Galaxy Properties for 2000 Type Ia Supernovae}

\newcommand{\uhawaii}{\affiliation{Department of Physics and Astronomy, University of Hawai`i at M{\=a}noa, Honolulu, Hawai`i 96822}}
\newcommand{\stsci}{\affiliation{Space Telescope Science Institute, 3700 San Martin Drive Baltimore, MD 21218, USA}}
\newcommand{\lbnl}{\affiliation{E.O. Lawrence Berkeley National Laboratory, 1 Cyclotron Rd., Berkeley, CA, 94720, USA}}
\newcommand{\ANUrsaa}{\affiliation{Research School of Astronomy and Astrophysics, The Australian National University, Canberra, ACT 2601, Australia}}
\newcommand{\ANUcga}{\affiliation{Centre for Gravitational Astrophysics, College of Science, The Australian National University, ACT 2601, Australia}}
\newcommand{\lancaster}{\affiliation{Physics Department, Lancaster University, Lancaster LA1 4YB, United Kingdom}}
\newcommand{\ucberkeley}{\affiliation{Department of Physics, University of California Berkeley, Berkeley, CA 94720, USA}}

\author[0000-0001-9664-0560]{Taylor J. Hoyt}
\ucberkeley
\lbnl
\email{taylorjhoyt@gmail.com}

\author[0000-0001-5402-4647]{David Rubin}
\uhawaii
\lbnl
\email{drubin@hawaii.edu}

\author{Greg Aldering}
\lbnl
\email{galdering@lbl.gov}

\author{Saul Perlmutter}
\lbnl
\ucberkeley
\email{saul@lbl.gov}

\author{Andrei Cuceu}
\lbnl
\email{ACuceu@lbl.gov}

\author{Ravi Gupta}
\lbnl
\email{MISSING}

\begin{abstract}
Photometrically derived distances of Type Ia supernovae (SNe~Ia) rely on a $\sim$5\% empirical correction based on host galaxy properties, e.g., global stellar mass. 
Unbiased cosmology inference therefore requires the self-consistent determination of host properties across the full range of redshifts probed, which we undertake here for approximately 2000 SNe in the Union3 compilation (now Union3.1).
We use homogeneous, optical--infrared photometry from the DESI Legacy Imaging Surveys to infer global galaxy properties using the stellar population synthesis and SED-fitting code \texttt{Prospector}. 
We find that the host masses of $z<0.10$ SNe in Union3 were on average overestimated, while the opposite was true for $z<0.15$ SNe in Pantheon+. 
After correction, the two studies' average distance modulus estimated for low-redshift SNe, previously $>0.03$~mag discrepant, come into 0.01~mag agreement.
Updating the UNITY SN analysis, we find the uncertainties on all standardization parameters shrink to 0.6--0.9$\times$ their previous sizes.
For flat-$\Lambda$CDM, we find from SNe alone $\Omega_m = \OmValueThreeOne{}$ (a $-0.4\sigma$ shift from Union3). We then combine with measurements of Baryon Acoustic Oscillations and the Cosmic Microwave Background exactly as done by DESI DR2 and find for flat $w_0w_a$CDM, \woDESIuThreeOne{} and \waDESIuThreeOne, corresponding to $3.4\sigma$ evidence against a cosmological constant (down from $3.8\sigma$ per DESI-DR2+Planck+Union3).
Updating the DESI-DR2+Planck+SN combined probe analysis with the recent Dovekie recalibration of DES-SN5YR \citep{Popovic_2025arXiv250605471P} or the updated Pantheon+, we find $3.4\sigma$ (was $4.2\sigma$ before Dovekie) and $3.2\sigma$ (was $2.8\sigma$ before our correction to Pantheon+) evidence, respectively, against a cosmological constant---a significantly improved consistency between SN analyses.
\end{abstract}

\keywords{Type Ia supernovae; cosmology; dark energy; galaxy masses}

\section{Introduction} \label{sec:intro}

Type Ia supernovae (SNe~Ia) are the most reliable probes of cosmic luminosity distance. Their use as relative distance indicators yielded the first evidence that the universe's expansion is accelerating \citep{Riess1998, Perlmutter1999}, and continue to provide unique constraints on dark energy in the present landscape of precision cosmology \citep{Scolnic_2022, DES5YR, Rubin_2025, DESICollaboration2025}. SNe~Ia also constitute the final rung of the Hubble constant ``distance ladder,'' providing the highest precision constraints on the present day expansion rate \citep{Riess_2022, Freedman_2025}, making them a focal point of the ongoing Hubble Tension debate \citep{Freedman_2021, diValentino_2021, Freedman_2023}.

\subsection{Homogeneous Calibration to Accurately Probe Cosmology}

The inference of cosmological parameters from SNe~Ia hinges on our ability to calibrate the underlying data onto a single, homogeneous system across all probed redshifts. ``Data'' here primarily refers to the SN flux measurements, typically acquired by different teams, at different sites, and with different telescope-instrument systems. Efforts that aim to cross-calibrate independent sets of SN photometry include the SuperCal/Pantheon+ \citep{Scolnic_2015,Brout_2022_supercalif} and Union \citep{Kowalski_2008, Suzuki_2012,Rubin_2025} projects.

The importance of this photometric cross-calibration to tracing the dynamics of the universe cannot be overstated. The procedure has long been, and still is, a dominant source of uncertainty in local measurements of the Hubble constant. See, e.g., \citet{Riess_2005} for a discussion in regards to (not) combining photometry estimated from photographic plates with CCD photometry. Similarly, photometric cross-calibration is crucial in discerning between a cosmological constant or evolving dark energy in modern SN cosmology studies \citep{Rubin_2025, Popovic_2025arXiv250605471P}, which requires combining very low and high redshift SNe, necessarily observed with different telescopes, onto one Hubble diagram.

Notably, however, another piece of the multi-survey cross-calibration procedure, related to the properties of the SN host galaxies, can become a systematic uncertainty comparable in size to modern photometric calibration uncertainties, if not carefully and self-consistently taken into account.

\subsection{Properties of SN host galaxies and their role in standardization}
SNe~Ia occur in nature at relatively standard brightnesses. The dispersion in their observed luminosities is approximately 40\%, or about 20\% in distance. After correcting for the timescale at which their brightness evolves in time, referred to as the light curve decline rate or stretch \citep{Pskovskii1977, Phillips1993, Riess1996, Perlmutter1997}, as well as their observed colors \citep{Riess1996, Tripp1998}, this improves to about 15\% in luminosity, or 7\% in distance.

After correcting SN magnitudes as per their light curve properties, however, there remains in the residuals about the line of the best-fit distance-redshift relation a discontinuity, or ``step,'' equal to about 5--10\% in luminosity  that is correlated with the properties of the SNe's host galaxies \citep{Kelly2010, Sullivan2010, Lampeitl2010, Childress2014}. The parameter most commonly used is the total stellar mass \citep{Sullivan2010}, with other probes such as the color local to the SN location used \citep{Roman2018, Ginolin2025a}. Notably, it has been shown that the local specific star formation rate (LsSFR) is the galaxy property most strongly correlated with SN luminosities, suggesting the LsSFR is more closely aligned with the underlying cause of this residual, non-linear variation in SN brightness \citep{Rigault_2013, Rigault_2015, Rigault_2020, Briday_2022}.

The underlying physical reason for this residual dependence of SN~Ia luminosities on host galaxy properties is not well understood. Some findings suggest metallicity \citep{Childress2014}, others age \citep{Rigault_2013,Rigault_2015,Rigault_2020,Ginolin2025a}, and more recently dust, or a dust-like (exponentially distributed) color \citep{Brout_2021, Rubin_2025}. Recent studies, however, have shown that by accounting for the full spectrophotometric diversity of SNe, rather than just what can be captured by broadband photometry, drastically reduces, and possibly eliminates, this dependence on host galaxy properties \citep{boone2021a, boone2021b, Ganot_2025}. This suggests the host property-luminosity dependence observed to date is most likely a chance projection of astrophysical correlations between the progenitors and \textit{local} properties of SNe with the \textit{global} properties of their host galaxies. Nevertheless, significant effort continues to be invested in the light curve approach to SN Ia distance measurement \citep{Bellm_2019, Gris_2023, Rigault_2025, Kessler_2025, Rubin_2025_roman}, and so we devote this study to improving the accuracy of these corrections related to the SN host properties.

This dependence of galaxy properties with SN~Ia characteristics has also emerged as a significant, under-reported systematic discrepancy between independent analyses of the same SNe, thereby also impacting the combined interpretation of Baryon Acoustic Oscillations (BAO) and Cosmic Microwave Background (CMB) fluctuations with SNe \citep{Efstathiou_2025, Vincenzi_2025}. \citet{Efstathiou_2025} tracked down differences between inferred SN distance between the DESY5 and Pantheon+ analyses of the same SNe. \citet{Vincenzi_2025} expanded on that comparison and found that 0.01~mag of the discrepancy was due to a 0.3~dex average offset between the two groups' estimates of SN host galaxy stellar masses.\footnote{Another 0.024~mag was reported to be due to differences in modeling selection effects and the intrinsic scatter of SNe, though \citet{Popovic_2025arXiv250605471P} recently showed that 0.022~mag of this difference was in fact due to inaccuracies in the photometric calibration of the DES SN photometry, as well as a significant bug in the DES light curve fitting software.} As we will demonstrate in this article, some of these systematic discrepancies are in fact even larger, emphasizing that host galaxy mass estimates must be made a priority in the multi-survey cross-calibration efforts currently employed in the kinds of SN~Ia cosmology analyses that cannot eliminate the mass step by some other means such as with the spectrophotometric standardization models previously mentioned. In this study, we will tackle this issue in the context of the Union3 compilation, providing a near complete sample of robust, homogeneously-derived, full-posterior-sampled mass estimates for the host galaxies of approximately 2000 SNe~Ia.

\subsection{Union 3.1: Updating Host Masses in the Union3 compilation}

The Union3 compilation merged SN photometric data from 24 independent surveys for $\sim$2700 SNe ($\sim$2100 after data quality cuts) to perform a state-of-the-art cosmology inference within a Bayesian hierarchical framework over the redshift range $0.01 < z < 2.2$ \citep[][hereafter \citetalias{Rubin_2025}]{Rubin_2025}. Union3 re-calibrates onto the Pan-STARRS1 (PS1) magnitude system \citep{Tonry_2012} the intermediate redshift surveys that published the local (tertiary) standards of their observed SN fields, including SNLS, Pan-STARRS, and Foundation. This approach was not pursued for most of the low-redshift surveys because there is not consensus whether this recalibration process actually improves cross-calibration accuracy, with some studies reporting improvements \citep{Scolnic_2015, Scolnic_2022}, and others not \citep{Currie_2020}. Once a calibration for the surveys in Union was established, the characteristics of the SNe such as light curve decline rate and color are measured as per the SALT3 model \citep{Kenworthy_2021, Taylor_2023}, which were later used in the SN luminosity standardization and distance inference.

The SALT3 light curve parameters estimated from the homogenized Union3 photometry, along with information of the SN host galaxy such as mass and redshift, are then passed to the Unified Nonlinear Inference for Type-Ia cosmologY, or UNITY, the Bayesian hierarchical framework built to model complex aspects of SN cosmology inference. These aspects include: the selection effects associated with each aggregated survey, along with many other considerations such as the effects of intergalactic dust extinction integrated through foreground galaxies along the line of sight to distant SNe, and decomposing the distribution of peculiar velocities in the nearby universe into robust basis vectors. 

The latest versions of the light curve data and fits, referred to as Union3, as well as the cosmology modeling framework, UNITY1.5, were presented in \citetalias{Rubin_2025}. They found for flat $\Lambda$CDM a value of $\Omega_m = 0.356^{+0.028}_{-0.026}$ from the SN constraint alone, and a value of $0.313 \pm 0.006$ when combining with Planck CMB and BOSS BAO measurements.\footnote{They also combined with the low redshift BAO measurement of 6dF \citep{Beutler_2011}, but its statistical power at that redshift is small compared to that of the low redshift SNe.} When considering the $w_0$--$w_0$ parametrization of $w(z)$ in a flat universe, they found $w_0 = -0.744^{+0.100}_{-0.097}$ and $w_a = -0.79^{+0.35}_{-0.38}$ when combined with CMB and BAO.\footnote{Note the SNe on their own did not provide tight enough constraints in the $w_0$--$w_a$ plane for credible intervals to be reported.}

Underlying the \citetalias{Rubin_2025} cosmology measurements were host galaxy stellar mass estimates compiled directly from the SN literature, if published. This was the case for \nMassThree{} of the \nTotThree{} SNe (\fracWithMassThree{}\%) used in the Union3/UNITY1.5 cosmology analysis, which left \nNoMassThree{} missing an explicitly determined host galaxy mass estimate. In those cases, a default filler value of 11.0~dex was  adopted for SNe with $z<0.1$ and a value of 10.0~dex for SNe with $z > 0.1$. The larger default value at low redshifts was intended to account for the fact that many low redshift surveys were galaxy-targeted, rather than blind, searches. We will later show, however, that this assumption led to a sizable bias in the Union3 distances because most of the nearby SNe that were assigned this larger default value were in fact discovered by \textit{un-targeted} searches that have become more common in recent years, such as ASAS-SN \citep{Shappee_2012, Shappee_2014, Holoien_2017}, the La Silla QUEST survey \citep{Baltay_2013}, and the Pan-STARRS Survey for Transients \citep{Huber_2015, Chambers_2016}.

In this study, we will determine new stellar mass estimates for $\sim$2000 of the SNe in Union3, placing them onto a single, contiguous system. To achieve this, the new masses will be derived from a homogeneously calibrated set of galaxy photometry, a single model of galaxy spectral energy distributions (SEDs), as well as a consistent set of underlying assumptions on, e.g., the initial mass function (IMF) and the priors adopted for the parameters that describe the analytic star formation history (SFH). 

As briefly alluded to earlier in this section, this new homogeneous set of masses will diagnose a systematic offset in the Union3 masses previously adopted for low-redshift SNe, as well as in those used in the Pantheon+ analysis, explaining most of the discrepancy between Pantheon+ and Union3 in their respective reported distances to low-redshift SNe, as well as a significant portion of the discrepancy between Pantheon+ and DES-SN5YR identified by \citet{Efstathiou_2025}. We will demonstrate that correcting for the systematic offsets in the host masses significantly improves the consistency of SN cosmology analyses.

\setlength{\tabcolsep}{4pt}
\begin{deluxetable*}{llrrrrrrrrrc}
\tablecaption{SN host galaxy coordinates and unique identifiers in the DESI Legacy Imaging Surveys. \label{tab:host_coords}}
\tablehead{
\colhead{SN} &
\colhead{Sample} &
\colhead{R.A.$_{\mathrm{SN}}$} &
\colhead{Dec.$_{\mathrm{SN}}$} &
\colhead{R.A.$_{\mathrm{Host}}$} &
\colhead{Dec.$_{\mathrm{Host}}$} &
\colhead{$\delta r$} &
\colhead{$\delta r$} &
\colhead{Brick ID} &
\colhead{Obj. ID} &
\colhead{DR code} &
\colhead{in UNITY} \\
\colhead{} & \colhead{} & \colhead{(deg)} & \colhead{(deg)} & \colhead{(deg)} & \colhead{(deg)} &
\colhead{(arcsec)} & \colhead{(kpc)} & \colhead{} & \colhead{} & \colhead{} & \colhead{}
}
\startdata
03D1au & SNLS &  36.0432 &  -4.0375 &  36.0434 &  -4.0374 & 0.71 &   4.5 & 307479 & 15247 & 10000 & 1 \\
03D1aw & SNLS &  36.0616 &  -4.5172 &  36.0614 &  -4.5171 & 0.94 &   6.4 & 304605 & 15257 & 10000 & 1 \\
03D1ax & SNLS &  36.0973 &  -4.7208 &  36.0969 &  -4.7211 & 1.96 &  12.3 & 303169 & 19096 & 10000 & 1 \\
03D1bp & SNLS &  36.6572 &  -4.8388 &  36.6571 &  -4.8391 & 1.33 &   6.7 & 303172 &  4413 & 10000 & 1 \\
03D1dt & SNLS &  36.6300 &  -4.0523 &  36.6301 &  -4.0524 & 0.55 &   3.8 & 307482 &  5451 & 10000 & 0 \\
&&&&\vdots&&&&&&&
\enddata
\tablecomments{SN coordinates are repeated exactly as they were published in the original study and subsequently compiled in Union. Host galaxy centroids are in the coordinate frame of the DESI survey. $\delta r$ is the separation of the SN on the sky from the centroid of its host, listed in angular (arcsec) and physical (kpc) units. Brick ID, Obj. ID and Rel. (Tractor release code) together uniquely identify the associated host galaxy in the LS Tractor catalogs: Brick ID is fixed across data releases, whereas Obj. ID is assigned per Tractor run, so it must be paired with the release code: 9011 for DR9 (northern hosts) and 10000 for DR10 (southern hosts); a code of 10002 marks DR10 bricks that were re-processed in 2023 to fix the SUB\_BLOB source-extraction bug. UNITY is 1 for SNe that pass the cosmology selection cuts and enter the UNITY fit, 0 otherwise. The full machine-readable table is accessible via the journal.}
\end{deluxetable*}

We will also present several smaller updates to the Union database, including a refined merging of duplicate SNe and a fix to the foreground reddening corrections (which had no impact on the measured cosmology), among other minor changes. In addition, the UNITY model will be updated from version 1.5 to 1.7, which includes an improved outlier model and an explicit fit parameter for the fiducial mass at which the luminosity step is placed. The cosmological parameters will then be redetermined. This version of the analysis and its associated cosmological parameter estimates are labeled Union3.1 (data and light curve fits) and UNITY1.7 (standardization and cosmology). In a companion paper, \citet{Rubin_x1split} present an improved standardization that models SNe as being drawn from two distinct populations of the SALT $x_1$ parameter, referred to as UNITY1.8.

This paper is organized as follows. In \autoref{sect:galphot} we present our association of each SN with a host galaxy and the accompanying photometry. In \autoref{sect:prospector} we describe our inference of SN host galaxy properties via SED fitting with \texttt{Prospector} and explore the results in \autoref{sect:hostprops}. In \autoref{sect:otherupdates} we describe other, more minor updates to the Union3 database. In \autoref{sect:sncosmology} we present the new cosmology and SN standardization results based on the new mass estimates. In \autoref{sect:combined} we combine with other cosmology probes and present updated $w_0$--$w_a$ constraints, before summarizing and concluding in \autoref{sect:conclusion}.

\section{SN Host Galaxy Information}\label{sect:galphot}

Inferring a galaxy's mean properties from broadband photometry is typically performed using stellar population synthesis (SPS) codes to generate trial spectra via the superposition of various single stellar population (SSP) SEDs. Adding UV coverage to restframe optical photometry of a galaxy helps to constrain recent star formation, and combining those with mid-infrared (MIR) coverage helps to break part of the dust-metallicity-age degeneracy \citep[see, e.g.,][and references therein]{Leroy_2019}. In this study, we will not produce new photometry of SN host galaxies, but instead rely on established, well-calibrated, and homogeneously reduced sets of galaxy photometry. In this section we will describe our process for determining host galaxy coordinates, as well as the published sources of photometry that we will use later to infer galaxy properties via SED-fitting.

\subsection{Host Galaxy Identification and Centroiding}

To begin building our SN host galaxy catalog, we examined image cutouts from DR10 (DR9 in the North) of the DESI Legacy Imaging Surveys \citep[][hereafter referred to as just Legacy Surveys, or LS]{Dey_2019} to assign a host to each SN in Union and estimate its centroid on the LS coordinate system. For targets with $\delta\leq30$~deg, this corresponds to imaging from the DECam Legacy Survey \citep[DECaLS,][]{Blum_2016}, and for targets with $\delta>30$~deg the combined Beijing-Arizona Sky Survey \citep[BASS,][]{Zou_2017} + Mayall z-band Legacy Survey \citep[MzLS,][]{Silva_2016}. We inspected $g$-band images for $z<0.3$ SNe, and $r$-band images for $z>0.3$ SNe. After retaining only unambiguous host associations, we determined centroids for approximately 1300 SN host galaxies, which are presented in \autoref{tab:host_coords}. In a few dozen cases, we associated a host galaxy to a SN for which the source survey had not reported host properties. We discuss those cases in detail in \autoref{app:new_hosts}.

\subsection{Sources of Photometry}
The primary set of galaxy photometry comes from the LS DR10 object catalog, which we supplement with GALEX FUV and NUV photometry. For the high redshift SNe discovered in the GOODS fields with HST \citep{Riess_2007}, we assemble multi-band photometry of host galaxies published by \citet{Thomson_2011}. The remainder of this section describes these source catalogs and our procedure for matching into them.

\subsubsection{The DESI Legacy Imaging Surveys}

The newly determined host galaxy coordinates are used to match into the LS DR10 Tractor catalog database \url{https://datalab.noirlab.edu/data/legacy-surveys}. The LS Tractor reports de-blended, aperture-matched fluxes photometered from LS optical $gr(i)z$ and unWISE infrared W1,2,3,4 imaging \citep{Lang_2014, Dey_2019}. SDSS-SN published their host galaxy coordinates in \citet{Sako_2018}, and those were used to match into the Tractor. Critical radii equal to $1.0''$, $1.0''$, $0.6''$, $0.6''$, and $0.3''$ were adopted to match into the LS Tractor catalog for the low-redshift, SDSS, Pan-STARRS, SNLS, and DESY3 samples, respectively. We provide the query into the LS Tractor catalog in \autoref{app:queries}.

In \autoref{tab:host_coords}, we present the SN host galaxy information. Column (1) contains the SN names, columns (2,3) the SN coordinates, (4,5) the newly assigned host coordinates, (6,7) the on-sky separation between the SN and its host in arcsec and kpc, respectively, and (8,9) the brick and object ID that together match to a unique entry in the LS Tractor catalog. 

When more than one source was identified within the matching radius, the brightest source was adopted only if it was at least 2~mag brighter than all other candidate matches. This was done to ensure that, when a very large galaxy is fragmented by the Tractor's deblending algorithm, we chose the aperture containing the majority of the galaxy light, and that the fragmentation had only negligibly reduced the measured galaxy flux. This procedure was needed for only 14 cases in total, twelve for particularly nearby SNe with redshifts $z<0.05$ (which will be more resolved than more distant host galaxies), twice in the SDSS sample, and zero times in all other samples. We now discuss three notable cases that provide useful checks on this quality control procedure.

First, in the case of SN2006qm, the galaxy self-fragmentation flag was tripped and controlled for per the 2~mag threshold. Interestingly, however, this galaxy was not fragmented in the LS DR9 photometry, providing us the means to directly test and validate our countermeasures against self-fragmentation significantly impacting the galaxy photometry. In DR9, the unique source at the host location has a magnitude $g=18.65$~mag and color $g-r = 0.39$~mag. In DR10, the source we have assigned as the host as per the 2~mag rule has a magnitude $g=18.68$~mag and color $g-r = 0.38$~mag, while the secondary (presumably false) nearby source has $g=21.27$~mag and $g-r = 0.17$~mag. So the galaxy photometry with and without the seemingly errant fragmentation is nearly unchanged (0.03~mag difference, corresponding to about 0.01~dex in mass), indicating that the procedure for handling fragmentation of bright, large galaxies is reliable.

In the case of ASASSN-15il, the galaxy light was not fragmented into multiple sources, but the actual SN occurred both during the observing window of the LS imaging campaign and within $1''$ of the galaxy center; fortunately, the same protocol adopted to mitigate self-fragmentation effects works here to also make sure co-temporal SN light does not significantly bias that of the host galaxy. The SN was detected in the LS catalog with an average magnitude of $g=19.2$~mag, and the galaxy $g=15.9$~mag, indicating that the SN was only bright for a small fraction of the LS images and thus contributed little flux to the final, averaged stacks.

\begin{figure*}
    \centering
    \includegraphics[width=0.98\linewidth]{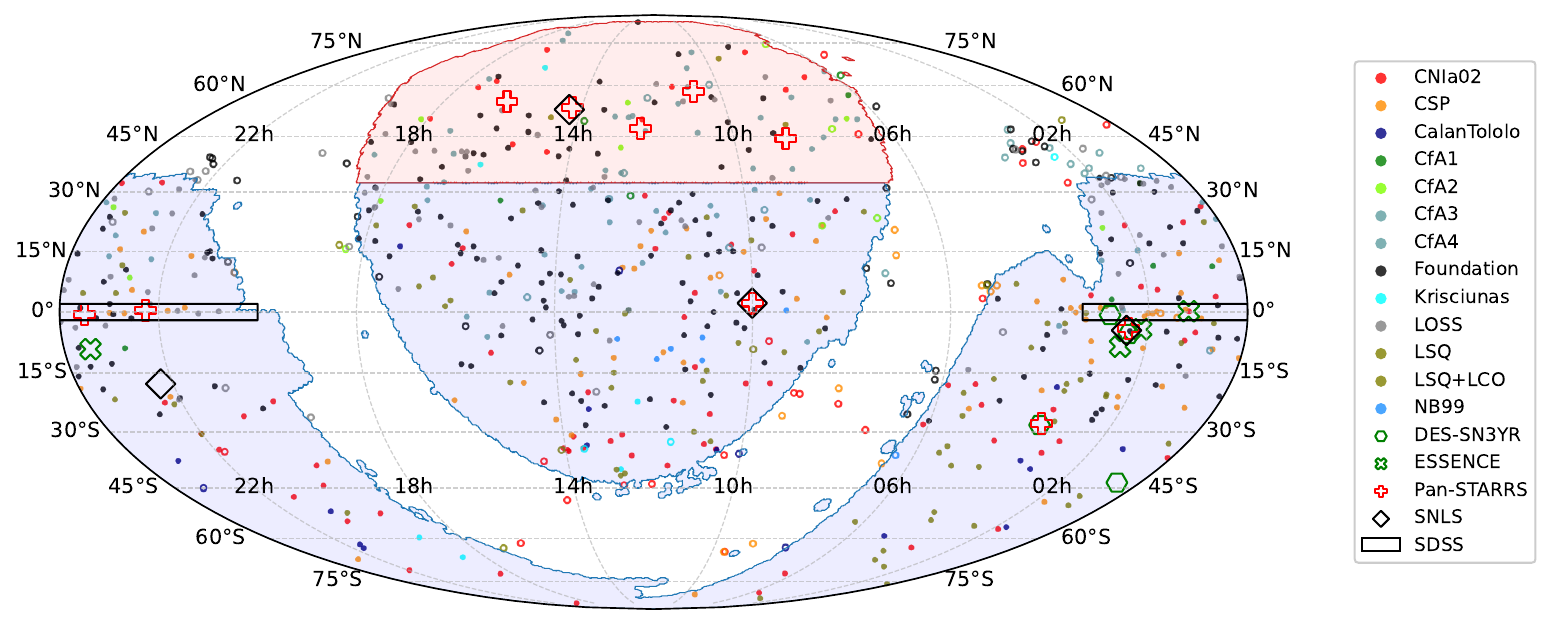}
    \caption{Positions on the sky for the \nTotLow{} low-redshift ($z < 0.15$) SNe in Union3.1 that passed UNITY cosmology cuts (small circle markers). \nLSTPlow{} (\fracLSTPlow{}\%) have been assigned a host galaxy contained in the Legacy Surveys (LS) DR10 Tractor catalog (small filled circles). The majority of the low-redshift SNe for which we did not find a counterpart (small open circles) were not covered by the BASS+MzLS (red-shaded region) or the DECaLS (blue-shaded region) DR10 footprints. In the intermediate redshift bin ($0.15 \lesssim  z \lesssim 0.90$), \nLSTPmid{} of \nTotMid{} (\fracLSTPmid{}\%) SNe were assigned to a source in the LS DR10 catalog. The fields on the sky (not individual SNe) targeted by those intermediate-redshift surveys are marked (large unfilled polygons).}
    \label{fig:skyPos_lowz}
\end{figure*}

Finally, the attempt to assign an LS object as the host of SN2006fw was the only case that failed the 2~mag brighter requirement and so we do not assign to it an entry from the LS Tractor catalog, nor do we attempt to determine a host mass. Upon visual inspection there are indeed two seemingly distinct sources, one redder and one bluer, superposed on each other. This could either be a superposition in the radial direction, or an oversized bulge that is being deblended from a disk. 

In all, we are able to associate a host galaxy in the LS DR10 Tractor catalog to $\sim$2000 Union3 SNe, $\sim$1500 of which pass our cosmology selection cuts (previously defined in \citetalias{Rubin_2025}). For completeness, and to facilitate future analyses that may prefer a different set of selection criteria, we also include in \autoref{tab:host_coords} host galaxy information for $\sim$500 additional SNe in Union3 that did not pass the cosmology selection cuts. For the remainder of this article we will only discuss the cosmology-selected SNe.

In \autoref{fig:skyPos_lowz} we plot in equatorial coordinates the distribution of all SNe observed by a survey we have classified as low-redshift ($z\lesssim0.15$), color-coded per their source survey. The DECaLS and BASS+MzLS footprints included in the LS DR10 release are drawn as blue and red shaded regions, respectively. The locations of the large, untargeted intermediate-redshift surveys compiled in Union ($0.15 \lesssim z \lesssim 0.9$) are marked by open rectangles. It is clear that the homogeneous, aperture-matched OIR photometry of the LS Tractor catalog provides a robust foundation upon which we can build a \textit{Union} of SN host stellar mass estimates, thereby reducing systematics in the correlation of SNe~Ia properties with the masses of their hosts.

Host galaxy photometry in the LS Tractor catalog was found for \nLSTPlow{} of the \nTotLow{} (\fracLSTPlow{}\%) low-redshift SNe and \nLSTPmid{} of the \nTotMid{} (\fracLSTPmid{}\%) intermediate redshift SNe that passed cosmology selection cuts. Incompleteness in the low redshift sample of SNe is driven primarily by the LS DR10 footprint excluding a conservative portion of the Galactic plane. Indeed, of the \nNoLSTPlow{} low-redshift SNe for which we have not identified an LS DR10 counterpart, 72 are outside the DR10 footprint.\footnote{Note LS DR11 is expected to expand closer into the Plane, which should fill in the remaining gaps in our coverage at low redshifts.} In the intermediate redshift sample, the dropouts are a result of the dedicated SN surveys probing much deeper in their respective deep fields than the all-sky imaging of LS.

\subsubsection{Revised Catalog of GALEX Ultraviolet Sources}

We then supplemented the LS DR10 OIR photometry with UV coverage from the Revised Catalog of GALEX Ultraviolet Sources, or GUVcat AIS \citep{Bianchi_2017}, which we queried from Vizier \citep{Bianchi_2017_vizier} via the \texttt{pyVO} TAP module \citep{ref:pyvo}. The matching radius into GUVcat was set to 5~arcsec. During the host identification process, a flag was assigned to a SN if its associated host was located near other potential hosts. These flagged SNe were then inspected by eye on GALEX and unWISE color stacks provided at \url{https://www.legacysurvey.org/viewer}. The corresponding photometry of likely blends in GALEX was then discarded. If after this pruning multiple matches were still found within the $5''$ search radius, the highest SNR source was taken to be the host. This instance only occurred ten times for the $\sim2000$ host galaxies that we searched for a GALEX counterpart and thus negligibly impacts the final dataset of UVOIR photometry. We provide the query used to match into GUVcat in \autoref{app:queries}.

Only 586 (394) of the LS-identified host galaxies included in the UNITY cosmology analysis had a NUV (FUV) counterpart in GALEX. We assume the non-matches to be non-detections and assign them flux values of $0 \pm 1\sigma$, where the $1\sigma$ limit is estimated as one-fifth the quoted $5\sigma$ sensitivities of 19.9 and 20.8~mag \citep{Morrissey_2007}, or 7.96 and 3.48~$\mu$Jy as the upper limit constraints for the FUV and NUV, respectively.

\subsubsection{Photometry of GOODS SN host galaxies}

For the high redshift GOODS SNe \citep{Riess_2007}, which lie beyond the reach of the ground-based LS imaging, we can take advantage of the rich data archive associated with the GOODS fields. We adopt the host galaxy photometry for GOODS SN host galaxies from \citet[][hereafter TC11]{Thomson_2011}. The bandpasses (and corresponding instruments) included in their analysis are: F438W, F606W, F775W, F850LP (HST/ACS/WFC), $JHK_s$ (CFHT/WIRCAM and VLT/ISAAC), $U$ (KPNO/Mayall/Mosaic), the [3.6], [4.5], [8.0]\micron{} bands of Spitzer/IRAC, and the [24]\micron{} band of Spitzer/MIPS. At $z\sim 1$, these bandpasses provide restframe coverage from the UV to the bluest/warmest of dust emission, similar to the coverage we have in our low-redshift host SEDs.

\begin{figure*}[ht!]
    \centering
    \includegraphics[width=0.7\linewidth]{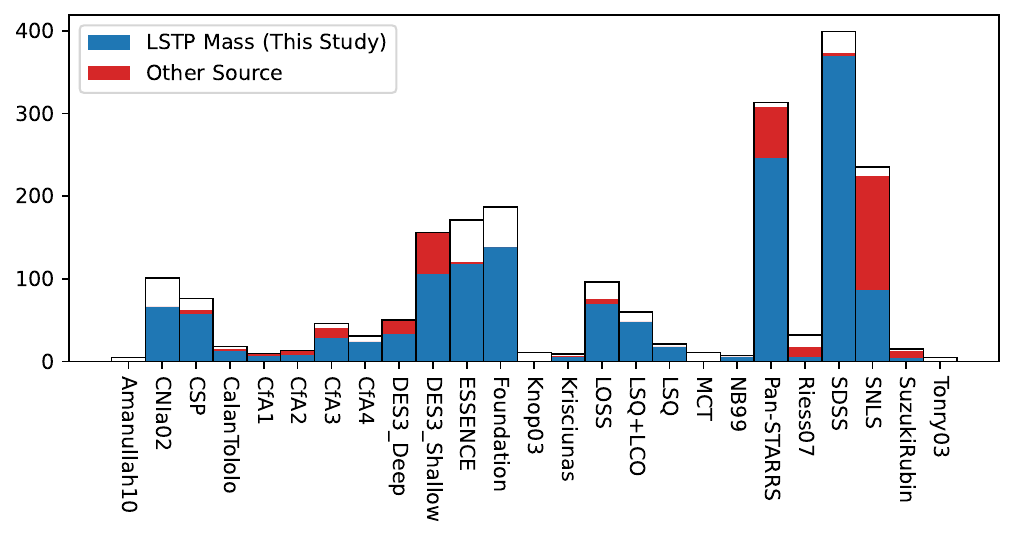}
    \caption{Per-survey breakdown of the number of Union3.1 SNe whose host mass comes from the new Legacy Survey Tractor + \texttt{Prospector} (LSTP) analysis (blue patches), another published source (red patches), or for which we have no host information (clear/white patches). In \autoref{fig:massComp_summary}, we will directly compare the LSTP mass estimates with these Other Sources. The LSTP masses are supplemented with $\sim$300 from other published sources, primarily for the mid-$z$ surveys that probed much deeper in their stare fields than the all-sky LS imaging, particularly in the case of SNLS (DESI LS is about a magnitude shallower in the north than in the south and SNLS is the deepest northern SN survey.}
    \label{fig:massSource_perSurvey}
\end{figure*}

\subsection{Additional Sources of Host Galaxy Information}

We were not able to assign a LS DR10 source as the host for all of the SNe in Union3.1. We therefore supplement our new LS host galaxy catalog with 29 from \citet{Neill_2009} for low-redshift SNe and 8 from \citet{Meyers_2012} for the SCP high-$z$ sample. We also add 137, 66, 62, and 6 hosts from the those published by SNLS, DES, and PS1, and SDSS for their respective surveys. In Section 6, \autoref{fig:massComp_summary}, we will show that these supplementary host mass sources agree with our homogeneously derived ones at the level of 0.2~dex. To our knowledge, of those surveys only SDSS published accurate galaxy coordinates, so we can only supplement our catalog with host masses as given by the others.

\begin{deluxetable*}{lrrrrr}
\tablecaption{Host galaxy statistics per survey. \label{tab:masses_per_survey}}
\tablehead{
\colhead{Survey/group} &
\colhead{$N_{\mathrm{SNe}}$} &
\colhead{$N_{\mathrm{LS}}$} &
\colhead{$N_{\mathrm{other}}$} &
\colhead{$N_{\mathrm{NoHost}}$}
}
\startdata
\noalign{\vskip 2pt}
CalanTololo     &   18 &   14 ( 77.8\%) &    2 ( 11.1\%) &    2 ( 11.1\%) \\
CfA1            &   10 &    7 ( 70.0\%) &    2 ( 20.0\%) &    1 ( 10.0\%) \\
Krisciunas      &    9 &    6 ( 66.7\%) &    2 ( 22.2\%) &    1 ( 11.1\%) \\
CfA2            &   14 &    9 ( 64.3\%) &    5 ( 35.7\%) &    0 ( \phantom{1}0.0\%) \\
CfA3            &   46 &   32 ( 69.6\%) &   11 ( 23.9\%) &    3 ( \phantom{1}6.5\%) \\
CfA4            &   31 &   27 ( 87.1\%) &    0 ( \phantom{1}0.0\%) &    4 ( 12.9\%) \\
CSPI            &   79 &   65 ( 82.3\%) &    4 ( \phantom{1}5.1\%) &   10 ( 12.7\%) \\
SCP Low-z       &    7 &    6 ( 85.7\%) &    0 ( \phantom{1}0.0\%) &    1 ( 14.3\%) \\
LOSS            &   99 &   77 ( 77.8\%) &    7 ( \phantom{1}7.1\%) &   15 ( 15.2\%) \\
Foundation      &  187 &  161 ( 86.1\%) &    0 ( \phantom{1}0.0\%) &   26 ( 13.9\%) \\
CNIa02          &  101 &   80 ( 79.2\%) &    0 ( \phantom{1}0.0\%) &   21 ( 20.8\%) \\
LSQ+CSP         &   21 &   19 ( 90.5\%) &    0 ( \phantom{1}0.0\%) &    2 ( \phantom{1}9.5\%) \\
LSQ+LCO         &   60 &   56 ( 93.3\%) &    0 ( \phantom{1}0.0\%) &    4 ( \phantom{1}6.7\%) \\
\noalign{\vskip 2pt}
\hline
\noalign{\vskip 2pt}
Low-$z$ totals    &  682 &  559 ( 82.0\%) &   33 ( \phantom{1}4.8\%) &   90 ( 13.2\%) \\
\noalign{\vskip 2pt}
\hline
\noalign{\vskip 3pt}
SCP Mid-$z$       &   11 &    3 ( 27.3\%) &    0 ( \phantom{1}0.0\%) &    8 ( 72.7\%) \\
SNLS            &  235 &   87 ( 37.0\%) &  137 ( 58.3\%) &   11 ( \phantom{1}4.7\%) \\
SDSS            &  399 &  375 ( 94.0\%) &    4 ( \phantom{1}1.0\%) &   20 ( \phantom{1}5.0\%) \\
ESSENCE         &  171 &  118 ( 69.0\%) &    2 ( \phantom{1}1.2\%) &   51 ( 29.8\%) \\
Pan-STARRS      &  313 &  246 ( 78.6\%) &   62 ( 19.8\%) &    5 ( \phantom{1}1.6\%) \\
DESY3           &  206 &  139 ( 67.5\%) &   66 ( 32.0\%) &    1 ( \phantom{1}0.5\%) \\
\noalign{\vskip 2pt}
\hline
\noalign{\vskip 2pt}
Mid-$z$ totals    & 1335 &  968 ( 72.5\%) &  271 ( 20.3\%) &   96 ( \phantom{1}7.2\%) \\
\noalign{\vskip 2pt}
\hline
\noalign{\vskip 3pt}
HZT             &    5 &    0 ( \phantom{1}0.0\%) &    0 ( \phantom{1}0.0\%) &    5 (100.0\%) \\
HZT GOODS       &   32 &    6 ( 18.8\%) &   12 ( 37.5\%) &   14 ( 43.8\%) \\
SCP High-$z$      &    5 &    0 ( \phantom{1}0.0\%) &    0 ( \phantom{1}0.0\%) &    5 (100.0\%) \\
MCT             &   11 &    0 ( \phantom{1}0.0\%) &    0 ( \phantom{1}0.0\%) &   11 (100.0\%) \\
SCP Cluster     &   15 &    4 ( 26.7\%) &    9 ( 60.0\%) &    2 ( 13.3\%) \\
\noalign{\vskip 2pt}
\hline
\noalign{\vskip 2pt}
High-z totals   &   68 &   10 ( 14.7\%) &   21 ( 30.9\%) &   37 ( 54.4\%) \\
\noalign{\vskip 2pt}
\hline \hline
\noalign{\vskip 3pt}
Totals          & 2085 & 1537 ( 73.7\%) &  325 ( 15.6\%) &  223 ( 10.7\%) \\
\noalign{\vskip 3pt}
\enddata
\tablecomments{The columns correspond to: (1) name of the sample in Union3, (2) number of SNe that pass the cosmology selection cuts, (3) the number of SNe with a host identified in the LS Tractor catalog, (4) the number of SNe whose host information was compiled from other published sources, in most cases from the same data release that the SN light curves were compiled, (5) the number of SNe that are missing a mass estimate, which UNITY does not distinguish from a ``hostless'' SN. For columns (3-5), the fraction of each relative to column (2) is provided in parentheses as a percentage. Note that many SNe were observed by multiple surveys, but for bookkeeping purposes Union assigns one ``unique'' survey per SN.}
\end{deluxetable*}

\subsection{Summary}

In \autoref{tab:masses_per_survey}, we show the fraction of the host galaxy information compiled for each survey that belongs to three classifications: an object in the LS DR10 Tractor catalog (LS), from some other published study (Other), or we have no host information (which is treated equivalently to a ``hostless'' identification). In all, Union3.1 is only missing host galaxy information for 13.2\%, 7.2\%, and 54.4\% of low-, mid-, and high-redshift SNe, respectively. This amounts to only \fracNoMassThreeOne{}\% of the SNe in Union3.1 missing information about their host galaxies, a marked improvement over the \fracNoMassThree{}\% in the previous Union3 database. The SNe without host mass information in Union3.1 are assigned a default value of $10.0 \pm 1.0$~dex. Furthermore, the majority of the Union3.1 host galaxies will have their properties determined homogeneously, rather than as the heterogeneous mix of systems in Union3. We now discuss the inference of galaxy properties for the Union3.1 SN hosts.

\section{Inferring Host Galaxy Properties with Prospector} \label{sect:prospector}

Inferring properties of a galaxy from broadband photometry relies on modeling the observed spectral energy distributions (SEDs). In this study we will use the \texttt{Prospector} code to infer the properties of the SN host galaxies identified in the previous section \citep{Johnson_2021}. Many of the published masses currently compiled in Union used different SED modeling codes and underlying assumptions such as the adopted IMF or priors on dust extinction, which can lead to significant and oftentimes systematic deviations between galaxy stellar mass estimates. 

For example, if one assumes a Salpeter IMF to infer the mass for some set of galaxy photometry, the returned mass will be on average 0.2~dex larger than if one were to fit the same photometry with either a Kroupa or Chabrier IMF instead, which would agree with each other to 0.03~dex \citep{Lower_2020}. By redetermining the majority of the masses in Union3 using one homogeneous set of photometry and a single set of SFH model assumptions, we will minimize such offsets, which can also manifest as increased dispersion for a large, heterogeneous mix of these assumptions. 

We adopt throughout a delayed exponential (often shortened to delay-$\tau$) plus burst model of star formation \citep{Gavazzi_2002}, which has been shown to perform sufficiently well at capturing galaxy SFHs \citep[see][for a review]{Iyer_2017}. Additionally, \citet{Lower_2020} used \texttt{Prospector} to assess the accuracy and stability of host masses as predicted by different SFH models. Their baseline ``truth'' values were set by galaxies simulated with \texttt{SIMBA} and mock SEDs computed via \texttt{POWDERDAY}. Masses were then inferred from the mocks with one of four different SFH models in \texttt{Prospector}: constant, delay-tau, delay-tau+Burst, and nonparametric. Most pertinent to our analysis, they found the delay-$\tau$+burst model to be systematically offset from truth by $-0.34$~dex with an RMS of 0.11~dex. Crucially, however, this offset in the masses inferred from the delay-$\tau$ model will not propagate into the distance-redshift relation because the host masses will have been derived using the same delay-$\tau$ SFH model. In \autoref{app:priors}, we detail the priors and fixed constraints adopted for the SFH model.

With an SED model in place, we now correct the galaxy photometry compiled in the previous section for foreground extinction as per a CCM89 \citep{Cardelli1989} law. The de-reddened host SEDs were then input to \texttt{Prospector} to infer the parameters corresponding to our adopted delay-$\tau$ plus burst model. 

\begin{figure}
    \centering
    \includegraphics[width=0.99\linewidth]{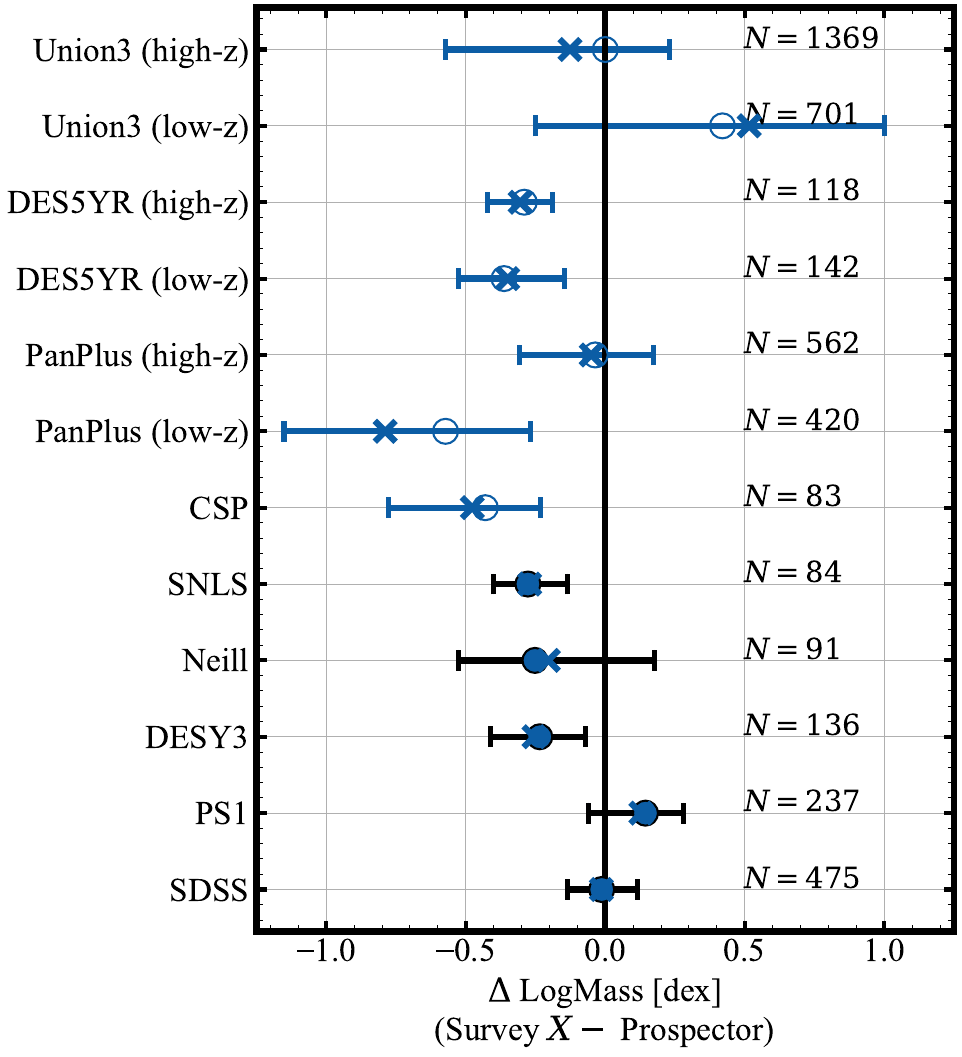}
    \caption{Comparison of the LSTP stellar masses determined in this study with other studies or compilations. Filled circles represent the median mass offset between LSTP and a published mass source that we use to fill gaps in the LSTP masses. X markers represent the mean, while error bars represent the 68th percentile. The error on the mean of each offset is smaller than the size of the marker. The mix of mass sources we use in Union3.1 to supplement the LSTP masses demonstrate generally good agreement at the $\leq 0.25$~dex level. Notably, the Union3 and Pantheon+ compilations exhibit large, redshift dependent offsets of 0.6 to 0.8~dex between their low-redshift and high-redshift samples. The DES-SN5YR masses are systematically offset from LSTP, but are internally self-consistent across redshifts. $N$ is the number of hosts in common between the LSTP sample and the named published catalog, not the number of Union3 SNe from that survey (see text for details).}
    \label{fig:massComp_summary}
\end{figure}

Sampling of the posteriors was performed using the \texttt{dynesty} nested sampler as implemented in \texttt{Prospector} building on the \texttt{demo\_params.py} template script provided in the Prospector repository. To build more robust final posteriors, we increased the target $N_{eff}$ from 10,000 to 30,000, the number of live points from 100 to 400, and the number of walks from 25 to 48. We also decreased the \texttt{nested\_dlogz\_init} and \texttt{nested\_posterior\_thresh} parameters (the latter seems unique to Prospector's implementation of \texttt{dynesty}) from 0.05 to 0.02, as per the \texttt{Prospector} documentation for producing higher quality posteriors. The typical number of effective samples per object was $16901^{+6413}_{-2969}$ (90\% CI), corresponding to a typical sampling ratio $N_{eff}/N_{tot} = 0.735^{+0.046}_{-0.054}$. 

For all fits we inspected, the marginalized posteriors for the mass, dust, metallicity, star-formation decay rate, and the age of the onset of star formation did not resemble the adopted priors. On the other hand, the posteriors for the two parameters associated with the star-forming burst were almost always prior dominated. This is not surprising because spectral line information is typically needed to infer the properties of a burst. In any case, the burst parameters had little to no impact on the final inferred mass. We present example fits for representative galaxies in \autoref{app:sed_fits}, including corner plots to explicitly show parameter correlations.

As pointed out by \citet{Carnall_2019}, the adoption of a parametric SFH model alone imposes strong priors on the complex, real SFHs of observed galaxies. Importantly, however, they find the stellar mass to be the parameter least impacted by prior assumptions ($\leq0.1$~dex variations in stellar mass), and the influence on our relative SN distance measurements is reduced further because we are adopting the same parametric SFH model for all of the host mass estimates. That is, our goal is not to produce stellar mass estimates for SN host galaxies that are accurate in an absolute sense, but that they are determined under a self-consistent set of assumptions under which an \textit{absolute} bias is expected to cancel.

\begin{figure*}
    \centering
    \includegraphics[width=0.95\linewidth]{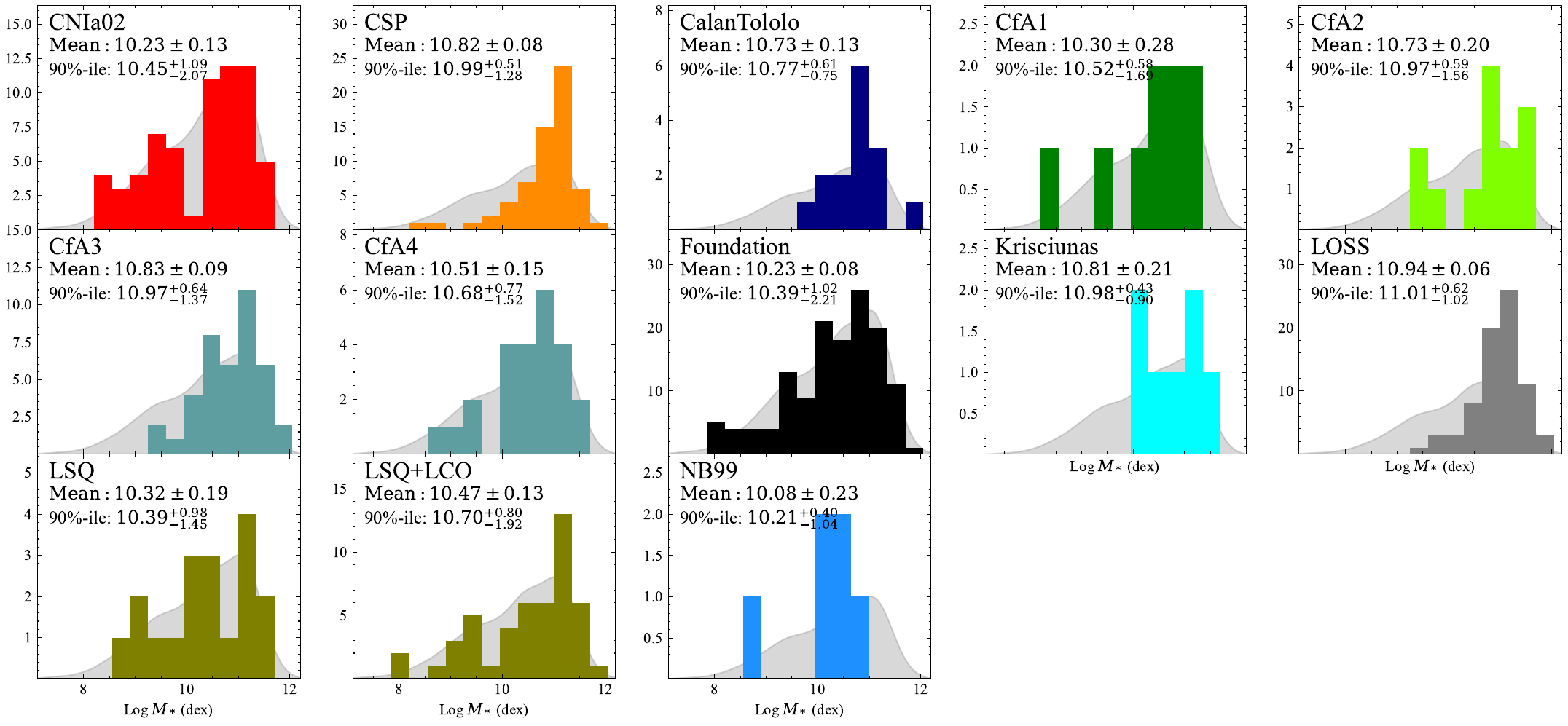}
    \caption{Distributions of host masses for low-redshift $(z<0.15)$ surveys in Union3.1. The bins have width 0.35~dex. The gray shaded curve is a Gaussian KDE representation of the host mass distribution for all SNe in Union3.1. Two summary statistics are included: (1) the mean mass and a bootstrapped error on the mean, (2) the 90th percentile interval of masses. The mass distributions of galaxy-targeted surveys such as CfA, CSP, Krisciunas, and LOSS peak at large values between 10.8 and 11.0~dex, while NB99, LSQ, CNIa02, and Foundation peak at much lower values between 10.2 and 10.4~dex, because those surveys observed SNe discovered by untargeted searches such as ASAS-SN, LSQ, and PSST.
    }
    \label{fig:massHists_lowz}
\end{figure*}

\begin{figure}
    \centering
    \includegraphics[width=0.88\linewidth]{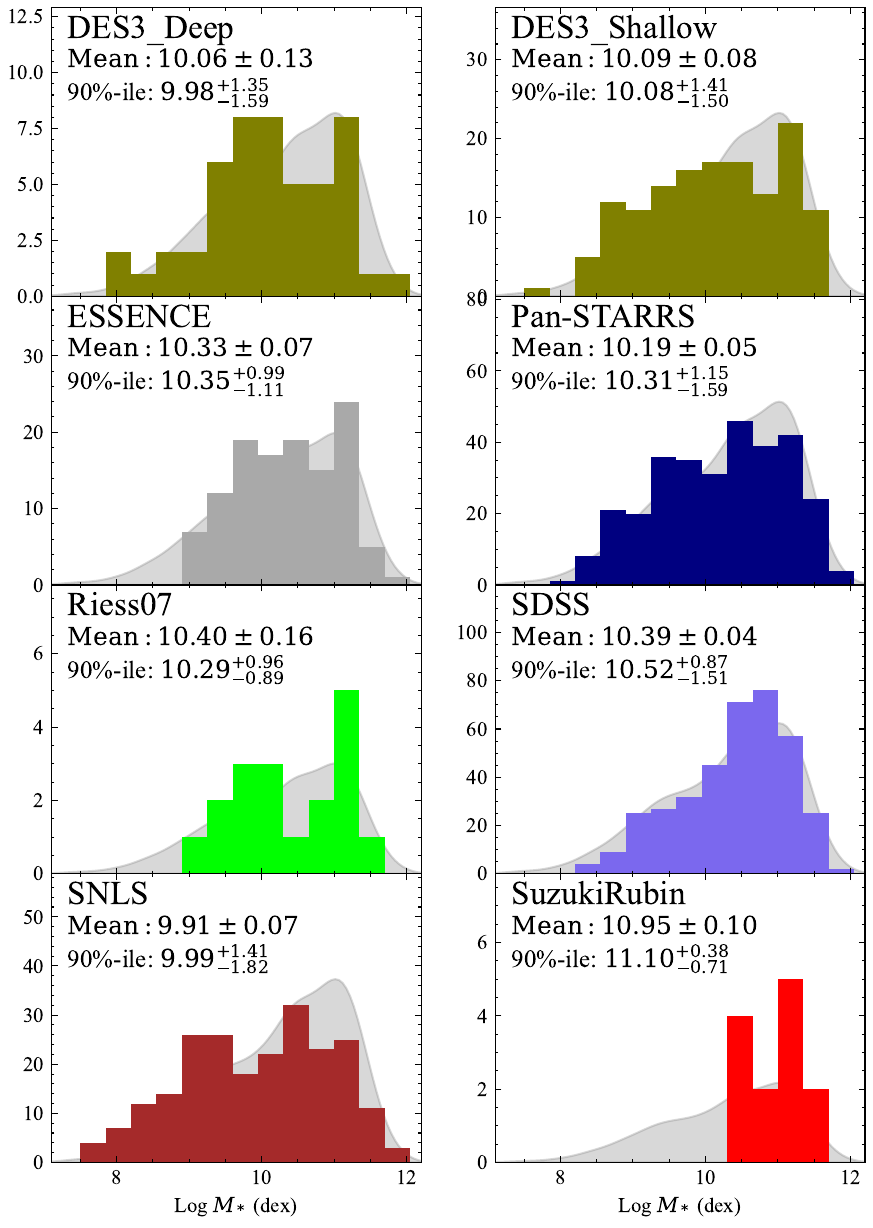}
    \caption{Same as \autoref{fig:massHists_lowz}, but for the $z \geq 0.15$ surveys in Union3.1. HZT (T03), SCP (K03 and N09), SCP (A10), and MCT (R18) are not pictured here because directly determined host masses do not exist for those samples. SNe in the SuzukiRubin sample come from a galaxy cluster search and are thus found exclusively in massive hosts.}
    \label{fig:massHists_interHighz}
\end{figure}
In the end, we determine host galaxy properties for $\sim$2000 SN host galaxies ($\sim$1500 after cosmology selection cuts), which we refer to as the LS Tractor+\texttt{Prospector} (LSTP) sample. As mentioned in the preceding section, we supplement this compilation with $\sim$300 from other published sources, primarily from the mid-$z$ surveys that probed much deeper in their stare fields than the all-sky LS. The sources of each survey's host masses are visualized in \autoref{fig:massSource_perSurvey}, with blue representing masses sourced by this study's LSTP analysis, red for masses from external published sources, and unfilled for SNe with no host information. This demonstrates that the majority of the SNe in Union3.1 have been placed on a homogeneous system of mass estimates. We now examine the properties of the Union SN host galaxies.

\section{The Properties of Union SN Host Galaxies} \label{sect:hostprops}

In this section we explore the SN host galaxy properties derived in the previous section. In 4.1, we compare the new LSTP masses to several external sets of host galaxy mass estimates, including the previous Union3 compilation of masses. In 4.2, we present the final distributions of host masses for surveys contained in the new Union3.1 database. Finally, in 4.3, we will investigate the properties of SNe selected from hosts located in sub-regions of the global sSFR-$M_*$ plane.

\subsection{Internal and External Mass Comparisons}

In \autoref{fig:massComp_summary}, we compare the new LSTP masses with several external sets of published host mass estimates, including the previous Union3 compilation, DES-SN5YR, Pantheon+, CSP \citep{Uddin_2020}, SNLS, \citet{Neill_2009}, DESY3, Pan-STARRS1, and SDSS. The per-SN mass comparisons underlying each point are presented in \autoref{app:mass_compares}. 
Each data point compares the LSTP masses against one published host-mass catalog, and $N$ is the number of hosts common to the LSTP catalog and that external catalog, irrespective of which survey Union assigns the SN to (e.g., \autoref{tab:masses_per_survey}) or whether the SN passes the cosmology cuts. Because the published mass catalogs and Union's survey assignments partition the SNe differently, and several catalogs span many of Union's surveys, the $N$ in this figure are not expected to match, and in general exceed, the per-survey counts of Table 2. For example, the SDSS row contains 475 hosts drawn from \citet{Sako_2018}, whereas Table 2 assigns only 399 cosmology-sample SNe to SDSS (375 with an LSTP mass). Statistics in each row are computed after a conservative $4\sigma$ clip, which removes between 0 and 6 per cent of hosts depending on the catalog.

Filled markers represent comparisons of LSTP with external published sources of masses that we adopt in Union3.1 for SNe without an LSTP mass, i.e., the ``Other'' sources in \autoref{fig:massSource_perSurvey} and \autoref{tab:masses_per_survey}. There is generally good agreement ($<0.2$~dex) between LSTP and the additional sources of mass estimates. 

Open markers represent comparisons to other studies or compilations that are not used anywhere in the new Union3.1 analysis, but make for generally interesting comparisons. In particular, for Pantheon+, Union3, and DES-SN5YR, the three SN cosmology analyses that are being discussed extensively for potentially pointing to dynamic dark energy, we break the samples down into low- and high-redshift subsamples. Doing so allows us to search for any redshift-dependent offsets that would also bias cosmology inference. Indeed, we immediately see that the Pantheon+ and Union3 compilations have large offsets between their low- and high-redshift samples, but in opposite directions. 

The average difference in mass between the high and low-redshift Union3 samples is 0.4 to 0.6~dex depending on the mean or median calculation. The same quantity for Pantheon+ is $-0.5$ to $-0.8$~dex. We will later show that these mass offsets introduced $+0.3\sigma$ and $-0.6\sigma$ biases into the $\Omega_m$ estimates from Union3 and Pantheon+, respectively, and were the dominant source of disagreement over the two analyses' conflicting evidence regarding evolving dark energy.

\subsection{Final Host Mass Distributions}

With the LSTP masses and those compiled from other sources validated at the $\sim$0.2~dex level, we now examine the distribution of host masses for all Union3 SNe for which we have one, which constitutes 89\% of the SNe in Union3.1.

In \autoref{fig:massHists_lowz} and \autoref{fig:massHists_interHighz}, we plot the distributions of host masses for all SNe with an explicit host mass in the Union3.1 compilation, broken down by each survey, for low-redshift and mid-to-high-redshift surveys, respectively. We plot behind each distribution the corresponding distribution of masses for all Union3.1, represented by a Gaussian kernel density estimate (KDE). Note the majority of the low-redshift surveys skew toward high mass on account of being targeted/follow-up surveys. Exceptions include the mass distributions seen for the NB99, LSQ, CNIa and Foundation samples, which targeted SNe discovered by all-sky monitoring surveys such as ASAS-SN, PSST, and LSQ. The masses of the high redshift surveys in \autoref{fig:massHists_interHighz} are more evenly distributed toward lower mass on account of being designed around untargeted, deep ``stare'' fields. The notable exception is the SuzukiRubin sample, which is entirely high mass because that program searched for SNe by targeting rich galaxy clusters. Missing from \autoref{fig:massHists_interHighz} are distributions for the HZT/Tonry03, SCP/Amanullah10, and MCT/Riess18 high redshift SN samples, for which we do not have a host mass for any SN.

\subsection{Correlations between the Properties of SNe and their Hosts}

In this section we will examine more than just the global stellar masses of SN host galaxies in an attempt to better understand how the properties of SNe~Ia correlate with those of their hosts.

\begin{figure}
    \centering
    \includegraphics[width=0.98\linewidth]{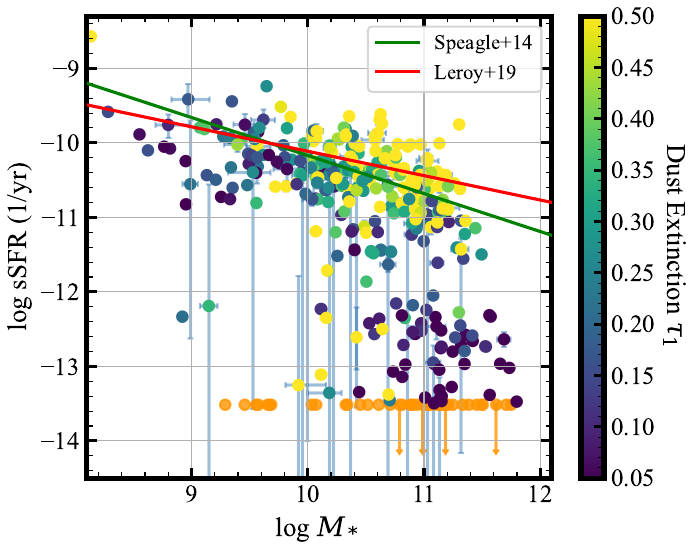}
    \caption{The log of the specific star formation rate (sSFR) plotted against the log of the total stellar mass for the 375 SN host galaxies with full UVOIR wavelength coverage. The median redshift of this sample is $z=0.033$. Points are colored according to the inferred dust content, parametrized by $\tau_1$ in Prospector/FSPS. Two estimates of the slope of the star-forming main sequence at low redshift, from \citet{Speagle_2014} and \citet{Leroy_2019}, are plotted as solid lines. Passive, massive galaxies with little dust content cluster in the lower-right of the plot, clearly separated from the star-forming main sequence that spans the top half of the figure. We assume that values  $\log sSFR < -13.5$ returned by \texttt{Prospector} are best represented by an upper limit on the star formation and plot them as such (orange circles with downward pointing arrows). Error bars are plotted for only every five datapoints to reduce clutter on the plot.}
    \label{fig:galaxy_ms}
\end{figure}

\begin{figure*}[ht!]
    \centering
    \includegraphics[width=0.37\linewidth]{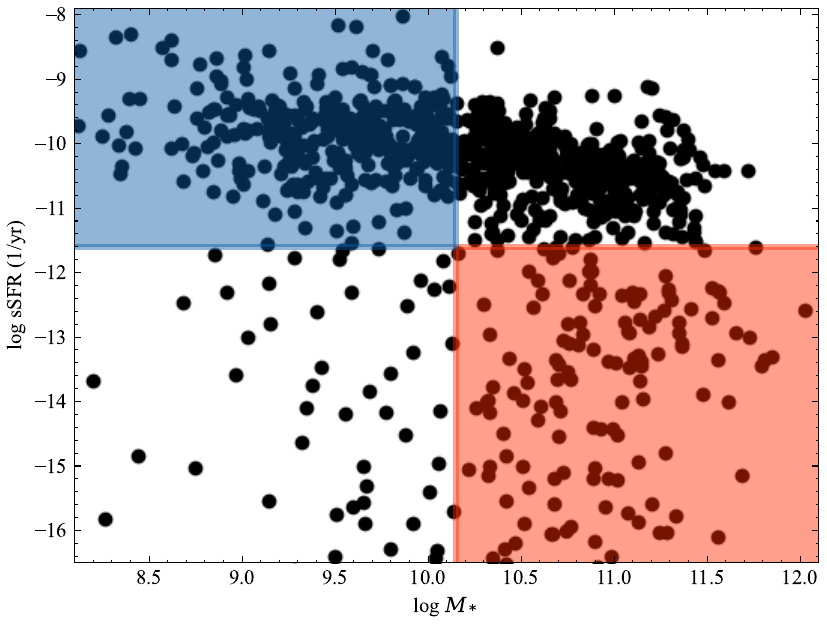}
    \includegraphics[width=0.52\linewidth]{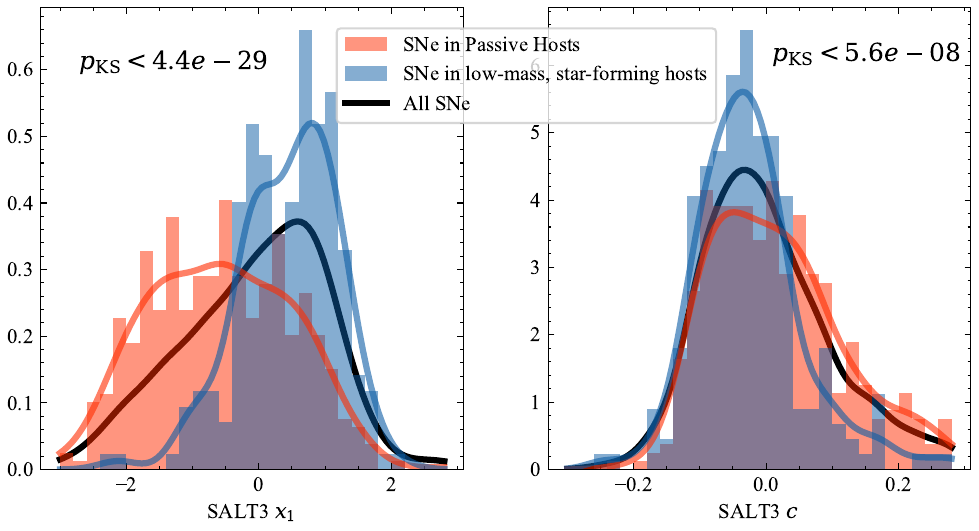}
    \caption{Correlations of SN light curve parameters with the positions of their host galaxies in the sSFR-$M_*$ plane. \textit{Left:} The log of the sSFR in units of yr$^{-1}$ ($\log$~sSFR) is plotted vs. the log of the stellar mass ($\log M_*$) for the LSTP SN host galaxies. Low-mass, star-forming host galaxies are enclosed by a blue-shaded rectangle and massive, quiescent hosts by a red-shaded one. \textit{Middle:} The distributions of SALT3 $x_1$ (light curve width) for SNe hosted by the galaxies selected in the left panel, colored to match their corresponding selection rectangles. The $x_1$ values of SNe hosted by passive galaxies are distributed more negatively and broadly than those hosted by low-mass, star-forming galaxies. \textit{Right:} The same as the middle panel but for the SALT3 $c$ parameter. The colors of SNe hosted by quiescent galaxies are redder and exhibit an extended exponential tail, despite the light from passive galaxies being minimally extincted by dust (see \autoref{fig:galaxy_ms}). This suggests the red colors of these predominantly fast-declining SNe are produced by local effects and not, e.g., interstellar dust situated along the line of sight. The $p$-values from Kolmogorov-Smirnof (KS) similarity tests are printed on the middle and right panels.}
    \label{fig:host_prop_distributions}
\end{figure*}

We begin with \autoref{fig:galaxy_ms}, which plots the log of the specific star formation rate (sSFR) against the log of the total stellar mass $M_*$ for 375 galaxies in our sample that had valid values in all photometric bandpasses from the UV to IR. The median redshift of this sample is $z=0.033$. The passive, massive galaxies are clearly separated from the star-forming galaxy main sequence. We also plot two estimates from the literature for the slope of the star-forming main sequence in the nearby universe, from \citet{Speagle_2014} and \citet{Leroy_2019}, demonstrating good agreement with our results using \texttt{Prospector}. This two-dimensional space is one important projection of the many-dimensional space of correlations occupied by galaxy properties. We aim to use the general astrophysics contained in the sSFR--$M_*$ to guide an attempt to better understand the observed properties of SNe~Ia.

In the leftmost panel of \autoref{fig:host_prop_distributions}, we again plot the sSFR-$M_*$ plane, this time with two sub-regions marked by blue and red shaded rectangles, respectively. The former is intended to select star-forming galaxies on the low-mass end of the SN mass step, while the latter is intended to select passive galaxies on the high-mass end of it. In the middle and right panels, we plot the resulting distributions of the SALT3 $x_1$ (stretch or decline rate) and $c$ (color) parameters when selecting for only those SNe contained within the blue and red selection boxes on the left. As can be seen, SNe that occur in these two distinct types of host galaxies also form very distinct populations in both light-curve parameters $x_1$ and $c$. 

The SNe hosted in massive, passive galaxies have decline rates that are on average much faster than those of SNe hosted by low-mass, star-forming galaxies. The distribution of $x_1$ for SNe in passive hosts has an average $<x_1> \simeq-0.6$ and a standard deviation equal to 1.1. For SNe in low-mass, star-forming hosts, the sample mean is $ <x_1>= 0.6$ with standard deviation $\sigma=0.7$. We evaluate how distinct the two distributions are from each other with a Kolmogorov-Smirnov (KS) test and find a $p$-value equal to $4\times10^{-29}$, corresponding to a significance of $11.2\sigma$, indicating the two samples are not drawn from the same parent population.

In terms of the $c$ distributions, we can see that SNe found in passive, massive galaxies, which should nominally exhibit little to no extinction due to interstellar dust (see \autoref{fig:galaxy_ms}), exhibit an extended tail to red colors. This extended red tail must then be sourced by effects local to the SN, e.g., intrinsic color (which of course cannot be observationally separated from circumstellar dust around the progenitor). That redder SNe are found in ostensibly dust-free environments provides evidence against the hypothesis that the host property dependence observed in SN~Ia standardization can be explained entirely by variations in dust properties \citep[e.g.,][]{Brout_2021}. A more measured interpretation of the apparent dependence of $\beta$ on mass is the possibility of a chance confluence of three factors: (1) within the SALT light curve fitting paradigm, fast-declining ($x_1 < -0.5$) SNe exhibit an exponential distribution that skews toward redder colors, (2) the optical color-luminosity relation of fast declining SNe coincidentally resembles an interstellar-dust-like relation with a nonstandard $R_V$, and (3) fast-declining SNe preferentially occur in high-mass host galaxies. This broad qualification is important to make when attempting to better understand SNe Ia and how we standardize them, which remains a purely empirical endeavor.

We will now describe some minor updates made to the Union database since \citetalias{Rubin_2025} before we formally establish the version 3.1 data release and begin the SN cosmology analyses.

\section{Miscellaneous Updates to Union} \label{sect:otherupdates}

In addition to the new host masses derived above, we describe in this section minor updates and fixes to the Union3 database. The conclusion of this section then formally establishes the Union3.1 compilation.

\subsection{Foreground Reddening}

In Union3, the foreground extinction correction based on the \citet{Schlegel_1998} maps had been accidentally rescaled twice by the 0.86 renormalization factor derived in \citet{Schlafly_2011}. Importantly, however, UNITY, as part of its self-calibration checks, had independently identified this bug and its effect was straightforwardly marginalized out of the cosmology analysis. This update reintroduces into the cosmology analysis five SNe that previously failed the UNITY selection cuts because they had SALT3 $c$ values greater than 0.3, which are now below that threshold due to the larger extinction correction; these are LSQ11ot, LSQ13cwp, SN2017fzy, SN2018chl, and psc040780. In the other direction, one SN, PSN~J21312375+4336312, falls out of the UNITY selection cuts after its foreground reddening $E(B-V)_{MW}$ increased to 0.322~mag.

\subsection{SDSS-SN Host Mass Parsing}
We found a bug in how host masses from the SDSS-SN tables presented in \citet{Sako_2018} were parsed in Union3. This caused the masses of 191 SDSS SNe to swap with other SNe in the SDSS sample and led to no host mass being assigned to 230 SDSS-SN SNe that should have had one. As a result, those 230 were assigned default mass values of $11.0 \pm 1.0$~dex if $z<0.1$ and $10.0 \pm 1.0$~dex if $z>0.1$. By directly comparing with the FSPS masses in the \citet{Sako_2018} table, we can see that this led to a net underestimation of the masses for the entire SDSS sample of $-0.24$~dex ($-0.09$~dex for the 191 and $-0.43$~dex for the 230) relative to what it would have been with the published \citet{Sako_2018} masses. The RMS in the comparison with \citet{Sako_2018} is equal to 0.8~dex. For completeness, we compute the same statistics from comparing the new LSTP masses to \citet{Sako_2018} and find a median difference of 0.01~dex and RMS equal to 0.12~dex. Considering \texttt{Prospector} is an FSPS-based code, and given that we adopted several assumptions used by \citet{Gupta_2011}, the excellent agreement is not surprising.

Therefore, we can conclude that minimal distance bias was incurred as a result of this bug, but significant added dispersion was introduced into the SDSS portion of the Hubble Diagram. Indeed, the unexplained dispersion for the SDSS sample was equal to 0.11~mag in \citetalias{Rubin_2025}, which has decreased to 0.09~mag with the new LSTP masses (now comparable to Pan-STARRS, SNLS, and DES3-Shallow), emphasizing the importance of using accurate mass estimates in SN analyses. Note that, given the consistency between the SDSS FSPS masses and this study's LSTP ones, fixing the parsing bug alone would have yielded the same improvement in the unexplained dispersion of the SDSS Hubble diagram, i.e., this is not an issue with the published SDSS-SN measurements.

\subsection{Previously Published Masses added to Suzuki/Rubin Cluster Search Sample}
For Union3, \citetalias{Rubin_2025} adopted mass values equal to $10.0\pm5.0$~dex for all of the SNe discovered via a high redshift cluster search \citep{Suzuki_2012}. However, \citet{Meyers_2012} published direct mass estimates for ten host galaxies in that sample, which we have now incorporated into Union3.1 along with four for which we derived an LSTP mass. The mean (median) offset in mass for these galaxies between the Union3 default values and those now used in Union3.1 is equal to $0.84$~dex (0.90~dex), with an RMS dispersion of $0.49$~dex.

Fortunately, one of the cluster SNe, SCP05D6, has both an LSTP mass and one from \citeauthor{Meyers_2012}, allowing us to check the two sets of masses are on roughly the same system, to within the uncertainties. For SCP05D6, \citet{Meyers_2012} report a log stellar mass of $11.4 \pm 0.3$~dex and we derived an LSTP value equal to $11.20_{-0.03}^{+0.04}$~dex. The two estimates are in very good agreement, which is notable considering \citeauthor{Meyers_2012} derived theirs from photometry in just two HST bandpasses, and that our LSTP mass estimate was based on optical ground-based, plus unWISE W1 and W2, photometry of this high-redshift galaxy at $z=1.3$.

\subsection{Duplicates}

In Union3, a search for duplicated SN entries was only undertaken for low-redshift ($z<0.1$) SNe. The search was done in R.A., Decl. and redshift space, i.e., a three-dimensional match. The tolerance on differences in redshift was set to a strict $\Delta z \leq 0.001$ because Union3 explicitly requires host galaxy redshifts for these nearby SNe, and galaxy redshifts tend to be much more precise than 1 part in a thousand (closer to 1 in ten thousand).

Upon further inspection, however, we found four additional duplicate SNe in Union3. Three were genuinely missed by the internal Union3 search routine, while the fourth was missed due to a typographical error in one of the source survey tables. We will discuss the fourth in detail in the following paragraph. For the first three, we list their names here, along with their previously unique survey associations, as ordered pairs: (SDSS/SN2005fr, ESSENCE/m001); (SDSS/SN2006mn, ESSENCE/q125); (LOSS/SN2012bh, Pan-STARRS/psc370356). The first two were too distant to have been checked for possible duplicates ($z>0.1$), while the last pair was missed on account of the host redshifts reported by LOSS and Pan-STARRS differing by 0.0013, which exceeded the $\Delta z \leq 0.001$ tolerance for identifying duplicates. These SNe have now been merged for this study, with the PanSTARRS host galaxy information adopted.

The fourth duplicated pair is LOSS/SNF20080514-002 and CfA4/SN2008cm. However, the name SN2008cm is associated with a southern hemisphere SN, which could not have been observed by CfA4 from the Fred Lawrence Whipple Observatory (FLWO). We find that in the tables presented in \citet{Hicken_2012}, SNF20080514-002 has been mislabeled as SN2008cm. This mislabel propagated into the tabulation of the following erroneous quantities in their Table 1: an association with host galaxy NGC~2369, a Milky Way extinction of $E(B-V)= 0.1139$ and a heliocentric redshift $z_{hel} = 0.0111$. The corresponding correct quantities, associated with SNF20080514-002, are: host UGC~8472, MW extinction of 0.033~mag (from SFD98 maps), and $z_{hel} = 0.0219 \pm 0.0001$ (from NED). For this update to Union, we remove the name SN2008cm entirely and merge its associated CfA4 light curve data with the LOSS light curve data of SNF20080514-002. The host redshift for this SN as reported by LOSS is $z=0.021939 \pm 0.000013$, consistent with the value reported in NED \citep{https://doi.org/10.26132/ned1}.

\subsection{ESSENCE updates}

In Union3, the metadata information for ESSENCE SNe, such as redshifts and foreground extinction, were adopted exactly as enumerated in the headers of the light curve data files in their published data release. Now, we update the redshifts to the values published in Table 6 of the paper that accompanied the light curve data release \citep{Narayan_2016}. This updated the host galaxy redshifts of 115 ESSENCE SNe and added new host redshifts for 22 more. The redshift of one SN, s353, shifted significantly from $z_{Gal} = 0.580$ to $z_{Gal} = 0.5956$. For the other 114 SNe, the median, mean, and standard deviation in the residuals of the host galaxy redshift changes were, respectively, $(0 \pm 5)\times10^{-5} $, $(6 \pm 15) \times 10^{-5}$, and $0.0006 \pm 0.0001$, with uncertainties estimated via bootstrapping. We also identified 16 SNe that had their host redshifts mislabeled as SN redshifts in the headers of the light curve files. The story was similar for the 76 SNe that did not have a host galaxy redshift. The median, mean, and standard deviation in the residuals of the changes to the SNID redshifts were, respectively, $(0 \pm 8)\times10^{-5} $, $0.0016 \pm 0.0014$, and $0.012 \pm 0.003$, with uncertainties again estimated via bootstrapping. With an average redshift in ESSENCE of 0.44, these adjustments to the redshifts at the few-tenths of a percent level in RMS, and much smaller in the average, have very little impact on cosmology.

We also update the foreground extinction for the ESSENCE SNe to be queried directly from the SF11 renormalization of the SFD98 dust maps to be consistent with how foreground extinction is compiled for the rest of the Union database. Previously, the parsing of the extinction coming directly from the light curve headers had placed the ESSENCE SNe on the original SFD98 normalization of Galactic extinction. When taking into account the double renormalization discussed in 5.1, this scaled the MW extinction values tabulated for ESSENCE SNe up by $1/0.86^2$ relative to the rest of Union3. This update causes SN d058 to now be masked from the cosmology analysis due to its SALT3 color increasing from 0.288 to 0.300, just outside the UNITY selection cut.

Two of the ESSENCE SNe, n246 and r317, have in Table 6 of \citet{Narayan_2016} SNID redshifts that are anomalously discrepant with the associated host redshift. n246 has a $z_{\mathrm{SNID}} = 0.503 \pm 0.005$ and $z_{\mathrm{gal}} = 0.7055 \pm 0.0005$, while r317 has $z_{\mathrm{SNID}} = 0.736 \pm 0.005$ and $z_{\mathrm{gal}} = 0.3361 \pm 0.0004$. It is likely these discrepant redshifts are due to a chance projection of a background or foreground galaxy close to the location of the SN on the sky, while the actual SN host may be too faint to detect. For now, we retain use of the host galaxy redshifts to remain consistent with how we treat the rest of the ESSENCE sample. In the future, it may be prudent to either drop these two SNe entirely or to switch to their SN-derived redshifts.

Finally, while inspecting the SDSS mass parsing bug discussed earlier in this section, we serendipitously identified ten cross-matches between ESSENCE and the SDSS-SN survey. These cross-matches were previously missed in Union3 because their SDSS data were not included on account of being in their photometric classification sample, and Union explicitly only compiles spectroscopically classified SNe. However, we have the spectroscopic classification from ESSENCE and can now fold in and merge with the SDSS photometry. The newly matched pairs between ESSENCE and SDSS are: (m062, 3825); (q022, 13898); (q069, SN2006md); (q102, SN2006mh); (r185, 16362); (x033, 18942); (x017, 19523); (y137, 20560); (y134, 20882); (y145, 21615), where the SDSS CID is written in cases where an IAU name was not identified.

\subsection{Summary of Minor Updates}

We detailed in this section several minor updates and fixes made to the Union3 database, which we summarize here:

\begin{enumerate}
    \item \textbf{Milky Way Reddening Renormalization.} We fixed an accidental double application of the SF11 renormalization, which only impacted sample selection---and not the actual SN standardization or cosmology---because UNITY marginalized the bug out thanks to an explicit dust map renormalization parameter. This added five SNe to the cosmology sample and removed one.
    \item \textbf{SDSS Mass Parsing.} A bug in the parsing of the SDSS-SN host masses was identified and fixed, though the masses were replaced entirely with new LSTP masses, anyways. This decreased the unexplained dispersion in SDSS from $0.11 \pm 0.01$~mag to $0.09 \pm 0.01$~mag.
    \item \textbf{Masses for Red Sequence Hosts of SCP Cluster SNe.} Published host masses from \citet{Meyers_2012} were incorporated for nine of fifteen SNe in the SCP Cluster sample for which we could not derive an LSTP mass. A single SN overlapped between the \citeauthor{Meyers_2012} and LSTP samples and was used to show the two sets of masses agree to no worse than the $0.3$~dex level. With 13/15 host masses now accounted for, the average mass in this sample increased from 10.0~dex to 10.9~dex.
    \item \textbf{Identification of Duplicated SNe.} Four new duplicated SNe were found in Union3, one of which required correcting an error in the name and subsequent host association of an SN reported by CfA4.
    \item \textbf{Various Updates to ESSENCE Sample.} Metadata for ESSENCE SNe were updated from being parsed out of the headers of the light curve data files to being pulled from Table~6 of \citet{Narayan_2016}, producing minor shifts to the redshifts and foreground extinction values. We also identified ten new matches with SNe in the SDSS sample that were previously excluded from Union due to being photometrically classified by SDSS (but spectroscopically classified by ESSENCE).
\end{enumerate}

At this stage, we have now formally established the next iteration of the Union compilation that incorporates all of the above updates and label it version 3.1.

\section{SN Cosmology Results} \label{sect:sncosmology}

In this section we will use the now-defined Union3.1 database to update the \citetalias{Rubin_2025} SN cosmology analysis. That study used the previous Union3 as input to the hierarchical Bayesian UNITY model that simultaneously solves for parameters that describe, e.g., selection effects for each aggregated survey, the global distributions of SALT light curve parameters, their evolution with redshift, the SN luminosity standardization coefficients, and an adopted cosmological model that can predict a luminosity distance-redshift relation.

We briefly introduce here the foundational equations that describe how SN distance measurements are used as tracers of cosmological information.

The distance modulus is simply the luminosity distance mapped onto a logarithmic scale conventionally defined in astronomy to be, 
\begin{equation}
    \mu = m-M = 5 \log \left(\frac{d_L}{10\mathrm{pc}}\right)
\end{equation}

We then define the comoving distance, simplified here for the case of a flat $\Lambda$CDM universe,
\begin{equation}
    d_{c} = \frac{c}{H_0}\int_z^0 dz \left[\Omega_{\Lambda} + \Omega_m(1+z)^{-3} \right]^{-\frac{1}{2}}
\end{equation}
where $z$ is the cosmological redshift measured relative to the rest-frame of the observable universe, as traced by the CMB, and so is referred to as $z_{\mathrm{CMB}}$. Additionally, we will also consider the definition of the luminosity distance based on the $w(z) = w_0+w_az/(1+z)$ parameterization, which we do not repeat here \citep[see][]{Linder_2003}.

The luminosity distance is then related to the comoving distance as $d_L=d_c(1+z)$ where the redshift $z$ in this case is measured relative to the observer's frame, or the heliocentric redshift $z_{\mathrm{hel}}$.

We then repeat here an equation from \citetalias{Rubin_2025} that encapsulates both how UNITY corrects the brightnesses of SNe to improve their performance as distance indicators and how those corrected distances are used in, e.g., a likelihood/loss function to perform a statistical parameter analysis. The equation used to generate the predicted observable (a brightness) for each SN is,
\begin{align}
    m_{B}^{\mathrm{pred}} =& - \alpha x_{1}^{\mathrm{true}} + \beta_B c_B^{\mathrm{true}} \nonumber \\ 
    +&~[P_{\mathrm{eff}}^{\mathrm{high}} \beta_{R,\mathrm{high}} +  \nonumber (1-P_{\mathrm{eff}}^{\mathrm{high}})\beta_{R,\mathrm{low}}] c_R^{\mathrm{true}} \nonumber \\
    -&~\delta(0)P_{\mathrm{eff}}^{\mathrm{high}} +M_B + \mu(z,\mathrm{{cosmology}}) \label{eq:mb_pred}
\end{align}
where $P_{\mathrm{eff}}^{\mathrm{high}}$, defined explicitly in Equation 7 of \citetalias{Rubin_2025}, is intended to capture possible redshift evolution in both the luminosity host mass step $\delta(0)$, as well as the mass-dependent split of the red color coefficients $\Delta \beta_R$.

Here, the latent representations of observed quantities, $m_{B  }^{\mathrm{pred}}$, $x_{1}^{\mathrm{true}}$, $c_B^{\mathrm{true}}$, $c_R^{\mathrm{true}}$, and $P_{\mathrm{eff}}^{\mathrm{high}}$ are inferred for each SN in UNITY, while the standardization coefficients $\beta_B$, $\beta_{R,\mathrm{high}}$, $\beta_{R,\mathrm{low}}$, $\delta(0)$, and $\delta_h$, as well as whatever parameters describe one's adopted cosmological model, are single-valued for all SNe.\footnote{Note that $M_B$ (and $H_0$) subtracts out of SN cosmology, which is based purely on \textit{relative} distances between SNe. The absolute calibration of $M_B$ via some other means of absolute distance measurement (e.g., Cepheids or tip of the red giant branch) necessarily sets the value of $H_0$, given a set of already-estimated, relative SN distances.}

Before proceeding with the SN cosmology analysis, we first implement an adjustment to the outlier model of UNITY1.5, the version last published in \citetalias{Rubin_2025}, following Appendix~B of that same paper. The fixed model is officially referred to as UNITY1.6. We then add as an explicit fitted model parameter the fiducial mass at which the SN standardization equation bifurcates into either a low- or high-mass regime, denoted $\log M_{\mathrm{step}}$, with red SNe hosted by galaxies on the lower mass side of the step requiring a larger color correction than those hosted by galaxies on the higher mass side, i.e., $\beta_{R, \mathrm{low}} > \beta_{R, \mathrm{high}}$. In previous versions, the location of the host mass step was held fixed at $\log M_{\mathrm{step}} = 10.0$~dex. We also update the UNITY framework to support a determination of the Hubble constant, $H_0$, given an externally determined set of distance moduli vector and covariance matrix. A follow-up study will carry out a blinded, UNITY $H_0$ analysis \citep{TaylorUNITYHubble}. These two changes culminate in what we refer to as UNITY1.7, which we use for the remainder of this article. See \cite{Rubin_x1split} for an expanded discussion, as well as the introduction of a newer, two-population model of $x_1$ standardization as part of UNITY1.8.

\subsection{Parameter Constraints}

We begin with the results for modeling SNe~Ia and their standardization as described by \autoref{eq:mb_pred} before finishing with the SN constraints on $\Omega_m$ for a flat $\Lambda$CDM model. The 68\% confidence intervals for the key parameters are tabulated in \autoref{tab:fit_param_intervals}. Column 1 is the parameter name, and the next three columns present the results for three different variations of the SN analysis: Union3.1 and UNITY1.7 (this study), Union3+UNITY1.7 (the previous data of \citetalias{Rubin_2025} as input to the same UNITY1.7 model we use here), and Union3+UNITY1.5 (the exact combination published by \citetalias{Rubin_2025}). 

Throughout this subsection we will contrast the three sets of results, with an emphasis on the impact of the new LSTP host masses (Union3 vs. Union3.1). We will quote $\sigma$ shifts in parameters relative to the previous constraints' values and uncertainties (which are never smaller than the latest uncertainties), rather than using the uncertainty of the latest, final value (always the same or smaller than previous constraints). Changes in uncertainties will be quoted as simple fractions, e.g., the new uncertainty is 0.8$\times$ the previous one.

For the $x_1$-dependent correction coefficient $\alpha$, we find $\alpha = \alphavalueThreeOne$, which is a $0.5\sigma$ shift from the value found by \citetalias{Rubin_2025}, and a $-0.9\sigma$ shift relative to the intermediate Union3+UNITY1.7 result, indicating oppositely signed shifts to $\alpha$ stem from the updates to the input data and the model. Notably, however, the uncertainty on $\alpha$ decreased from 0.009 to 0.007, or $0.77 \times$ the uncertainty on either of the two Union3 results. This clarifies that the improvement in $\alpha$ standardization comes from the new LSTP masses.

\newcommand{\foutliervalueThreeOne}{\ensuremath{0.017\pm 0.004}\xspace}

\newcommand{\betaRlowvalueOneFivePub}{\ensuremath{4.15^{+0.35}_{-0.29}}\xspace}
\newcommand{\betaRhighvalueOneFivePub}{\ensuremath{3.07^{+0.18}_{-0.16}}\xspace}
\newcommand{\betaBvalueOneFivePub}{\ensuremath{2.59^{+0.43}_{-0.37}}\xspace}
\newcommand{\alphavalueOneFivePub}{\ensuremath{0.168 \pm 0.009}\xspace}
\newcommand{\foutliervalueOneFivePub}{\ensuremath{0.031^{+0.010}_{-0.011}}\xspace}
\newcommand{\OmValueOneFivePub}{\ensuremath{0.356^{+0.028}_{-0.026}}\xspace}
\newcommand{\deltahvalueOneFivePub}{\ensuremath{0.80^{+0.14}_{-0.26}}\xspace}
\newcommand{\deltazerovalueOneFivePub}{\ensuremath{0.031^{+0.019}_{-0.018}}\xspace}
\newcommand{\stepmassvalueOneFivePub}{\ensuremath{\ldots}\xspace}

\newcommand{\foutliervalueThree}{\ensuremath{0.018\pm 0.004}\xspace}

\renewcommand{\arraystretch}{1.12}
\begin{deluxetable}{l|c|cc}
\tablecaption{68\% confidence intervals for key UNITY parameters inferred for flat $\Lambda$CDM from three combinations of recent Union and/or UNITY versions. The Union3.1 column (bolded) contains this study's results. \label{tab:fit_param_intervals}}
\tablehead{
\multicolumn{1}{c|}{Parameter} &
\colhead{\textbf{Union3.1}} & 
\multicolumn{2}{|c}{Union3 (\citetalias{Rubin_2025})} \\
\multicolumn{1}{c|}{} &
\colhead{UNITY1.7} &
\multicolumn{1}{|c}{UNITY1.7} &
\colhead{UNITY1.5}
}
\startdata
$\alpha$                & \alphavalueThreeOne      & \alphavalueThree      & \alphavalueOneFivePub      \\
$\beta_{B}$             & \betaBvalueThreeOne      & \betaBvalueThree      & \betaBvalueOneFivePub      \\
$\beta_{R}$ (low)       & \betaRlowvalueThreeOne   & \betaRlowvalueThree   & \betaRlowvalueOneFivePub   \\
$\beta_{R}$ (high)      & \betaRhighvalueThreeOne  & \betaRhighvalueThree  & \betaRhighvalueOneFivePub  \\
$M_{\mathrm{step}}$     & \stepmassvalueThreeOne   & \stepmassvalueThree   & \nodata                    \\
$\delta_{0}$            & \deltazerovalueThreeOne  & \deltazerovalueThree  & \deltazerovalueOneFivePub  \\
$\delta_{h}$            & \deltahvalueThreeOne     & \deltahvalueThree     & \deltahvalueOneFivePub     \\
$f_{\mathrm{outlier}}$  & \foutliervalueThreeOne   & \foutliervalueThree   & \foutliervalueOneFivePub   \\
$\Omega_{m}$            & \OmValueThreeOne         & \OmValueThree         & \OmValueOneFivePub         \\
\enddata
\tablecomments{Intervals are reported as the median with 68\% credible intervals unless otherwise noted. The location of the mass step $M_{\mathrm{step}}$ was constrained to be 10.0~dex in UNITY1.5.}
\end{deluxetable}

For $\beta_B$, the color correction coefficient for the bluest SNe, we find $\beta_{B}=\betaBvalueThreeOne$. There was no significant change when going from Union3 to Union3.1 in the UNITY1.7 model, indicating that host masses, and any changes made to them, do not impact the standardization of blue SNe. Interestingly, the change from UNITY1.5 to UNITY1.7 yielded a dramatic change to the $\beta_B$ parameter. The central value shifted by $-1.2\sigma$ and the uncertainty became 0.55$\times$ what it was previously. This is due to the fix to the prior on the outlier model described in the Appendix of \citetalias{Rubin_2025} and implemented in UNITY1.7.

For $\beta_{R,\mathrm{low}}$, the color coefficient for red SNe hosted by low-mass galaxies, we find $\beta_{R, \text{low}} = \betaRlowvalueThreeOne$. This value is a $+0.6\sigma$ and $+0.9\sigma$ shift relative to, and has an uncertainty that is 0.72$\times$ and 0.59$\times$ smaller than, the Union3+UNITY1.7 and Union3+UNITY1.5 results, respectively.

For $\beta_{R,\mathrm{high}}$, the color coefficient for red SNe hosted by high-mass galaxies, we find $\beta_{R, \text{high}} = \betaRhighvalueThreeOne$. This constraint is a $+1.2\sigma$ shift relative to, and has an uncertainty that is 0.74$\times$ smaller than, the Union3+UNITY1.5 result. Note, however, that the entirety of this shift in the central value came from the update to the UNITY model, though the new LSTP masses did reduce the uncertainties. 

The location of the mass step was found to be $M_{\mathrm{step}} = \stepmassvalueThreeOne$, 0.11~dex larger than the value of 10.0~dex adopted in UNITY1.5, before it was added as an explicit free parameter. The Union3 and Union3.1 results are identical in both their central values and the full widths of their 68\% intervals. However, the asymmetry of the marginalized constraint was reduced from $^{+0.061}_{-0.040}$ in Union3 to $^{+0.044}_{-0.053}$ in Union3.1, likely due to the new, homogeneously determined LSTP masses replacing the heterogeneous compilation of Union3.

For the size of the luminosity step that remains in the SN~Ia Hubble diagram \textit{after} accounting for a mass-dependent split on $\beta_R$, referred to as $\delta(0)$, we find $\delta(0) = \deltazerovalueThreeOne$~mag. This is a $+0.3\sigma$ shift relative to Union3+UNITY1.5 in \citetalias{Rubin_2025} and equivalent to the Union3+UNITY1.7 result. Similar to the other standardization parameters, the uncertainty decreased sequentially across versions of the analysis, with the nominal Union3.1+UNITY1.7 uncertainty being 0.76$\times$ the size of the corresponding uncertainty in Union3+UNITY1.5.

For the redshift evolution of the host mass dependence we found $\delta(z\rightarrow \infty) = 0.73 \pm 0.21$. The uncertainty on $\delta(z\rightarrow \infty)$ appears to have increased slightly, but this is just from the mode of the posterior pulling further from the upper bound at a value of 1.

Finally, in terms of cosmology we find for flat $\Lambda$CDM a value of $\Omega_m = \OmValueThreeOne$. This is a $0.4\sigma$ decrease relative to the previous \citetalias{Rubin_2025} Union3 study. The majority of this change is caused by the new LSTP masses for low-redshift SNe, which are a significant improvement in terms of accuracy and consistency with the rest of the sample than the values that were previously used. 

\begin{figure}
    \centering
    \includegraphics[width=0.9\linewidth]{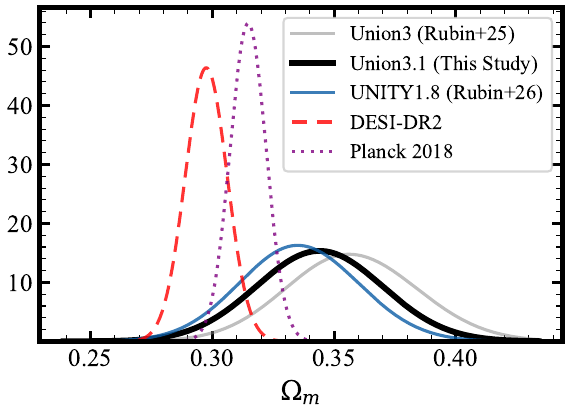}
    \caption{Constraints on $\Omega_m$ for flat-$\Lambda$CDM. The results from three Union+UNITY variants are plotted: this study, referred to as Union3.1 (thick black), the previous Union3 analysis \citep[thin gray;][]{Rubin_2025}, and a new UNITY1.8 model presented in a companion paper \citep[thin blue; see][]{Rubin_x1split}. Also plotted for reference are constraints from DESI-DR2 BAO \citep[red dashed;][]{DESICollaboration2025} and Planck CMB \citep[purple dotted;][]{PlanckCollaboration2020}.}
    \label{fig:om_pdfs}
\end{figure}

In \autoref{fig:om_pdfs} we plot the marginalized constraints on $\Omega_m$ from this study as a thick black curve, from the previous Union3+UNITY1.5 result as a thin gray curve, and from a companion paper that introduces a significant update to the $x_1$ standardization model, referred to as UNITY1.8, as a thin blue curve \citep[see][for details]{Rubin_x1split}. For comparison, we plot the constraints from Planck CMB \citep{Planck_2020_simall} and DESI-DR2 BAO \citep[][shortened hereafter to just \citetalias{DESICollaboration2025}]{DESICollaboration2025} as purple dotted and red dashed lines, respectively. It can be seen that the agreement of Union+UNITY with high redshift probes has improved since 
\citetalias{Rubin_2025}.

A corner plot for all parameters from the Union3.1+UNITY1.7 result discussed in this section is provided in \autoref{app:lcdm_corner}. We do not consider here models beyond flat-$\Lambda$CDM, and we refer the reader to a companion paper for a more expansive exploration of cosmologies using the latest versions of Union and UNITY \citep{Rubin_x1split}.

Finally, as mentioned in Section 4, there is a $-0.6$ to $-0.8$~dex offset in the Pantheon+ masses for their low-redshift SNe relative to their high-redshift ones, similar in size but opposite in sign to the mass offset we saw in Union3. Based on this finding, we update the Pantheon+ low-redshift SNe and align them to the same mass system as their high-redshift sample. The exact procedure, fully consistent with their bias correction approach to the mass step, is described in \autoref{app:panplus_biascorr}. Using this corrected set of Pantheon+ SN distances we re-derive their constraint on $\Omega_m$ and find $\Omega_m=0.342\pm0.018$, a $0.6\sigma$ increase over their original published result of $\Omega_m=0.332\pm0.018$ \citep{Brout2022}, and in excellent agreement with the Union3.1 result.

\subsection{SN Distance Measurement}

\begin{figure}
    \centering
    \includegraphics[width=0.99\linewidth]{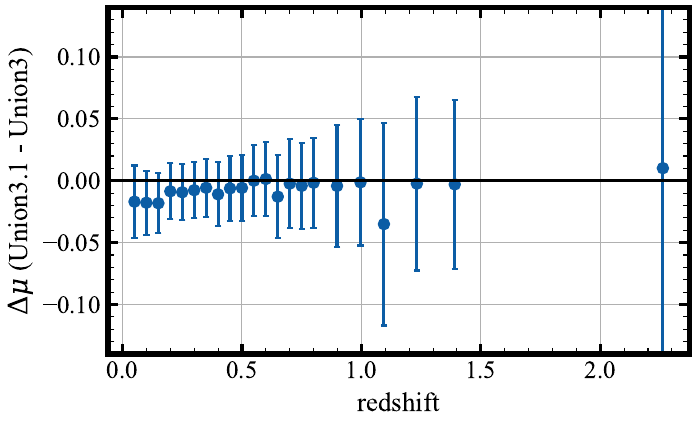}
    \caption{Comparison of distance moduli spline nodes from this study (Union3.1) and the previous Union3 result of \citetalias{Rubin_2025}. Details of the spline representation of the UNITY SN measurements are provided in \citetalias{Rubin_2025}. The offset in distance between the two Union analyses at low redshifts is due to the mass offset discussed in Section 4 that also drives the shifts seen in cosmological parameter estimates.}
    \label{fig:delta_dist_mod}
\end{figure}

With the SN analysis completed, including generation of the binned distance moduli as per the prescription presented in \citetalias{Rubin_2025} for Union3, we can now directly inspect how the SN distance measurements have changed in Union3.1.

In \autoref{fig:delta_dist_mod} we plot the difference in distance modulus between the Union3.1 and Union3 analyses as a function of redshift for the 22 spline nodes that both analyses generated in an identical way. This also allows for a 1:1 comparison of the two sets of distance moduli. The offsets seen in the lowest redshift bins drive the shift in $\Omega_m$ seen in the preceding section relative to the previous Union3 analysis. As we will see, this shift in the low-redshift distances also impacts the combined constraints on $w_0$--$w_a$ and the tentative evidence for evolving dark energy.

\section{Multi-probe Cosmological Constraints} \label{sect:combined}

In this section, we take the updated SN distance measurements and combine them for a multi-probe cosmology inference. We carry out this analysis in exactly the same way as done by \citetalias{DESICollaboration2025} for their nominal quoted BAO+CMB+SN results. This allows us to trace exactly how the new Union+UNITY analysis impacts cosmological inference, particularly in the $w_0$--$w_a$ parametrization of a time-varying dark energy equation of state. 

Specifically, the Planck CMB likelihoods adopted are: \texttt{Commander} for low-$\ell$ temperature \citep{Planck_2020_commander}, \texttt{simall} for low-$\ell$ polarization \citep{Planck_2020_simall}, and \texttt{CamSpecNPIPE} for high-$\ell$ \citep{Efstathiou_2021, Rosenberg_2022}. Also included is a combined Planck+ACT DR6 lensing power spectrum \citep{actdr6} and, of course, the \citetalias{DESICollaboration2025} BAO measurements.
For the SN data, we combined with either the new Union3.1+UNITY1.7 or the corrected Pantheon+ measurements presented in this study, as well as the new DES-Dovekie results \citep{Popovic_2025arXiv250605471P}. Note that Union and Pantheon+ share over 75\% of their SNe, so differences in their final parameter estimates are much more significant than the naive statistics would suggest.

We fit to the joint data only the commonly used $w_0w_a$ parametrization \citep{Linder_2003}, assume flatness, and sample the posterior using the \texttt{Cobaya} code \citep{Torrado_2021}. We note that testing alternative constructions of a time-evolving dark energy equation of state beyond $w_0w_a$ is beyond the scope of this SN-focused study, though it has been shown that some alternative classes of models of a time-varying $w(z)$ produce weaker tension with a cosmological constant \citep[see, e.g.,][]{desi_extended}. We define convergence to have taken place once the Gelman-Rubin statistic $\hat{R}$ falls below 1.01, again identical to that adopted in the \citetalias{DESICollaboration2025} combined probe analysis. The \texttt{Cobaya} code was also used compute the $\Delta\chi^2$ deviations from $\Lambda$CDM. The datasets, likelihoods, and fitted cosmological model were not chosen out of preference, but were intended to replicate exactly the choices made in \citetalias{DESICollaboration2025}, so we could isolate the contribution of SNe to the joint constraints.

\begin{figure}
    \centering
    \includegraphics[width=0.99\linewidth]{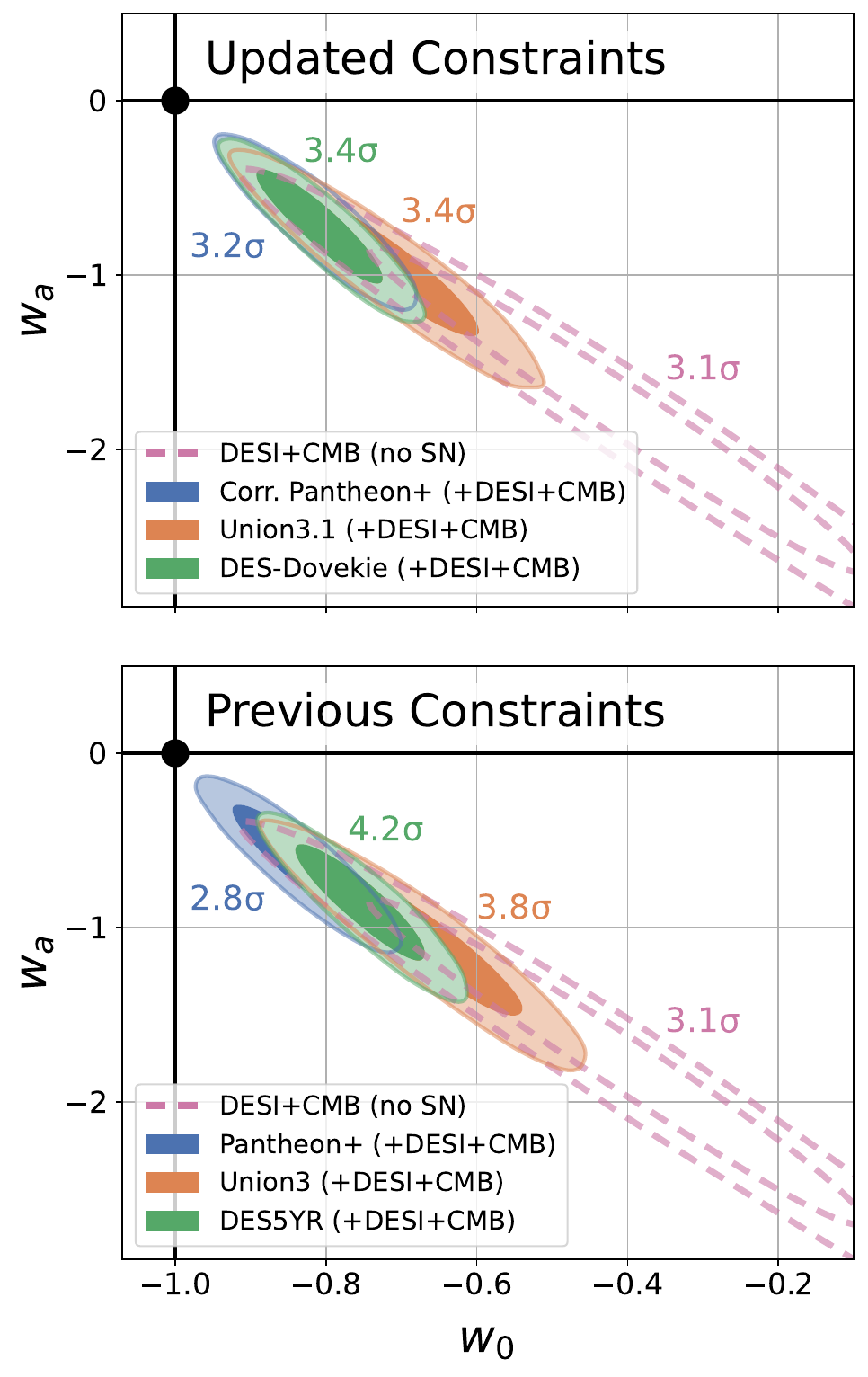}
    \caption{Contours in the $w_0$--$w_a$ plane based on the joint constraints from combining DESI DR2 BAO and Planck+(ACT Lensing) CMB with SNe. The cosmological constant case is marked by a black filled circle in both panels. \textit{Top:} $1,2\sigma$ contours when combining with the newly corrected Pantheon+ (blue) or Union3.1 (orange) SN distances presented in this study, or with the new DES-Dovekie SN measurements (green) presented in \citet{Popovic_2025arXiv250605471P}. The same for BAO+CMB alone is also plotted to represent the exact published \citetalias{DESICollaboration2025} results (purple unfilled and dashed ellipse) \textit{Bottom:} Original CMB+BAO+SN constraints published in \citetalias{DESICollaboration2025} for Pantheon+ (blue), Union3 (orange), and DES-SN5YR (green). The  BAO+CMB constraint is repeated. The same 75\% of the SNe in Union3 are also contained in Pantheon+, so the previous disagreement of $1\sigma$ when treating the uncertainties independently is larger if accounting for the SN data in common. The improved consistency among SN analyses is readily apparent; the range of evidence for $w_0w_a$ tightened from 2.8--4.2$\sigma$ to just 3.2--$3.4\sigma$.}
    \label{fig:desi_combined_cobaya}
\end{figure}

In \autoref{fig:desi_combined_cobaya} we present the 2$\sigma$ contours in the $w_0$-$w_a$ plane. In the top panel, we plot the new combined probe results that include Union3.1 (orange contour), the corrected Pantheon+ (blue contour), and DES-Dovekie (green contour). For reference, we include the DESI+CMB constraint without SNe (pink-purple dashed), to directly show the impact of including SN measurements. In the bottom panel, we reproduce the contours originally reported by \citetalias{DESICollaboration2025}, highlighting the $2.8\sigma$--$4.2\sigma$ discrepancies when combining with those earlier versions of the SN distance measurements. Written in the plot is the significance of each contour's deviation from the $\Lambda$CDM hypothesis, defined as $(w_0,w_a)\equiv(-1,0)$ and plotted as a black dot. 

In the case of Union3.1, the joint constraint on $w_0$-$w_a$ has moved toward $\Lambda$CDM. The $\chi^2$ distance from a cosmological constant is estimated to be $3.4\sigma$, a decrease from the $3.8\sigma$ reported by \citetalias{DESICollaboration2025}. 

In the case of Pantheon+, on the other hand, the joint constraint has moved further from $\Lambda$CDM and into closer agreement with CMB+DESI and Union. The $\chi^2$ distance from a cosmological constant is estimated to be $3.2\sigma$, an increase from the $2.8\sigma$ reported by \citetalias{DESICollaboration2025}.

The combination with DES-Dovekie moved significantly closer to $\Lambda$CDM and also into much better agreement with the other probes and SN combinations. The $\chi^2$ distance from a cosmological constant is estimated to be $3.4\sigma$, down from the $4.2\sigma$ reported by \citetalias{DESICollaboration2025}.
Note our reported constraints from the DES5YR-Dovekie combined-probe result differ from those reported originally in \citet{Popovic_2025arXiv250605471P} due to differing adopted CMB assumptions and datasets. In this study, we adopted the exact CMB constraints as adopted by \citetalias{DESICollaboration2025} for all combined constraints, to ensure any differences are due to the different SN analyses, and not from mixing different CMB assumptions into the comparison between SN analyses.

We additionally carry out the combined probe analysis with UNITY1.8, which was considered in the previous section for the SN-only $\Omega_m$ constraint. That updated UNITY model is presented in a companion paper \citep{Rubin_x1split}.

{\renewcommand{\arraystretch}{1.15}
\begin{deluxetable*}{ll|llll}
\tablecaption{Comparison of cosmological constraints from different SN analyses. \label{tab:sn_compare}}
\tablehead{
\colhead{Sample} &
\colhead{SN-only} & 
\multicolumn{4}{c}{SN+BAO+CMB} \\
\colhead{} &
\colhead{$\Omega_m$} &
\colhead{$w_{0}$} &
\colhead{$w_{a}$} &
\colhead{$\Delta\chi^{2}$} &
\colhead{$n_{\sigma}$}
}
\startdata
\hline
\noalign{\vskip 3pt}
\multicolumn{2}{c|}{\textbf{Original SN Results}} & 
\multicolumn{4}{c}{\textbf{Combined Constraints from \citetalias{DESICollaboration2025}}}\\
\noalign{\vskip 3pt}
\hline
Union3 \citep{Rubin_2025} & $0.356^{+0.028}_{-0.026}$ & $-0.667\pm 0.088$ & $-1.09^{+0.31}_{-0.27}$ & $-17.4$ & $3.8$ \\
Pantheon+ \citep{Brout2022} & $0.332\pm 0.018$ & $-0.838\pm 0.055$ & $-0.62^{+0.22}_{-0.19}$ & $-10.7$ & $2.8$ \\
DES-SN5YR \citep{DES5YR} & $0.352 \pm 0.017$ &  $-0.752\pm 0.057$ & $-0.86^{+0.23}_{-0.20}$ & $-21.0$ & $4.2$ \\
\hline
\noalign{\vskip 3pt}
\multicolumn{2}{c|}{\textbf{Updated SN Results}} & 
\multicolumn{4}{c}{\textbf{Updated Combined Constraints}}\\
\noalign{\vskip 3pt}
\hline
\textbf{Union3.1 (LSTP host masses)} & $0.344\pm 0.026$ & $-0.719\pm 0.084$ & $-0.95^{+0.29}_{-0.26}$ & $-14.4$ & $3.4$ \\
\textbf{UNITY1.8 (two-population $x_1$ model)} & $0.335^{+0.025}_{-0.024}$ & $-0.735\pm 0.081$ & $-0.92^{+0.29}_{-0.25}$ & $-14.0$ & $3.3$ \\
\textbf{Pantheon+ (corrected low-$z$ masses)} & $0.342^{+0.017}_{-0.020}$ & $-0.813\pm 0.055$ & $-0.68^{+0.22}_{-0.19}$ & $-13.2$ & $3.2$ \\
DES-SN-Dovekie \citep{Popovic_2025arXiv250605471P}\tablenotemark{a} & $0.330 \pm 0.015$ & $-0.806\pm0.055$ & $-0.73^{+0.23}_{-0.20}$ & $-14.3$ & 3.4\tablenotemark{b} \\
\enddata
\tablecomments{Cosmological constraints from the three primary SN cosmology experiments are presented in the top block of the table as published and in the bottom block of the table after recent updates---three of which were produced in this study and are \textbf{bolded}. A vertical line separates cosmological constraints inferred from SN measurements alone and a flat-$\Lambda$CDM universe (left) from those inferred by combining with CMB and BAO measurements in a flat-$w_0$-$w_a$ universe (right). In all cases the $w_0$-$w_a$ constraints quoted have been inferred under identical assumptions regarding the external probes. $(w_0,w_a)=(-1,0)$ corresponds to a cosmological constant.}
\tablenotetext{a}{Note that if just the bug in the SNANA implementation of the F99 reddening law identified in \citet{Popovic_2025arXiv250605471P} is corrected, and the Dovekie recalibration is \textit{not} applied (i.e., the original photometric calibration), the DES-SN5YR results became $\Omega_m = 0.369 \pm 0.017$ for SN-only in flat $\Lambda$CDM, a $+1.0\sigma$ increase over the original DES-SN5YR result. The Dovekie recalibration then shifted the DES-SN5YR constraint on $\Omega_m$ by $-2.3\sigma$, in the opposite direction of the SNANA bug fix, resulting in the net $-1.3\sigma$ shift to the DES-SN5YR result that is captured in this table.}
\tablenotetext{b}{Note that \citet{Popovic_2025arXiv250605471P} do not report $w_0$-$w_a$ constraints under the same assumptions used by \citetalias{DESICollaboration2025}. We have, however, re-run the DESI joint constraint analysis for consistency with the rest of this table, hence the discrepancy with the 3.2$\sigma$ significance level quoted in \citet{Popovic_2025arXiv250605471P}. We further note that this 0.2$\sigma$ stronger tension under the \citetalias{DESICollaboration2025} assumptions vs. those adopted by \citeauthor{Popovic_2025arXiv250605471P} is exactly consistent with the ``benchmarking'' tests undertaken in Section 11.1 of \citet{Popovic_2025arXiv250605471P}, which reported a 4.0$\sigma$ tension using the original DES-SN5YR and their choice of CMB assumptions, relative to the 4.2$\sigma$ reported by \citetalias{DESICollaboration2025} and their choice of CMB assumptions.}
\end{deluxetable*}
}

As can be seen from \autoref{tab:sn_compare}, these recent updates and corrections to SN measurements have brought SN cosmology into improved agreement. The full range of $\Omega_m$ constraints has shrunk by a factor of two, and the evidence against $\Lambda$CDM when combined with DESI BAO and Planck CMB has tightened from 2.8--4.2$\sigma$ to 3.2--3.4$\sigma$. Considering the three SN analyses (Union, DES, Pantheon) share much of the same data for their low-redshift measurements, the excellent consistency now attained is reassuring.

At this stage, there appears to be a growing consensus among SN cosmology analyses, though the estimation of uncertainties remains a topic of active discussion \citep[see, e.g.,][]{Rubin_2025}. It is important that we continue to carry out independent analyses of the same data, as well as collect data of multiple different cosmological probes, in order to identify new, and better quantify known, systematics going forward. As previously mentioned, photometric calibration remains the largest systematic uncertainty in SN cosmology experiments and more conservative uncertainty estimation is warranted, as suggested by the findings of \citetalias{Rubin_2025}. It is also notable, however, that host galaxy stellar mass estimates, an often underappreciated part of SN analyses, could bias cosmological inference at a level comparable to something as fundamental as photometric calibration.

\section{Discussion}

\begin{table*}
\centering
\caption{Host-galaxy mass measurement variants. Each variant modifies a single
aspect of the baseline Prospector analysis to test the sensitivity of the
inferred host masses as a function of redshift.}
\label{tab:mass_variants}
\setlength{\tabcolsep}{5pt}
\begin{tabular}{lll}
\hline
\hline
\parbox[t]{0.15\textwidth}{\textbf{Variant}} &
\parbox[t]{0.32\textwidth}{\textbf{Change from baseline}} &
\parbox[t]{0.44\textwidth}{\textbf{What does this variant test?}} \\
\hline
\parbox[t]{0.15\textwidth}{No rest-frame UV} &
\parbox[t]{0.32\textwidth}{
All rest-frame UV constraints on the galaxy SED are excluded.
} &
\parbox[t]{0.44\textwidth}{
Heterogeneous rest-frame wavelength coverage. At low redshift, galaxies often have strong GALEX constraints, while at high redshift the observed $g$ band shifts into the rest-frame UV. Intermediate redshifts can lack either constraint. This variant therefore tests whether redshift-dependent UV coverage introduces a redshift-dependent systematic in the inferred masses.
} \\
\hline
\parbox[t]{0.15\textwidth}{Emissivity of warm dust} &
\parbox[t]{0.32\textwidth}{
The warm-dust emissivity parameter $\Gamma$, referred to as \texttt{duste\_gamma} in FSPS and Prospector, is fixed to its default value of 0.01.
} &
\parbox[t]{0.44\textwidth}{
Adopted coefficient of warm dust emission. The
baseline analysis adopts a value of this parameter, motivated by visual inspection of the SED pulls as described in the Appendix.
} \\
\hline
\parbox[t]{0.15\textwidth}{Salpeter IMF} &
\parbox[t]{0.32\textwidth}{
A Salpeter IMF is adopted in place of the baseline Kroupa IMF.
} &
\parbox[t]{0.44\textwidth}{
Tests sensitivity to one of the largest astrophysical assumptions affecting
the absolute stellar-mass scale. In particular, this variant tests whether (1) an IMF-driven shift is comparable to the mass offsets observed in
Pantheon+ and Union3, and (2) a redshift-independent shift in the mass scale is absorbed by $M_{\mathrm{step}}$ and therefore cancels from cosmological inference.
} \\
\hline
\parbox[t]{0.15\textwidth}{Metallicity prior} &
\parbox[t]{0.32\textwidth}{
The physically motivated metallicity prior used in the baseline analysis is
replaced by a prior that is flat in log metallicity.
} &
\parbox[t]{0.44\textwidth}{
Tests sensitivity to the custom metallicity prior described in the Appendix,
which peaks near the Milky Way metallicity and has a slowly declining tail
toward metal-poor values, relative to an uninformed prior.
} \\
\hline
\end{tabular}
\end{table*}

\begin{figure*}[ht!]
    \centering
    \includegraphics[trim=0cm 0cm 0cm 1.7cm, clip=true, width=0.90\linewidth]{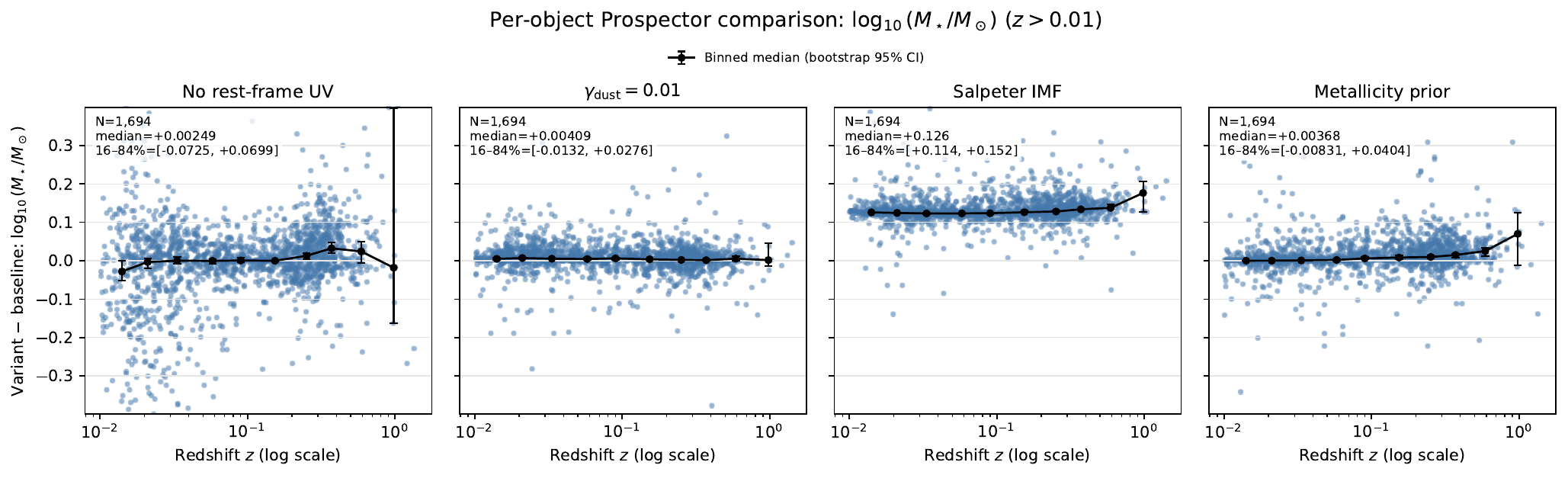}
    \caption{Mass measurement variant analysis results. Each of the four mass measurement variant analyses is compared to the baseline LSTP masses as a function of redshift. No measurable trend with redshift is observed.}
    \label{fig:variant_vs_redshift}
\end{figure*}

We saw above that redshift-dependent inconsistencies in SN host galaxy mass estimates led to biases in cosmological parameters of about $+0.4\sigma$ and $-0.6\sigma$ in $\Omega_m$ for the Union3 and Pantheon+ studies, respectively. In this section we expand on the source of these host mass systematics and show that they are not generically inherent to the measurement of, nor are they unavoidable when using, galaxy properties for SN distance standardization. We will show in this section that those large redshift-dependent discrepancies were in fact caused by internally inconsistent methodological choices.

To clarify this distinction between real, astrophysical limitations in measuring absolute galaxy masses and avoidable, methodological inconsistencies, we carried out four distinct re-measurements of all SN host galaxy masses in our LSTP sample, which we refer to as mass measurement variants. Each variant modifies a single key assumption in our adopted Prospector SED model. First, we tested the impact of dropping rest-frame UV coverage from the model entirely. The second mass measurement variant tested the impact of changing the dust emissivity parameter $\Gamma$ from our adopted, deliberately-elevated value of 0.05, to the FSPS default value of 0.01 (see Appendix \ref{app:dustegamma} for a detailed description). The third swapped our adopted Kroupa IMF \citep{Kroupa_2001} with a Salpeter one \citep{Salpeter_1955}. The fourth dropped our informed metallicity prior (described in Appendix \ref{app:metalprior}) in favor of an uninformed prior, i.e., a Jeffries prior which is uniform in the logarithm of the metallicity $Z$. We enumerate and describe each in more detail in \autoref{tab:mass_variants}.

\autoref{fig:variant_vs_redshift} summarizes the results of the mass measurement variant tests. Each panel from left to right plots the difference of each variant's set of mass estimates with the baseline Union3.1 masses as a function of redshift. This is the fundamental pair of axes that would diagnose if there is a potential systematic in the host galaxy mass estimation that could also propagate to the derived values for cosmological parameters. As can be seen, there is no redshift-dependent trend between any of the five sets of global stellar masses. The only notable shift from zero is in the Salpeter IMF variant, which exhibits a +0.13~dex increase in all stellar masses with a very tight dispersion of 0.02~dex. This is in agreement with expectation (see our discussion in Section 3), though our systematic shift is smaller (0.13~dex) than found by \citet{Lower_2020} which also used Prospector (very likely due to differences in input data such as spectra vs. photometry or differing wavelength coverage). Since this was the only variant that produced a measureable shift to the Union3.1 mass scale, we passed this set of masses to the full UNITY SN cosmology analysis.
  
We re-ran both the SN-only, $\Lambda$CDM (Section 6) as well as the multi-probe $w_0w_a$ analysis (Section 7), replacing only the baseline Union3.1 host masses with those from the Salpeter IMF mass measurement variant. The chains sampled to the same level of convergence as the baseline UNITY runs (all parameters having $\hat{R} < 1.01$). In both cosmology fits, we found the best-fit step mass parameter $M_{\mathrm{step}}$ increased by 0.133~dex, which is completely consistent with the directly computed median offset of 0.126~dex between the two sets of masses. This also highlights the importance of including the location of the mass step as a free parameter in SN cosmology analyses, i.e., $M_{\mathrm{step}}$. The next largest parameter shift was a $-0.28\sigma$ shift in $\beta_{R,low}$, and the sizes of all other shifts were $<0.2\sigma$. 

For the cosmological parameters, we saw for $\Omega_m$ in flat $\Lambda$CDM a shift of +0.0021, or $+0.08\sigma$. For the multi-probe constraints on flat $w_0w_a$CDM, we saw $\Omega_m$ shift by only $+0.0003$ ($+0.04\sigma$), $w_0$ by $+0.003$ ($+0.02\sigma$), and $w_a$ by $-0.01$ ($-0.02\sigma$). The very small impact on cosmology underscores that the derivation of ``true'' galaxy properties in an absolute sense is not directly relevant to SN~Ia cosmology, which only requires that the host galaxy properties be derived self-consistently as a function of redshift.

\begin{figure}
    \centering
    \includegraphics[width=0.9\linewidth]{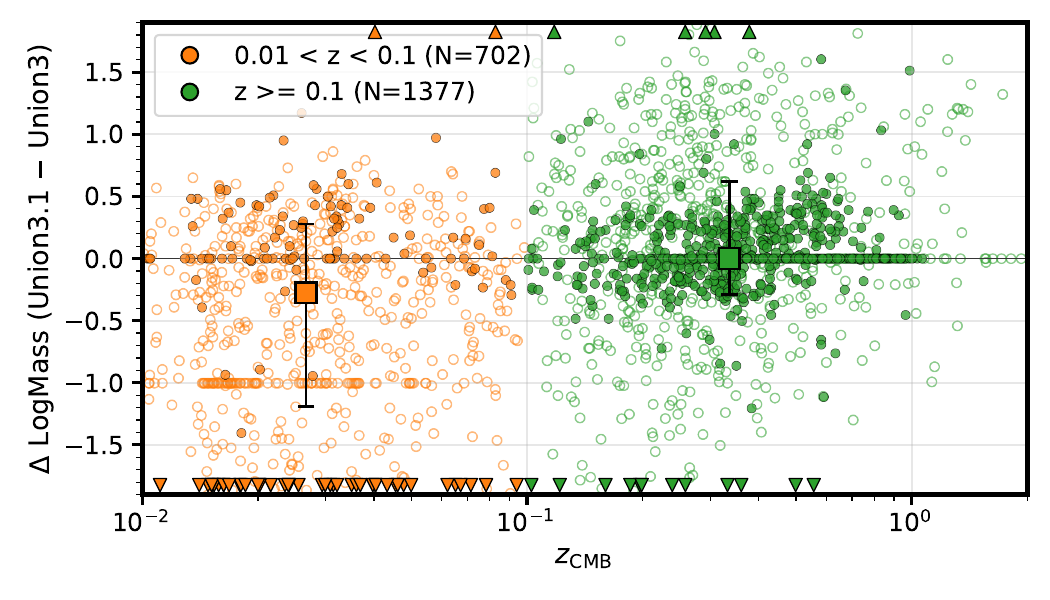}
    \includegraphics[width=0.9\linewidth]{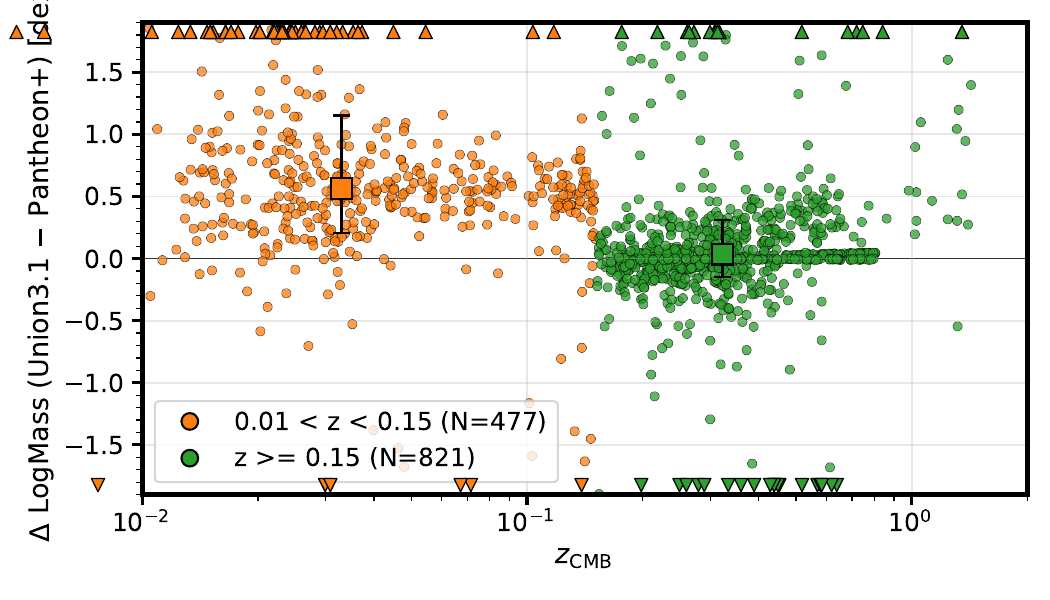}
    \caption{Comparing all Union3.1 masses (including LSTP, Other Sources, and fallback mass values) with either the Union3 (top) and Pantheon+ (bottom) host mass values used in those respective SN cosmology analyses. There is a clear discontinuity at $z=0.1$ and $z=0.15$ in Union3 and Pantheon+, respectively, which resulted in biased cosmological parameter inference. The size ($>0.5$~dex) and shape (a step) of the discontinuity, as well as the overall dispersion (full range $\gtrsim$ 1.0~dex), far exceed any variations seen in our extensive mass measurement variant analysis (full range $<0.4$~dex). The cause of these redshift-dependent inconsistencies are due to methodological assumptions adopted by either analysis that broke self-consistency of their respective mass scales as a function of redshift. Union3 overestimated their masses at low-redshift with respect to their high-redshift sample, while Pantheon+ underestimated them. Note the vertical axis limits are more extended than in \autoref{fig:variant_vs_redshift}.}
    \label{fig:deltaMass_pan_union}
\end{figure}

Finally, to contextualize the findings from the mass measurement variant analyses, we show in \autoref{fig:deltaMass_pan_union} a direct comparison between the masses used by either of the Union3 or Pantheon+, plotting the difference between the mass used by either and the LSTP masses presented in this study. The redshift-dependent mass offsets that biased the previous cosmology results are clear. Notice the size ($\sim$0.5~dex) and shape (discontinuous step at $z=0.1$ and $z=0.15$) do not come close to being explained by varying the assumptions of our underlying SED model.

Indeed, the explanations are much simpler and fall into the category of methodological errors or inconsistencies. Union3 adopted a loose fallback constraint of $11.0 \pm 1.0$~dex as the log of the host mass for only $z<0.1$ SN host galaxies that did not have a published mass in the literature, and $10.0 \pm 1.0$~dex for SNe with $z>0.1$. Pantheon+ determined new host masses with their own host galaxy photometry and SED fitting choices for only their host galaxies with $z<0.15$, but retained literature host mass measurements for $z>0.15$ SNe. These investigator choices---specifically the inconsistent treatment of low and high redshift SN host galaxies---in each analysis directly led to the unphysical discontinuities present in their respective mass scales. Going forward, to mitigate biases in cosmology, SN host galaxy properties used to standardize SN distances must be inferred from homogeneously derived galaxy photometry and modeled with one SED fitting code under an unchanging set of SED model assumptions.

\section{Summary and Conclusions} \label{sect:conclusion}

We have in this study presented new stellar mass estimates for the host galaxies of 2000 SNe contained in the Union3 database. 1300 of those included SN host galaxies whose coordinates had not been previously published. The stellar masses were derived using homogeneous, deblended, OIR photometry provided in the DESI LS DR10 Tractor catalog (supplemented with GALEX UV photometry) and the galaxy SED modeling code \texttt{Prospector}. This placed at least 75\% of all SNe in Union3 onto a single, homogeneous mass system, defined per the underlying photometry, galaxy SED modeling code, and adopted model assumptions. The version of Union3 that includes the new host masses, as well as some minor bug fixes and updates, is referred to and publicly distributed as Union3.1.

We reported several findings resulting from our analysis of SN host galaxy properties. First we showed that our SED modeling produced a star-forming main sequence consistent with published results \citep[e.g.,][]{Speagle_2014, Salim_2016, Salim_2018, Leroy_2019}. We then selected SNe from distinct sub-regions in the sSFR-$M_*$ plane, showing that SNe in massive and passive galaxies form a distinct population, in terms of both the SALT3 $x_1$ and $c$ parameters, from SNe observed in low-mass, star-forming host galaxies. 

SNe in passive galaxies exhibited an extended exponential tail toward red colors not seen in SNe hosted by low-mass star forming galaxies, providing further evidence against the argument that the reddening of SNe can be entirely explained by interstellar dust. Low-mass star-forming galaxies, meanwhile, were found to host few fast-declining SNe, forming a sharply peaked distribution in $x_1$ with mean $<x_1> \simeq 0.5$ and width $\sigma \simeq 0.7$.

We updated the SN analysis originally presented in \citetalias{Rubin_2025}, including some minor updates to the UNITY model, together referred to as the Union3.1+UNITY1.7 results. We summarize some of the salient findings below:

\begin{itemize}
\itemsep-0.1em 
  \item A value of $\log M_* = 10.115^{+0.044}_{-0.053}$~dex was found for the location of the host mass step, which is now included in UNITY as an explicit free parameter.
  \item The uncertainties on key SN standardization parameters became 0.7--0.9$\times$ smaller due to the new masses, except for that of the color correction coefficient of blue SNe.
  \item A $6\sigma$ measurement of a binary split in the color correction coefficient for red SNe, $\Delta \beta_{R}$, which is steeper for SNe in low-mass hosts; a marked increase from the $3\sigma$ measurement of this split in \citetalias{Rubin_2025}.
  \item The unexplained dispersion of 0.11~mag previously reported for SDSS-SN in Union3 decreased to 0.09~mag after updating the compiled host masses, which previously suffered a parsing bug in \citetalias{Rubin_2025}.
  \item From SN constraints alone, we found for flat $\Lambda$CDM, $\Omega_m = 0.344 \pm0.026$. This $0.4 \sigma$ decrease relative to \citetalias{Rubin_2025} was driven by a now-corrected systematic offset in the previous Union3 host masses for low redshift SNe.
  \item  A similar, but oppositely signed offset was found in the low-redshift sample of Pantheon+. After correction, their constraint on $\Omega_m$ increased by $0.6\sigma$ to $\Omega_m = 0.342 \pm 0.018$.
  \item A set of mass measurement variant analysis tests demonstrated that the Union3.1 cosmology results are unchanged when varying astrophysical assumptions of the adopted galaxy SED model.
\end{itemize}

We then revisited the combined probe analysis of \citetalias{DESICollaboration2025}, keeping the BAO and CMB assumptions exactly the same, and updating only the SN measurements, including the updated Union3.1 and Pantheon+ measurements presented here, as well as the recent DES-Dovekie recalibration \citep{Popovic_2025arXiv250605471P}. We found that the significance of the $w_0w_a$ tension against $\Lambda$ became far more consistent than before, with each of the three combinations shifting from the originally reported values in \citetalias{DESICollaboration2025} of ($3.8\sigma$, $2.8\sigma$ and $4.2\sigma$) to ($3.4\sigma$, $3.2\sigma$, and $3.4\sigma$). That is, \textit{all three} sets of SN cosmology analyses now provide consistent evidence for evolving dark energy, all of which are higher than the $3.1\sigma$ constraints from CMB and BAO alone.

The principal improvement achieved in this study is the use of a homogeneous set of profile-matched galaxy photometry with a single SED-fitting code to determine stellar masses for the majority of SN host galaxies in Union3. This corrected inconsistent methodological choices in the Union3 and Pantheon+ analyses that introduced oppositely signed, systematic offsets of $\sim$0.5~dex to the host masses of their low redshift SN samples relative to their high redshift ones, which then propagated as significant $\sim$$0.5\sigma$ biases in their respective cosmological parameter estimates. Encouragingly, once this methodological issue is handled, the Union3 and Pantheon+ SN cosmology results, previously in $1\sigma$ tension with each other, shift into $0.2\sigma$ consistency.\footnote{Recall the two analyses rely on the \textit{compilation} of published SN data and thus draw from the same light curve data sources. So the previous $1\sigma$ disagreement between them was much more significant than a naive comparison of parameter confidence intervals would suggest.}

As we move into the era of LSST and Roman, which aim to increase by an order of magnitude the number of SNe on the light-curve-standardized Hubble Diagram, systematics such as those identified in this and other recent studies will become several times more important to control. Ideally, one would prefer to use a method of SN distance measurement that \textit{eliminates} a systematic such as the host property dependence. which the planned Lazuli Space Observatory is aiming to accomplish \citep{Perlmutter_lazuli, Roy_2026arXiv260102556R}. In any case, it is imperative that continued effort be dedicated to accurately evaluating and/or removing sources of systematic error in SN~Ia distance measurements.

\begin{acknowledgments}

TJH is grateful for insightful discussions with Aliza Beverage and Dan Weisz regarding \texttt{Prospector}, and to Simone Ferraro for the same regarding the CMB. TJH acknowledges Schmidt Sciences for their support.

This work was supported in
part by the Director, Office of Science, Office of High Energy
Physics of the U.S. Department of Energy under Contract No. DE-AC02-
05CH11231.
Some results were obtained using resources
and support from the National Energy Research Scientific Computing
Center, supported by the Director, Office of Science, Office of
Advanced Scientific Computing Research of the U.S. Department of
Energy under Contract No. DE-AC02- 05CH11231.
Additional support was provided by NASA under the Astrophysics Data Analysis
Program grant 15-ADAP15-0256 (PI: Aldering).

The Legacy Surveys consist of three individual and complementary projects: the Dark Energy Camera Legacy Survey (DECaLS; Proposal ID \#2014B-0404; PIs: David Schlegel and Arjun Dey), the Beijing-Arizona Sky Survey (BASS; NOAO Prop. ID \#2015A-0801; PIs: Zhou Xu and Xiaohui Fan), and the Mayall z-band Legacy Survey (MzLS; Prop. ID \#2016A-0453; PI: Arjun Dey). DECaLS, BASS and MzLS together include data obtained, respectively, at the Blanco telescope, Cerro Tololo Inter-American Observatory, NSF’s NOIRLab; the Bok telescope, Steward Observatory, University of Arizona; and the Mayall telescope, Kitt Peak National Observatory, NOIRLab. Pipeline processing and analyses of the data were supported by NOIRLab and the Lawrence Berkeley National Laboratory (LBNL). The Legacy Surveys project is honored to be permitted to conduct astronomical research on Iolkam Du’ag (Kitt Peak), a mountain with particular significance to the Tohono O’odham Nation.

NOIRLab is operated by the Association of Universities for Research in Astronomy (AURA) under a cooperative agreement with the National Science Foundation. LBNL is managed by the Regents of the University of California under contract to the U.S. Department of Energy.

This project used data obtained with the Dark Energy Camera (DECam), which was constructed by the Dark Energy Survey (DES) collaboration. Funding for the DES Projects has been provided by the U.S. Department of Energy, the U.S. National Science Foundation, the Ministry of Science and Education of Spain, the Science and Technology Facilities Council of the United Kingdom, the Higher Education Funding Council for England, the National Center for Supercomputing Applications at the University of Illinois at Urbana-Champaign, the Kavli Institute of Cosmological Physics at the University of Chicago, Center for Cosmology and Astro-Particle Physics at the Ohio State University, the Mitchell Institute for Fundamental Physics and Astronomy at Texas A\&M University, Financiadora de Estudos e Projetos, Fundacao Carlos Chagas Filho de Amparo, Financiadora de Estudos e Projetos, Fundacao Carlos Chagas Filho de Amparo a Pesquisa do Estado do Rio de Janeiro, Conselho Nacional de Desenvolvimento Cientifico e Tecnologico and the Ministerio da Ciencia, Tecnologia e Inovacao, the Deutsche Forschungsgemeinschaft and the Collaborating Institutions in the Dark Energy Survey. The Collaborating Institutions are Argonne National Laboratory, the University of California at Santa Cruz, the University of Cambridge, Centro de Investigaciones Energeticas, Medioambientales y Tecnologicas-Madrid, the University of Chicago, University College London, the DES-Brazil Consortium, the University of Edinburgh, the Eidgenossische Technische Hochschule (ETH) Zurich, Fermi National Accelerator Laboratory, the University of Illinois at Urbana-Champaign, the Institut de Ciencies de l’Espai (IEEC/CSIC), the Institut de Fisica d’Altes Energies, Lawrence Berkeley National Laboratory, the Ludwig Maximilians Universitat Munchen and the associated Excellence Cluster Universe, the University of Michigan, NSF’s NOIRLab, the University of Nottingham, the Ohio State University, the University of Pennsylvania, the University of Portsmouth, SLAC National Accelerator Laboratory, Stanford University, the University of Sussex, and Texas A\&M University.

BASS is a key project of the Telescope Access Program (TAP), which has been funded by the National Astronomical Observatories of China, the Chinese Academy of Sciences (the Strategic Priority Research Program “The Emergence of Cosmological Structures” Grant \#XDB09000000), and the Special Fund for Astronomy from the Ministry of Finance. The BASS is also supported by the External Cooperation Program of Chinese Academy of Sciences (Grant \#114A11KYSB20160057), and Chinese National Natural Science Foundation (Grant \#12120101003, \#11433005).

The Legacy Survey team makes use of data products from the Near-Earth Object Wide-field Infrared Survey Explorer (NEOWISE), which is a project of the Jet Propulsion Laboratory/California Institute of Technology. NEOWISE is funded by the National Aeronautics and Space Administration.

The Legacy Surveys imaging of the DESI footprint is supported by the Director, Office of Science, Office of High Energy Physics of the U.S. Department of Energy under Contract No. DE-AC02-05CH1123, by the National Energy Research Scientific Computing Center, a DOE Office of Science User Facility under the same contract; and by the U.S. National Science Foundation, Division of Astronomical Sciences under Contract No. AST-0950945 to NOAO.

The technical support and advanced computing resources from University of Hawai`i Information Technology Services - Research Cyberinfrastructure, funded in part by the National Science Foundation CC* awards \#2201428 and \#2232862 are gratefully acknowledged. Support for this work was provided by NASA through grant number HST-AR-16631.001-A from the Space Telescope Science Institute, which is operated by AURA, Inc., under NASA contract NAS 5-26555.

This research has made use of the SIMBAD database,
operated at CDS, Strasbourg, France.

This research has made use of the NASA/IPAC Extragalactic Database (NED),
which is operated by the Jet Propulsion Laboratory, California Institute of Technology,
under contract with the National Aeronautics and Space Administration.

\end{acknowledgments}

\facilities{NED \citep{https://doi.org/10.26132/ned1}, SIMBAD \citep{Wenger_2000}}

\software{Astropy \citep{Astropy_2013, Astropy_2018, Astropy_2022}; 
Cobaya \citep{2019ascl.soft10019T, Torrado_2021}; 
Dynesty \citep{Speagle_2020, sergey_koposov_2025_17268284}; 
FSPS \citep{Conroy_2009, Conroy_2010ascl.soft10043C}; 
Jupyter/Lab \citep{2016ppap.book...87K}; 
Matplotlib \citep{Hunter:2007}; 
NumPy \citep{harris2020array}; 
Prospector \citep{Johnson_2021}; 
SciPy \citep{2020SciPy-NMeth}}

\clearpage

\appendix

\section{Previously ``Hostless'' SNe Newly Associated with a Host Galaxy} \label{app:new_hosts}

\subsection{SDSS}
\begin{figure}
    \centering
    \includegraphics[width=0.7\linewidth]{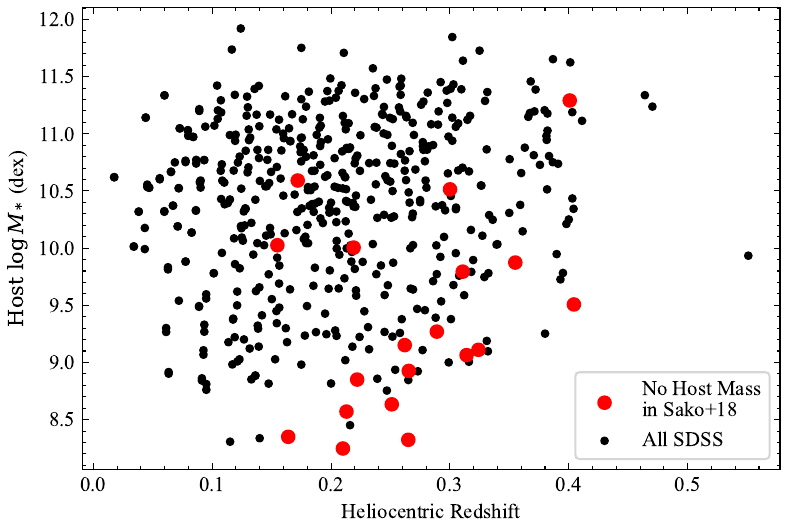}
    \caption{Completeness of SDSS host mass estimates. LSTP host mass plotted vs. heliocentric redshift for SDSS SNe compiled in Union3 from SDSS-SN \citep{Sako_2018}. SNe for which \citetalias{Sako_2018} reported host galaxy information are plotted as small black markers, while those hosts that were identified by us in the LS DR10 but not assigned a host centroid or mass by SDSS are plotted as red dots. 15 of the 18 new host associations we added for SDSS SNe lie along the detection limit of the existing host galaxies in \citetalias{Sako_2018}, indicating these were likely missed simply due to the SDSS imaging being shallower. The three low-redshift, high-mass hosts that \textit{should} have been detected by SDSS are examined in detail in the text of this Appendix.}
    \label{fig:sdss_mass_completness}
\end{figure}

We determined new host coordinates for 18 SNe in the SDSS-SN survey for which coordinates of the host or its associated properties were not published. From the \texttt{Prospector} SED fits (introduced later) we found the average mass of these hosts to be $9.33 \pm 0.19$~dex. This is consistent with the LS imaging and catalog being deeper than the SDSS DR8 imaging used for their host mass analysis. Three of these were low-redshift and high mass and thus should have, in principle, been in the \citetalias{Sako_2018} tables. These are: (SN2005hq, 11.3~dex, $z=0.4$), (SN2005iv, 10.5~dex, $z=0.3$), and (SN2006hk, 10.6~dex, $z=0.17$). 

We re-inspect the LS stacked, model, and residual images to understand why the host coordinate and photometry might be missing from the \citeauthor{Sako_2018} results, as well as to validate the likeliness of our new host identification, the robustness of our match into the LS Tractor catalog, and the quality of the associated photometry.

\begin{enumerate}
    \item The location of SN2005hq was equidistant to two hosts of comparable brightness. Our match into the LS Tractor catalog associated it with the host that has an object ID 3340 and DESI photometric redshift equal to $z_{phot} = 0.491 \pm 0.056$. The other host had a photometric redshift $z_{phot} = 0.162 \pm 0.165$. The spectroscopic redshift from SDSS is $z_{spec}=0.4008 \pm 0.0005$ and photometric SN redshift (prior-less) was $z_{SN,phot} = 0.406 \pm  0.0124$, indicating that the association with the higher redshift galaxy in the Tractor catalog is likely to be correct.
    \item Similarly, SN2005iv occurred between two possible hosts: one brighter and located $2.4''$ away; the other fainter and located $4.2''$ away. We associated the SN with the brighter, closer galaxy which has an LS object ID 7084. The brighter one is also separated by $1.2''$ from a comparably bright galaxy and the two appear to be cleanly deblended. The photometric redshift of object 7084 is $z_{phot} =0.276 \pm 0.092$, which is most consistent with the SDSS values $z_{spec}=0.3001 \pm 0.0005$ and $z_{SN,phot} = 0.2137 \pm  0.0058$, and so we are confident in that association.
    \item Finally, SN2006fk was separated on the sky by $1.2''$ from the galaxy we assigned as its host galaxy, which appears more than bright enough to have been measured by SDSS. It is possible that the very bright star 28~Aqr ($G\sim6$~mag) broke the SDSS co-add algorithm for the SDSS pointings that contained this SN's host galaxy.
\end{enumerate}

\subsection{PS1}

Of the 326 spectroscopic Pan-STARRS SNe compiled in Union3, 18 were assigned a host mass value of $-999$ in Table 2 of \citet{Jones_2018}. In our host identification described in Section 2, we newly assigned a host to twelve of these, and corroborate the original hostless (or very low mass) classification for the other six. Because we do not expect the DESI LS imaging to be deeper than PS1 Medium-Deep, we return to these cases and manually check them against external data to verify (or rule out) our new host galaxy association.

\begin{enumerate}
    \item We assign psc160039 with host galaxy LEDA 2311296. From the SN spectrum, PS1 assigned a redshift $z_{SN} = 0.27$, while LEDA2311296 was included in SDSS DR13 with a redshift $z = 0.267345 \pm 3.92e-5$ \citep{sdss_dr13}. Thus, the association with LEDA2311296 is confirmed. 
    \item We initially associated psc340334 with a nearby bright source, but upon re-inspection we find that was a foreground star with Gaia EDR3 source ID 854253949905430272, $G = 14.34$~mag, $\mu_{RA/Dec}$ values of 5.3 and $2.5 \pm 0.0$ mas/yr and parallax $\pi = 0.3 \pm 0.0$~mas, and thus reinstate its hostless designation. The SED fit quality was very poor, so this was unsurprising.
    \item We associated psc520062 with a nearby unnamed galaxy, observed by DESI as target 39633319012336875 with a reported $z = 0.3076$ \citep{DESI_DR1_galaxies}. PS1 reported a redshift derived from the SN spectrum $z_{SN} = 0.3$, confirming the association. 
    \item We associated psc550059 with a faint galaxy that is assigned brickid 326013 and objid 2039 in LS DR10 and a photometric redshift $z_{photo} = 0.435 \pm0.120$. The SN-derived redshift reported by PanSTARRS was $z_{SN} = 0.32$, so we accept the association. There is also a luminous red galaxy (LRG) separated from the SN by just 4~arcsec, but this LRG has an SDSS spectroscopic redshift $z=0.487$, ruling out its association with the SN. 
    \item We associated psc550137 with LS DR10 brickid 331707 and objid 328 which has a DR9 photometric redshift $z_{photo} = 0.497 \pm 0.104$. The SN redshift reported by PS1 is $z_{SN} = 0.421$, suggesting a good match with this object. This galaxy was also observed by the VIMOS VLT Deep Survey (VVDS) who assigned a spectroscopic redshift of $z=0.18$. However, this object was also assigned a redshift flag of 1 to reflect their lowest level of confidence in the quoted redshift, stated to be 50--75\% likely to be correct. 
    \item psc560027 was assigned to the (relatively low surface brightness) galaxy contained in LS DR10 with brickid 313231 and objid 4167. The SN was separated by 0.3~arcsec from the host galaxy and no other source was detected within 5~arcsec of the SN. Because this galaxy was quite diffuse it could have been missed by PS1 and since there was no alternative potential host, our updated association is likely correct. 
    \item We associated psc590005 with galaxy LEDA 1187963, which was observed by DESI who reported a redshift $z_{\mathrm{DESI}}=0.1743$ \citep{DESI_DR1_galaxies}. According to SIMBAD and the PS1 table published in \citet{Jones_2018}, the same association was made and they report a $z_{SN} = 0.175$ and $z_{Host} = 0.1749$. However, they did not assign a host mass, quoting it as $-999$, typically reserved for the SNe they have (presumably) classified as hostless (or perhaps an unsuccessful SED fit), i.e. the $z_{\mathrm{Host}}$ and host mass columns would typically both equal -999 if a PS1 SN is missing a host association. Furthermore, it is not included in the spectroscopic table published in \citet{Scolnic_2018}, despite being spectroscopically classified and given a host redshift in the master table. The SN sky position is located clearly in the outer disk of this galaxy, confirming its association with LEDA 1187963.
    \item The SN psc590191 landed near the core of galaxy COSMOS2015 0402113, which was observed by DESI as target 39627835563838289 which found $z_{\mathrm{DESI}} = 0.2198$ \citep{DESI_DR1_galaxies}. PS1 reported a $z_{SN} = 0.22$ and $z_{Host} =0.2209$, but, similar to the SN discussed immediately before this one, they report a host mass of -999. 
    \item We associated psc370356, now merged within Union3.1 with the LOSS-observed SN2012bn, with host galaxy UGC07228. From NED \citep{https://doi.org/10.26132/ned1}\footnote{The NASA/IPAC Extragalactic Database (NED) is operated by the Jet Propulsion Laboratory, California Institute of Technology, under contract with the National Aeronautics and Space Administration.} we see this host galaxy has redshift $z=0.025244 \pm 0.00006$ as per the Updated Zwicky Catalog \citep[UZC,][]{Falco_1999}. The SN is visually embedded in the outer disk of this nearby galaxy and the SN redshift reported by PS1 was $z_{SN} = 0.025$, confirming our new association.
    \item ps0909006 is likewise assigned a host mass of $-999$ in Table 3 of \citet{Jones_2018}, although that table lists a host position. We associate it with the galaxy contained in LS DR10 with brickid 338903 and objid 8790, separated from the SN by only 0.3~arcsec, which is the SDSS DR8 photometric source SDSS J221548.05+011105.0. The same association was made independently at the time of discovery, identified as Anon J221548+0111 \citep{Barbon_1999, Barbon_2010_vizier}. The LS DR9 photometric redshift of the host, $z_{photo} = 0.619 \pm 0.126$, is in mild ($2.7\sigma$) tension with the SN-spectrum redshift $z_{SN} = 0.284$ reported by PS1, but not outside the realm of typical photo-$z$ errors for faint galaxies.
    \item ps0909010 was assigned a host position by \citet{Jones_2018} but no host mass. We associate it with the galaxy contained in LS DR10 with brickid 307484 and objid 3936, a bright ($r = 20.7$~mag) S\'ersic-profile galaxy separated from the SN by 1.6~arcsec. Its LS DR9 photometric redshift $z_{photo} = 0.278 \pm 0.073$ is in excellent agreement with the SN-spectrum redshift $z_{SN} = 0.27$, and we accept the association.
    \item We associated ps0910018 with the LS DR10 source in brickid 307478 and with objid 5290, a bright ($r = 20.5$~mag) S\'ersic-profile galaxy 0.6~arcsec from the SN. Its LS DR9 photometric redshift $z_{photo} = 0.302 \pm 0.061$ is consistent with the SN-spectrum redshift $z_{SN} = 0.265$, and we accept the association.
\end{enumerate}

We therefore confirm our new host associations for eleven of these twelve Pan-STARRS SNe, and rule out our initial association of psc340334 with a foreground star.

\subsection{SNLS}

Of the 235 SNLS SNe compiled in Union3, 13 were assigned a host mass value of 0 in the SNLS host catalog. %
In our host identification described in Section 2, we newly assigned a host to two of these, and corroborate the original ``hostless'' classification for the other eleven. Since SNLS is much deeper than DESI LS, we investigate these two cases in greater detail to verify (or rule out) our new host galaxy association.

\begin{enumerate}
    \item We associated 04D3gt with the galaxy contained in LS DR9 with brickid 594364 and objid 2422, a de Vaucouleurs-profile source separated from the SN position by only 0.3~arcsec. Its LS DR9 photometric redshift is $z_{photo} = 0.508 \pm 0.023$, consistent within the typical photo-$z$ scatter with the SNLS-reported redshift $z_{SN} = 0.451$. No other LS source is detected within 5~arcsec of the SN. We therefore accept the new association.
    \item Similar to the case of psc340334. we initially associated 04D1aj with a bright source 7.1~arcsec north-west of the SN, but upon re-inspection we find that this is a foreground star with Gaia DR3 source ID 2488797350324724736, $G = 16.97$~mag, $\mu_{RA/Dec}$ values of $1.87 \pm 0.09$ and $-1.65 \pm 0.07$~mas/yr (a $22\sigma$ proper-motion detection) and parallax $\pi = -0.005 \pm 0.074$~mas, and thus reinstate its zero mass, hostless designation. The only other candidate host, an elongated galaxy at a similar separation from the SN, has an LS DR9 photometric redshift $z_{photo} = 0.528 \pm 0.040$, nearly $5\sigma$ from the SNLS-reported $z_{SN} = 0.721$, ruling out its possible association. The Tractor model for the star was an unresolved point source and the resulting SED fit returned an implausibly high and tightly constrained mass, so this was unsurprising.
\end{enumerate}

\section{SQL/ADQL Queries into LS Tractor and GALEX GUVcat} \label{app:queries}

In this appendix we provide template SQL/ADQL queries into the LS Tractor and GALEX GUVcat source catalogs used in this study. 

First, for querying into the publicly accessible LS Tractor catalog via the Astro Data Lab that is managed by NOIRLab\footnote{\url{https://datalab.noirlab.edu/data/legacy-surveys}},

\begin{verbatim}
    SELECT release,type,ls_id,brickid,objid,ra,dec,
    snr_g,mag_g,flux_g, flux_r, flux_i, flux_z, 
    flux_w1, flux_w2, flux_w3, flux_w4, 
    flux_ivar_g, flux_ivar_r, flux_ivar_i, flux_ivar_z, 
    flux_ivar_w1, flux_ivar_w2, flux_ivar_w3, flux_ivar_w4 
    FROM ls_dr10.tractor
      WHERE mag_g < 24
      AND q3c_radial_query(ra,dec,{ra0:f},{dec0:f},{radius:f})
\end{verbatim}

Then, for querying the GALEX GUVcat hosted at CDS \citep{Bianchi_2017},
\begin{verbatim}
    SELECT *
    FROM "II/335/galex_ais"
    WHERE 1=CONTAINS(
      POINT('ICRS', RAJ2000, DEJ2000),
      CIRCLE('ICRS', {r0:0.6f}, {d0:0.6f}, {rcrit:0.3f})
    )
\end{verbatim}
where the curly brackets in both queries represent \texttt{python} string formatting when looping over a set of RA/Dec. coordinates.

\section{Prospector Settings and Priors} \label{app:priors}

In the SED fitting to our galaxy photometry, we have assumed for most parameters their default values in \texttt{Prospector}. Since \texttt{Prospector} is built on FSPS, the parameter names and default values are mostly shared. First, we adopted a \citet{Kroupa_2001} IMF, indexed by the number ``2'' in \texttt{Prospector}. We then added a burst of star formation to the default delay-tau model in \texttt{Prospector} a for parametric SFH model, defined as \texttt{parametric\_sfh} and \texttt{burst\_sfh} in its internal library of Template models. The isochrone suites adopted---MIST, MILES, and \citet[][hereafter DL07]{Draine_2007}---are the FSPS defaults. We include dust emission and increase the emissivity, named \texttt{duste\_gamma}, from the default value, as described later in this section.

\subsection{SFH Parameter Priors} \label{app:metalprior}

\begin{figure}
    \centering
    \includegraphics[width=0.5\linewidth]{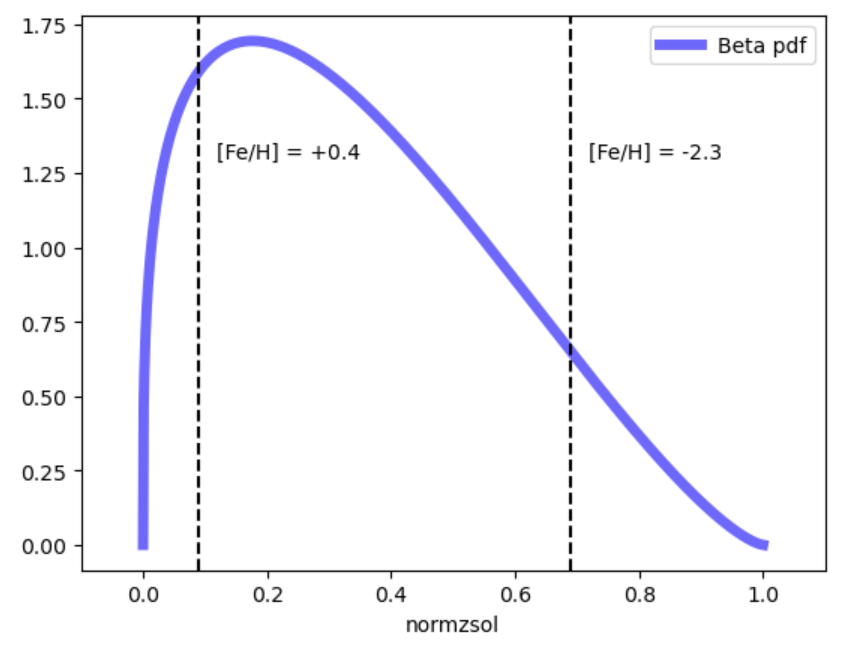}
    \caption{Adopted prior on the custom-defined normzsol parameter we used in \texttt{Prospector} fits. The strict bounds on the prior occur at values of $[Fe/H]$ between $-2.3$ and $+0.4$~dex, marked by vertical dashed lines.}
    \label{fig:metal_prior}
\end{figure}

We adopt uninformed flat priors for the mass  $\mathcal{U}\sim[10^7, 10^{12}]M_{\odot}$, dust $U\sim[0.0, 5.0]$, tage (as fraction of age of universe) $U\sim [0,1]$, and fage-burst $U\sim[0.01,0.99]$ parameters, and a log-uniform prior for the tau $\log \mathcal{U}\sim[0.1,10.0]$ and fburst $\log \mathcal{U}\sim[0.03,0.80]$ parameters. Metallicity is the only parameter for which we adopt an informed prior, shown in \autoref{fig:metal_prior}. We also implemented a custom normalization of the built-in zsol parameter of Prospector/FSPS defined as,

\begin{verbatim}
def norm_logzsol(logzsol=0):
    zsol_min, zsol_max = -3.7, 0.8
    norm = 1/(zsol_max - zsol_min)
    x_rescale = -(logzsol - zsol_max)*norm
    return x_rescale
\end{verbatim}

This was done to facilitate implementation of a physically informed prior shape on the metallicity parameter into \texttt{Prospector}, such as that used in \citet{Gallazzi_2005} or \citet{Childress_2013}, via the beta function that is already built into the \texttt{Prospector} infrastructure.

\subsection{Dust Emission} \label{app:dustegamma}

FSPS adopts the dust emission model of \citetalias{Draine_2007}, which assumes that dust grains are a mixture of silicate, graphite, and polycyclic aromatic hydrocarbons (PAHs). This model has three parameters, and we initially generated models with all three parameters fixed to their default values. However, when testing these models and comparing them against data we found that they were unable to match the high WISE W4 fluxes observed in some of our higher-inclination (likely dustier) spiral galaxies. One of the \citetalias{Draine_2007} parameters is $\gamma$, which represents the fraction of dust heated by starlight with intensity greater than some minimum value. The effect of varying $\gamma$ can be seen in \citetalias{Draine_2007}, their Figure 18. We find that increasing the value of $\gamma$ (\texttt{duste\_gamma} in FSPS) from the default value of 0.01 to 0.05 while also increasing the metallicity has the desired effect of increasing the model W4 flux while keeping the bluer portion of the SED unchanged. While allowing \texttt{duste\_gamma} to be a variable parameter would certainly improve the flexibility of our models, in an effort to keep the number of FSPS variables to a minimum, we opt instead to set \texttt{duste\_gamma} = 0.05 for all models.

\section{Example SED Fits from Prospector} \label{app:sed_fits}

\begin{figure}
    \centering
    \includegraphics[width=0.6\textwidth]{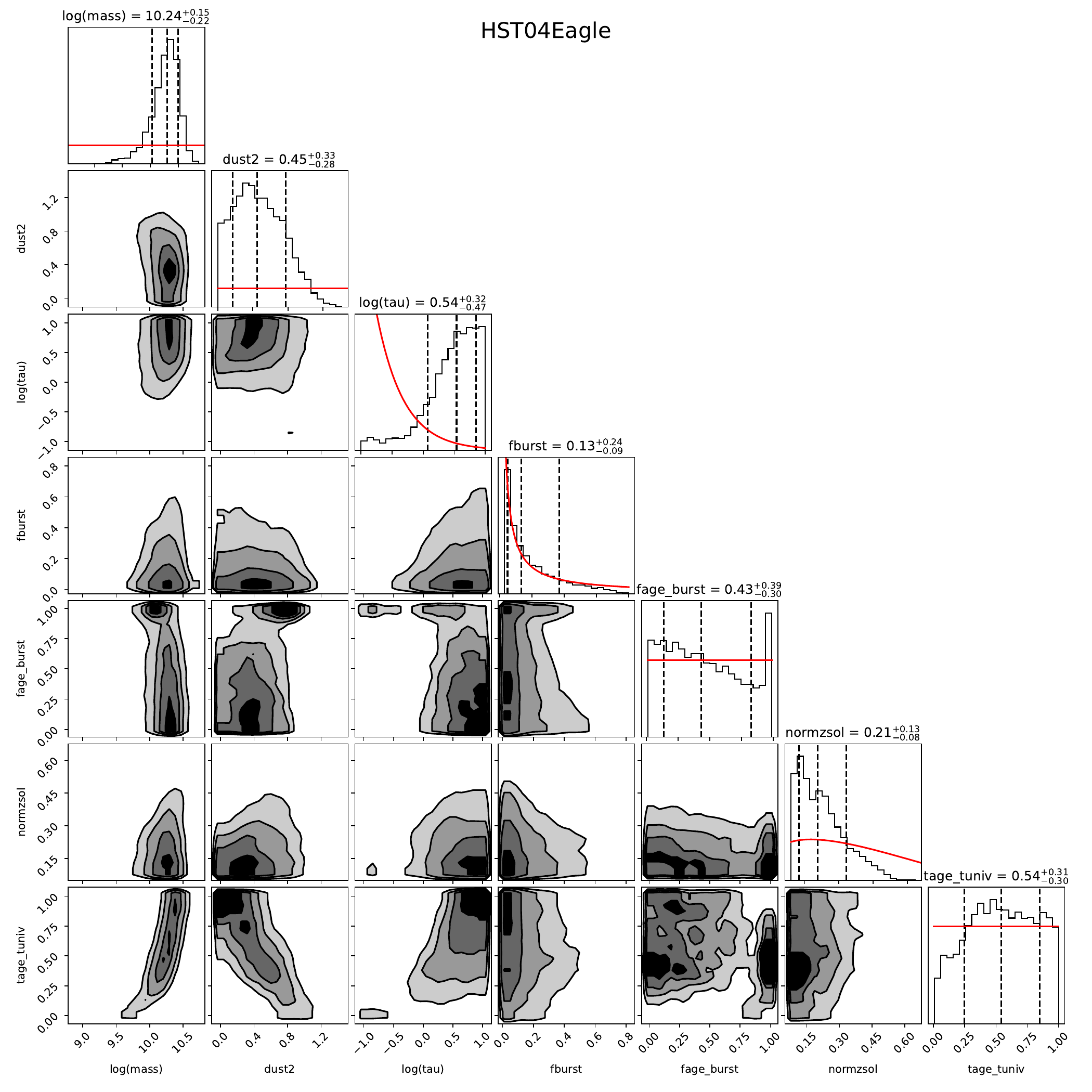}
    \includegraphics[width=0.98\textwidth]{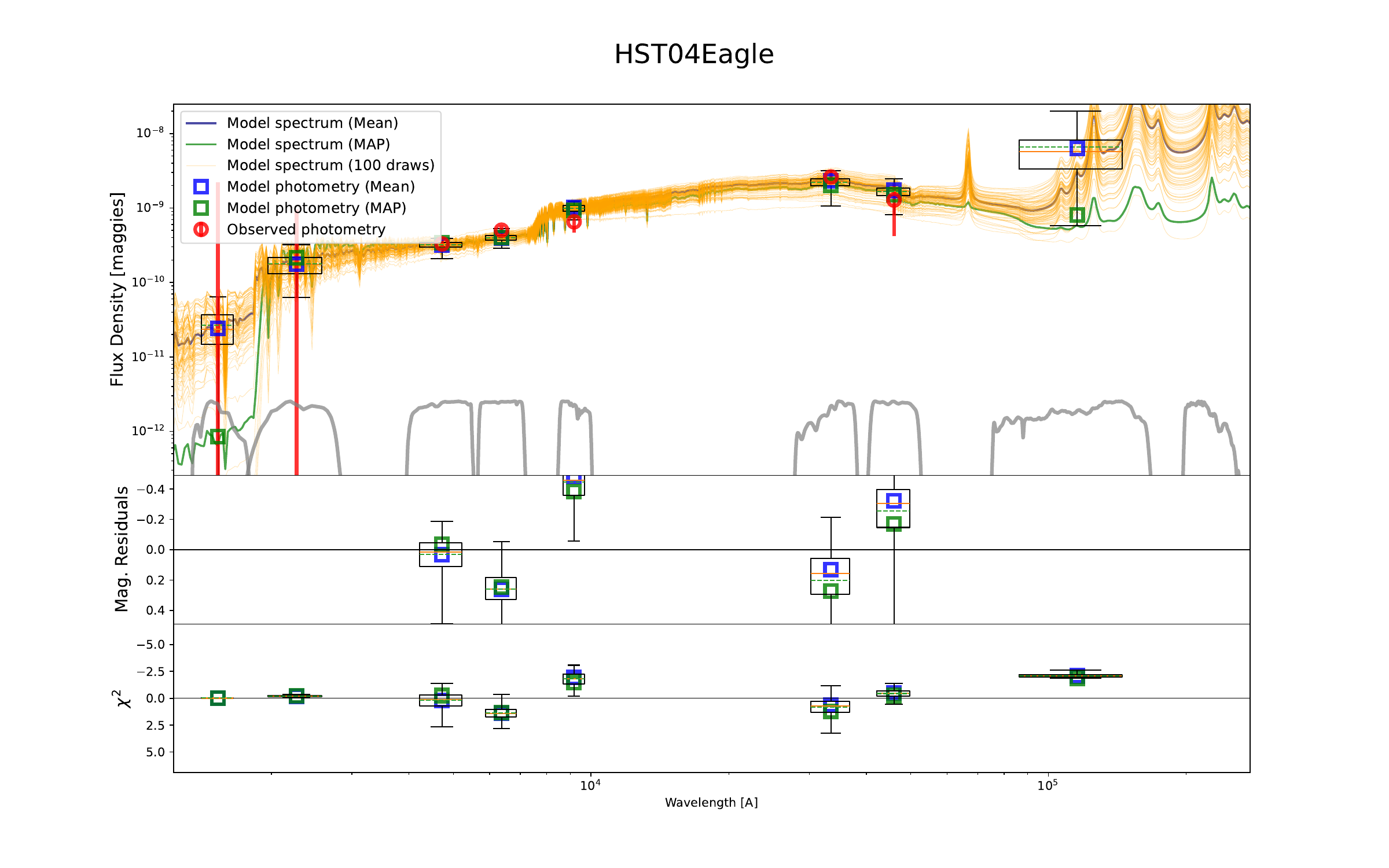}
    \caption{SED fit results for HST04Eagle, the highest redshift, low-mass star-forming galaxy in the LSTP sample at $z=1.02$. \textit{Top}: A corner plot of the sampling fit results. \textit{Bottom}: Direct comparison of model predictions with observed photometry.}
    \label{fig:eagle_fit}
\end{figure}

\begin{figure}
    \centering
    \includegraphics[width=0.6\textwidth]{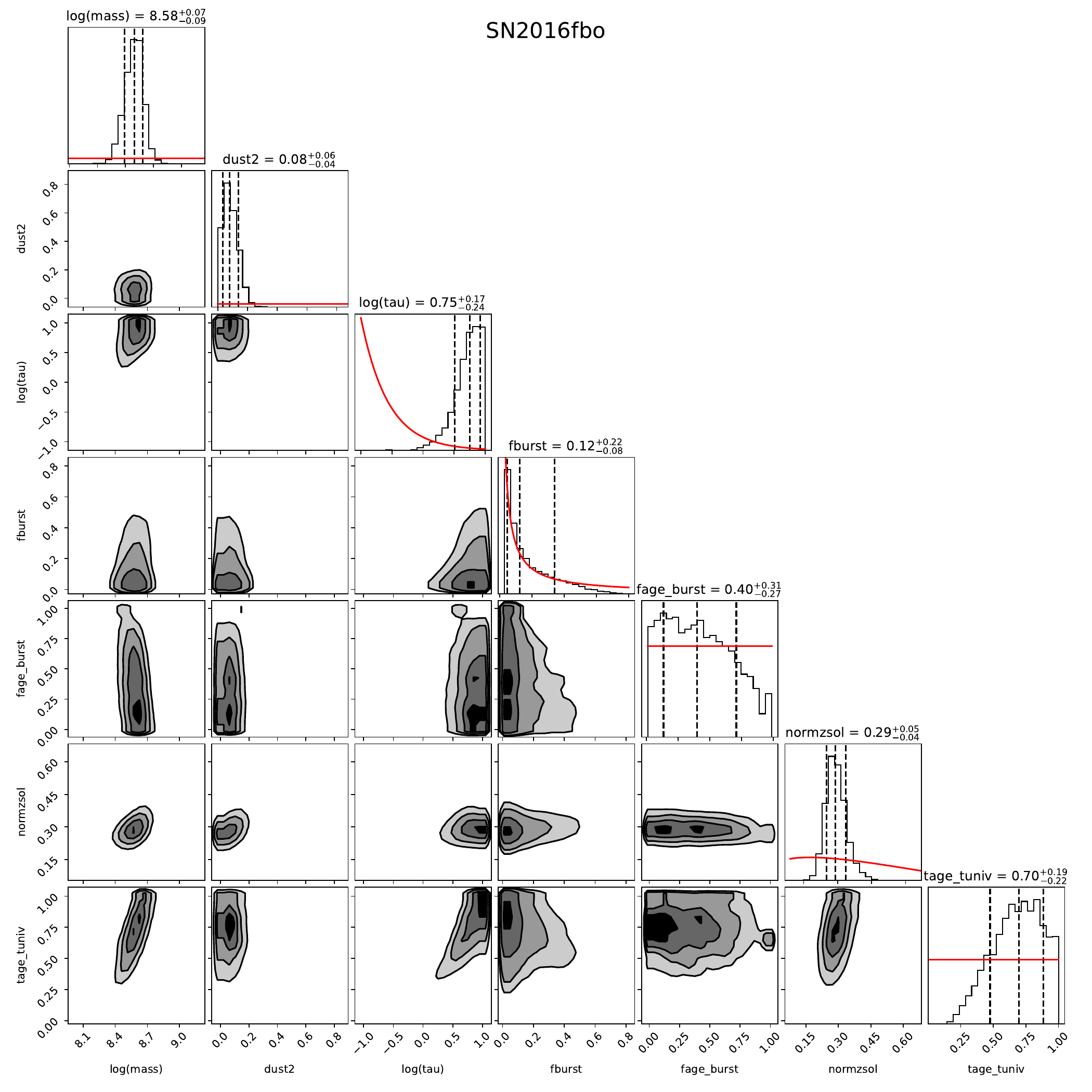}
    \includegraphics[width=0.98\textwidth]{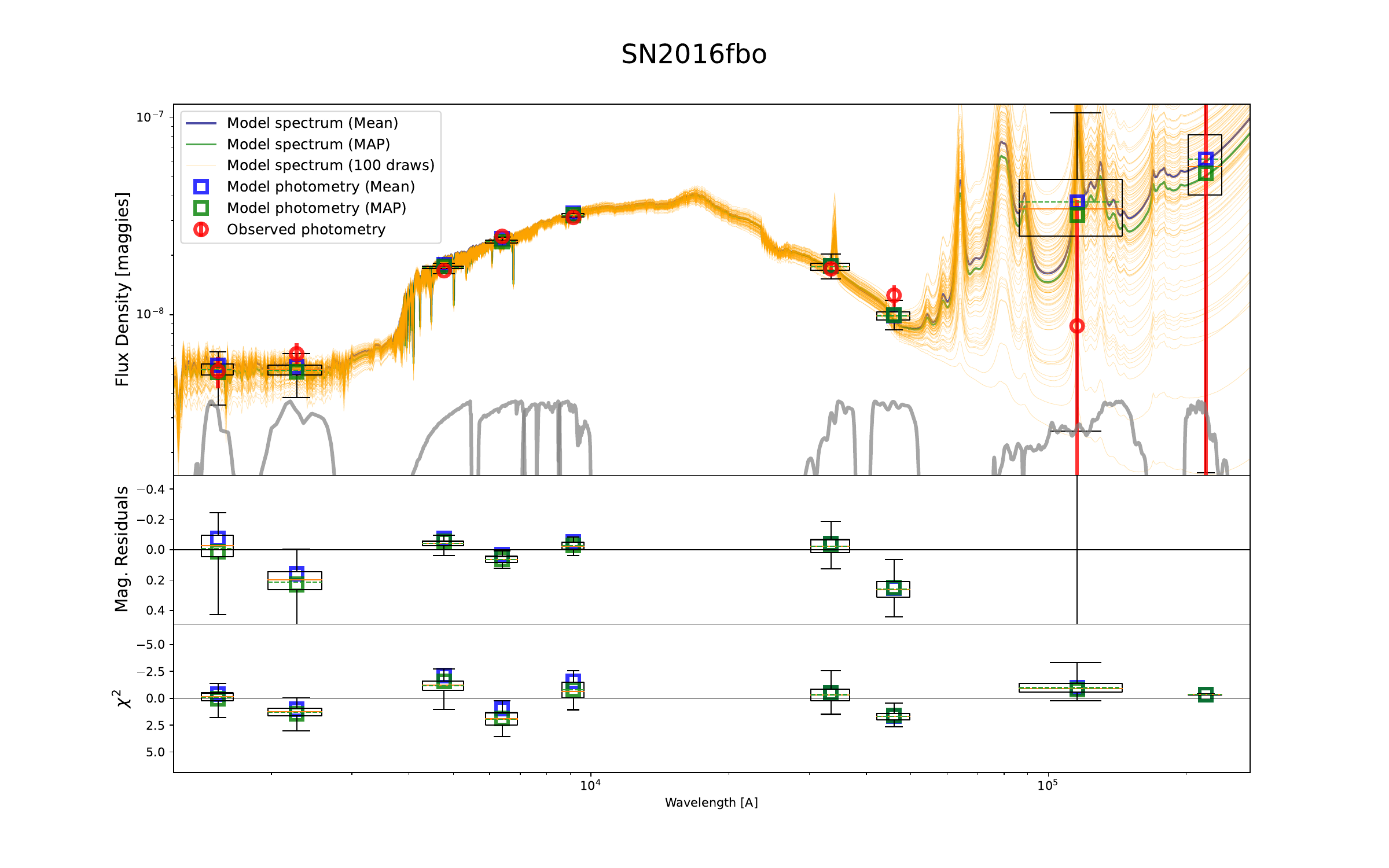}
    \caption{Same as \autoref{fig:eagle_fit} but for one of the nearby, low-mass, star-forming host galaxies at $z=0.03$.}
    \label{fig:nearby_lowmass_fit}
\end{figure}

\begin{figure}
    \centering
    \includegraphics[width=0.6\textwidth]{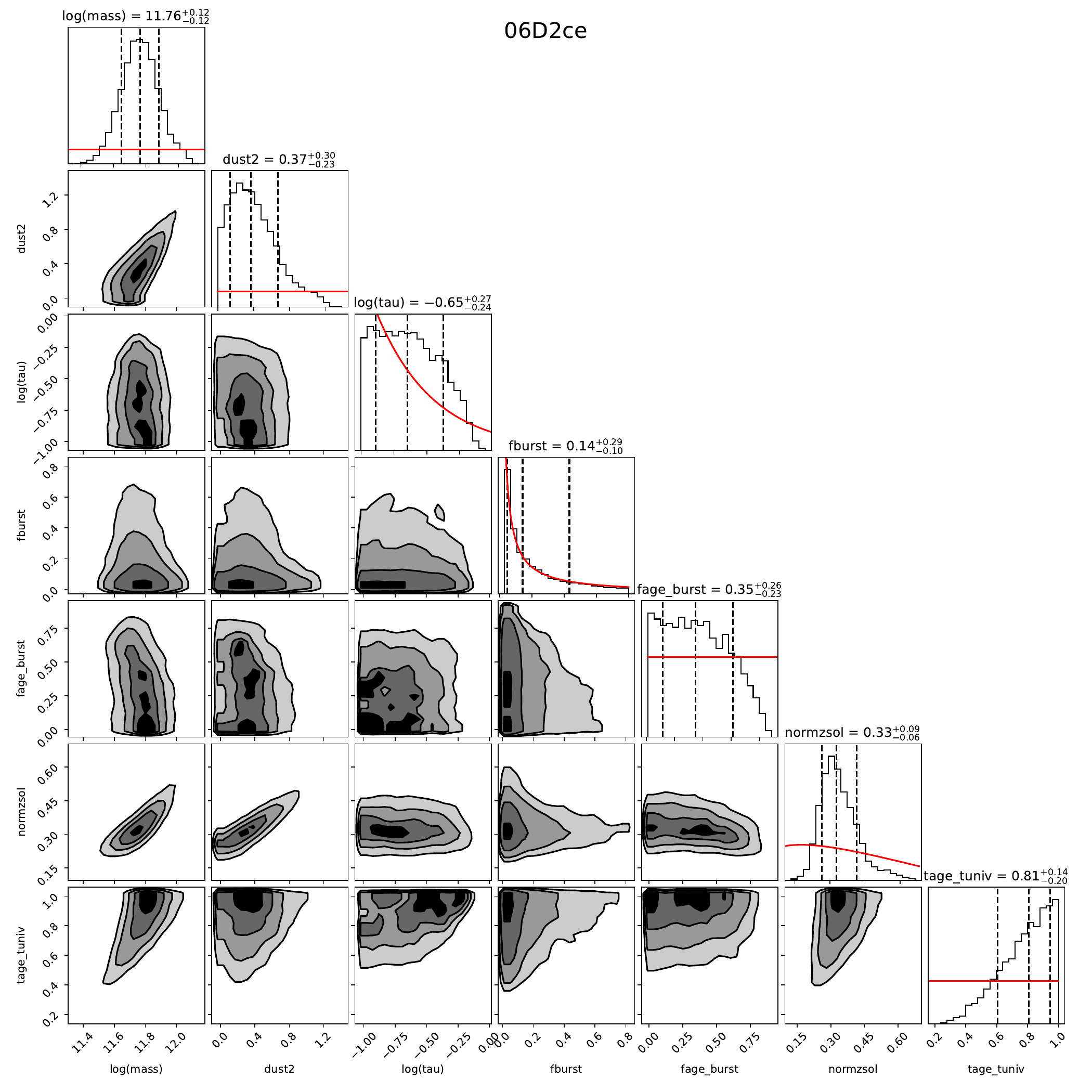}
    \includegraphics[width=0.98\textwidth]{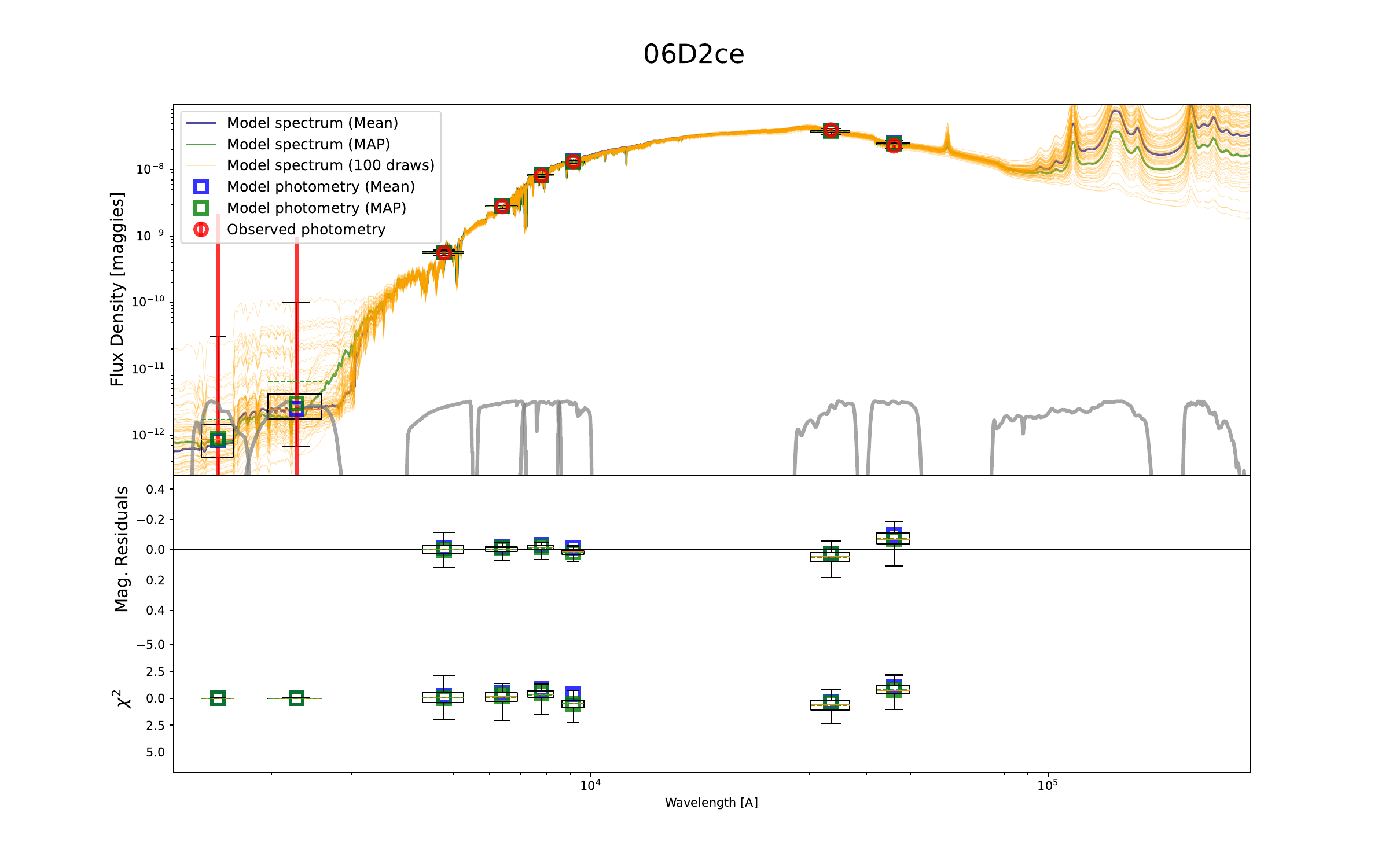}
    \caption{Same as \autoref{fig:eagle_fit} but for one of the higher redshift, massive, and passive host galaxies at $z=0.82$.}
    \label{fig:snls_passive_fit}
\end{figure}

\begin{figure}
    \centering
    \includegraphics[width=0.6\textwidth]{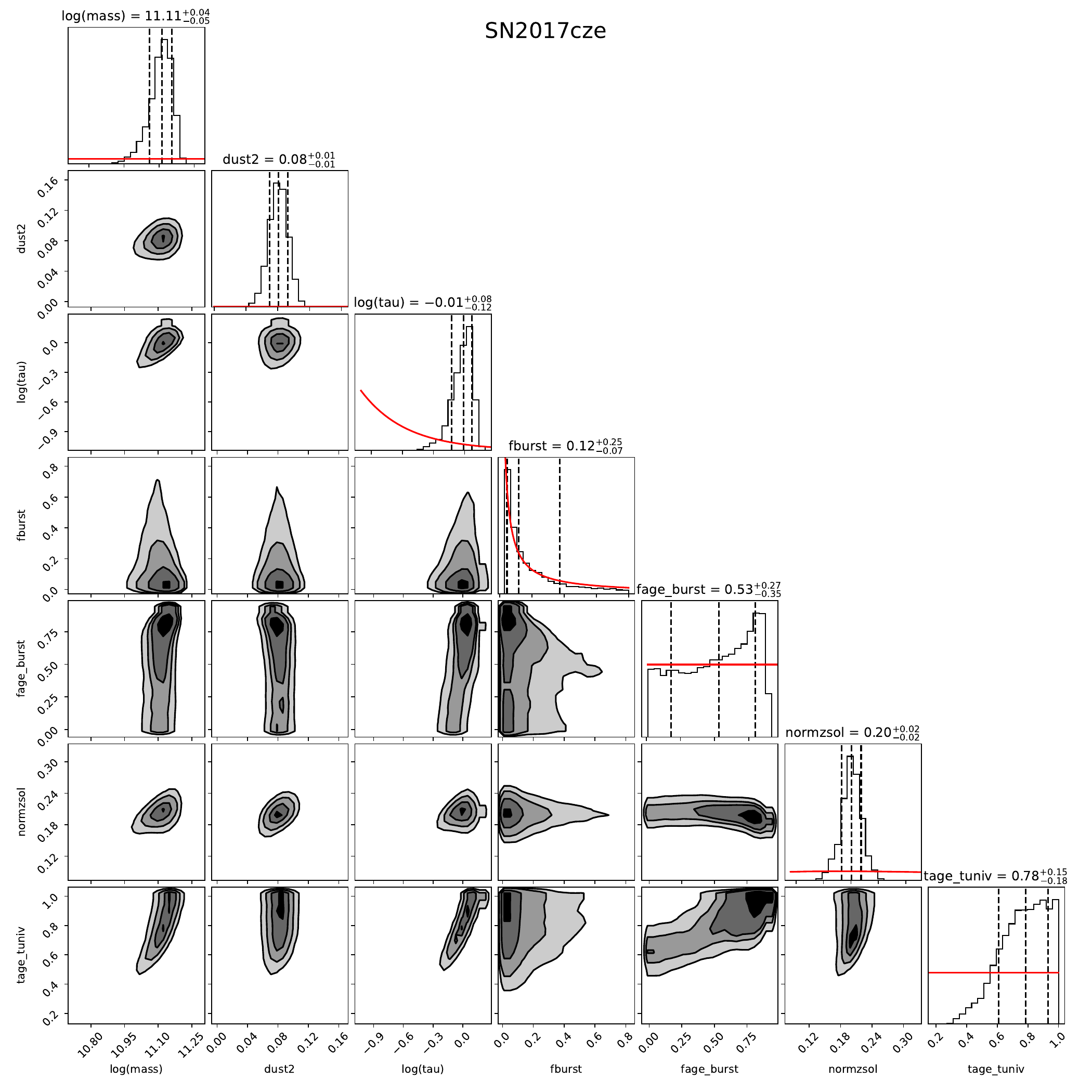}
    \includegraphics[width=0.98\textwidth]{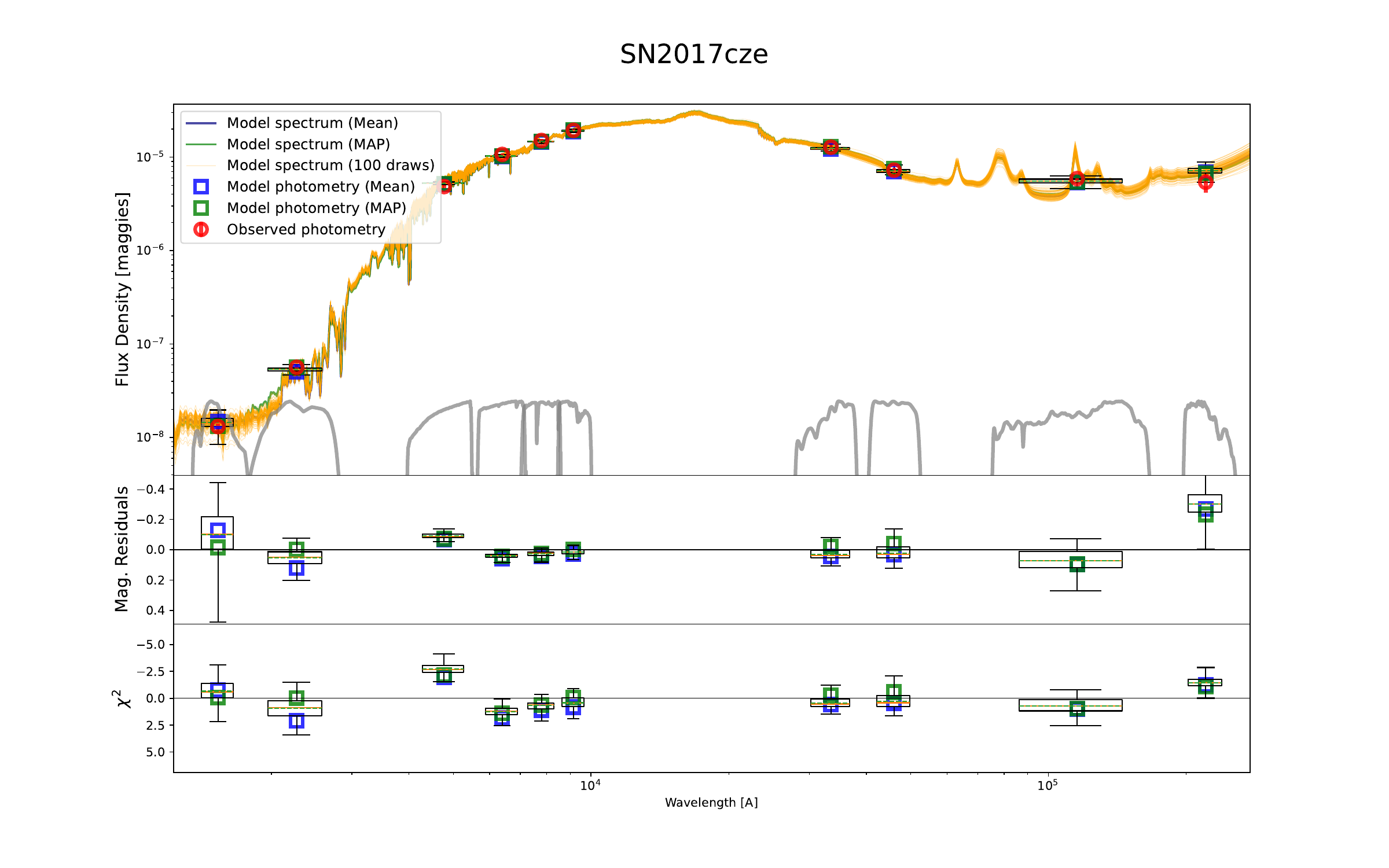}
    \caption{Same as \autoref{fig:eagle_fit} but for one of the nearby, massive, passive host galaxies at $z=0.014$.}
    \label{fig:nearby_massive_fit}
\end{figure}

In this Appendix, we present in Figures \ref{fig:eagle_fit}--\ref{fig:nearby_massive_fit} the fit results from \texttt{Prospector} for four Union3.1 SN host galaxies: one low- and one high- redshift pulled from each of the low-mass star-forming and the massive, passive sub-regions previously highlighted in \autoref{fig:host_prop_distributions}. 

At the top of each figure, we show the resulting corner plot for the fitted parameters, the log of the mass formed, the dust attenuation screening old stars, the log of the star formation decay rate, the fraction of the mass formed in a burst, the age at the time of the burst, the metallicity mapped onto a custom-defined parameter rescaling, and the age at onset of star-formation (reported as a fraction of the age of the universe).

At the bottom of each figure, we show the predicted SED along with the actual observed data. We plot the SED and synthesized photometry predicted by the MAP and the mean of the samples as green and blue squares, respectively. We also plot as orange curves the predicted SEDs from 100 weighted draws of the nested samples. In the panel immediately below the SED, we plot the absolute magnitude residuals, and below that the $\chi^2$ pulls of each data point. The zero of each set of residuals is set to the MAP result, though note we adopted the weighted median of the samples for the final parameter estimates.

\clearpage 

\section{Galaxy Mass Comparisons} \label{app:mass_compares}

As previously discussed, the accurate application of a mass step in SN cosmology relies on the mass estimates being on the same system. One of the leading causes for systematic offsets between independent sets of mass estimates is due different adopted stellar population synthesis models. And for Union3.1 we have elected to regularize onto the FSPS system. In this Appendix, we compare the LSTP masses presented in this study with mass estimates from various published analyses.

\subsection{Comparisons with Other Mass Estimates Used in Union3.1}

We compare the LSTP masses with masses from the ``Other'' sources used to fill gaps in the LSTP compilation, as described in the main text.

\begin{figure}
    \centering
    \includegraphics[width=0.8\linewidth]{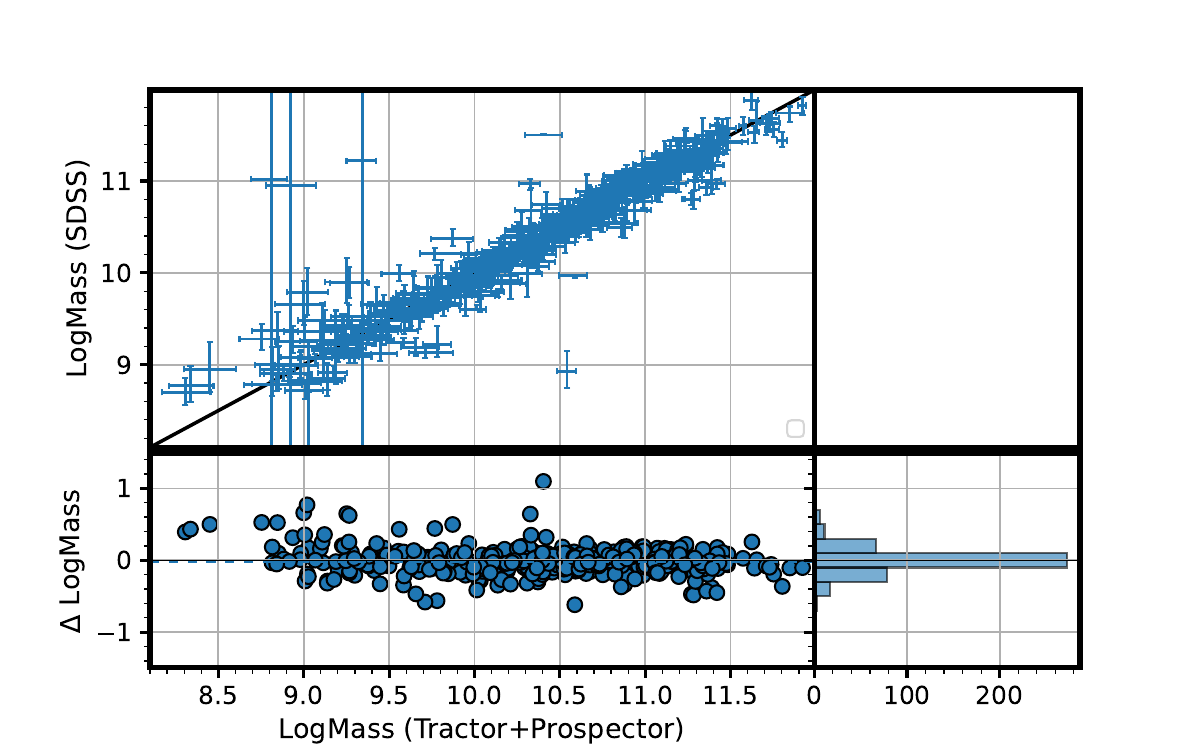}
    \caption{Comparison between SDSS FSPS-derived masses \citep{Sako_2018} and this study's LSTP masses. For the 475 SNe in common, the mean (median) agreement is $-0.014 \pm 0.008$~dex ($-0.013 \pm 0.006$~dex) with RMS $0.155 \pm 0.008$~dex.}
    \label{fig:sdss_mass_comp}
\end{figure}

\begin{figure}
    \centering
    \includegraphics[width=0.49\linewidth]{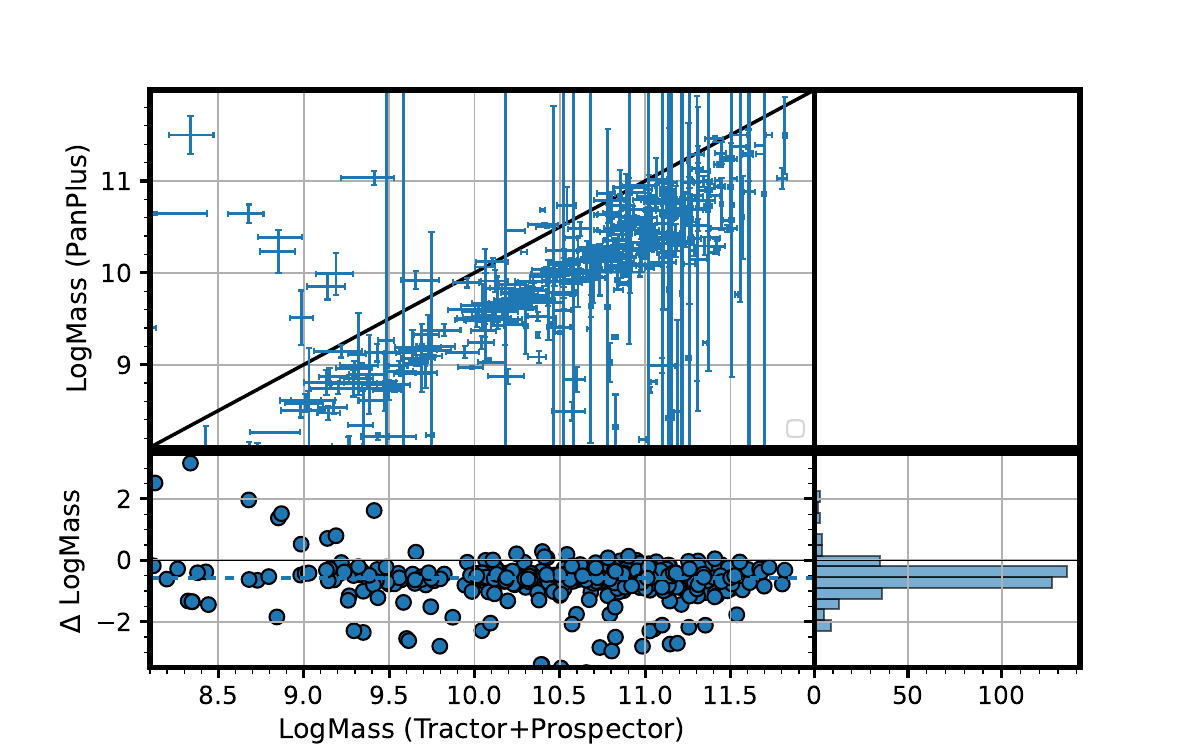}
    \includegraphics[width=0.49\linewidth]{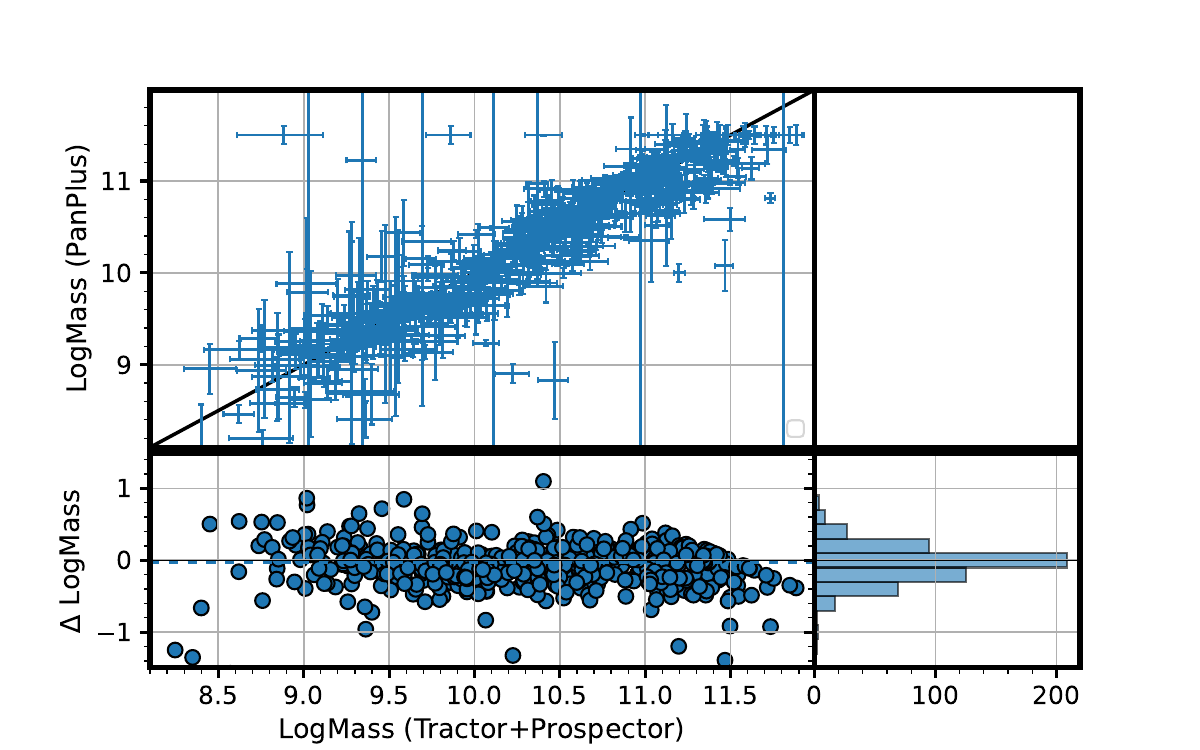}
    \caption{Same as \autoref{fig:sdss_mass_comp} but for the low-redshift ($z<0.15$, left panel) and high-redshift ($z>0.15$, right panel) Pantheon+ samples. \textit{Left:} Note the larger y-limits in the bottom panel than in other comparison plots. The mean (median) offset is $-0.797 \pm 0.053$~dex ($-0.571 \pm 0.015$~dex) with RMS $1.069 \pm 0.066$~dex for 422 SN host galaxies. \textit{Right:} The mean (median) offset is $-0.050 \pm 0.011$~dex ($-0.035 \pm 0.011$~dex) with RMS $0.254 \pm 0.009$~dex for 560 SN host galaxies.} %
    \label{fig:pplus_mass_comp}
\end{figure}

\begin{figure}
    \centering
    \includegraphics[width=0.8\linewidth]{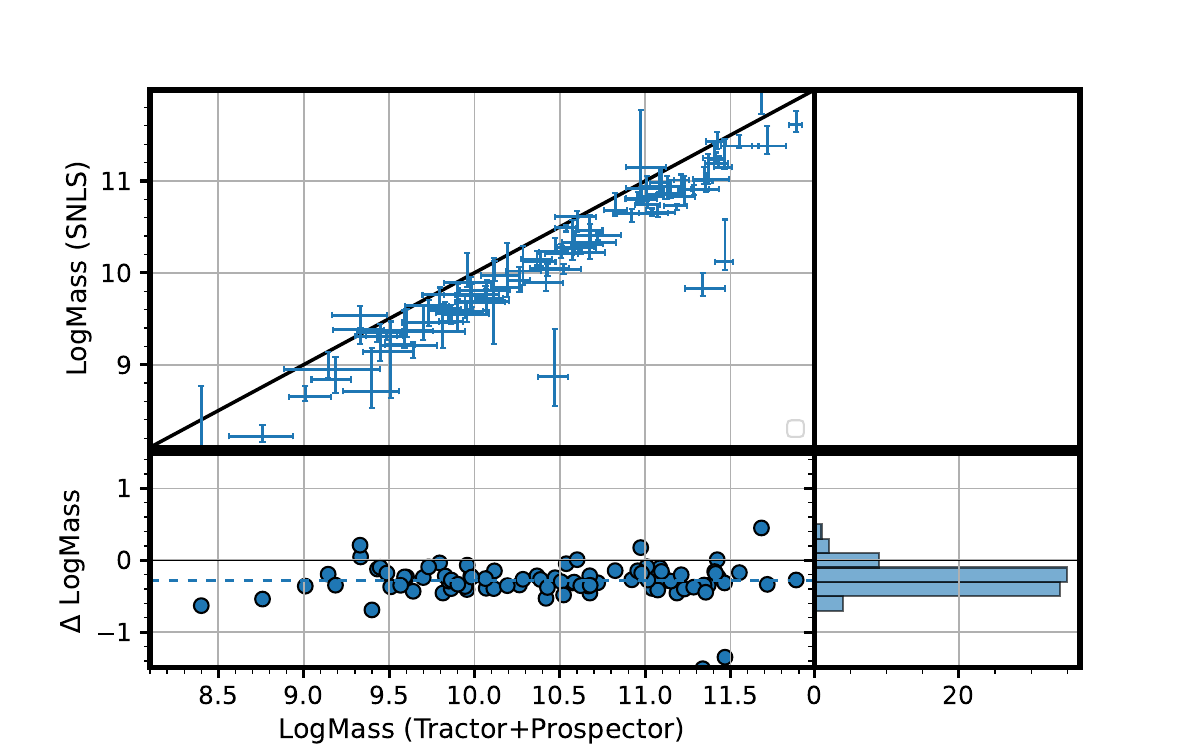}
    \caption{Same as \autoref{fig:sdss_mass_comp} but for SNLS. For the 84 SNe in common with the LSTP sample, the mean (median) agreement is $-0.271 \pm 0.017$~dex ($-0.277 \pm 0.019$~dex) with RMS $0.155 \pm 0.014$~dex.}
    \label{fig:snls_compare}
\end{figure}

\subsection{GOODS SNe}

\begin{figure}
    \centering
    \includegraphics[width=0.55\linewidth]{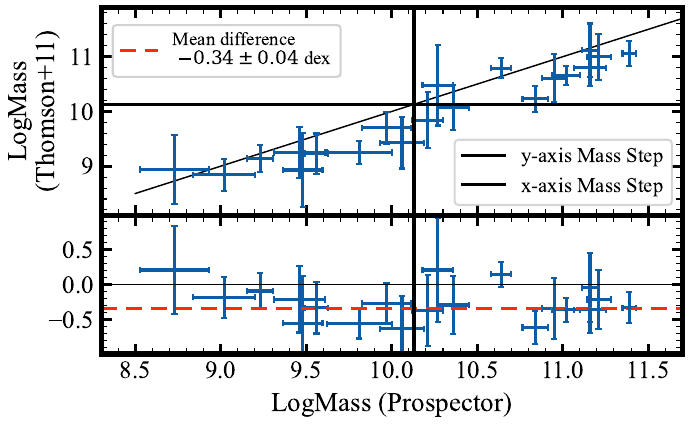}
    \caption{Comparison of this study's Prospector-derived mass estimates with those of \citet{Thomson_2011}. The same galaxy photometry was used for both sets of mass estimates, indicating the systematic offset is due primarily to differences in the adopted stellar population synthesis code and initial mass function. We discuss the precise differences in the text. Despite the offset, only one SN flips between a $1\sigma$ high vs. a $1\sigma$ low mass host, while another flips between lying at exactly the mass step, to $2 \sigma$ above it. The high-low classification of the rest is preserved between either set of masses.}
    \label{fig:goods_mass_compare}
\end{figure}

\citetalias{Thomson_2011} presented mass estimates of the GOODS-S SNe based on their galaxy photometry and a modified version of the population synthesis code GALAXEV \citep[][also commonly referred to as BC03]{Bruzual_2003}. As part of their SED fitting they allowed for a binary switch between a Chabrier or Salpeter IMF. 

We determined new masses using the same \citetalias{Thomson_2011} galaxy photometry and the same \texttt{Prospector} formulation used in the LSTP analysis (e.g., a Kroupa IMF and FSPS basis functions).

\subsection{LSTP Mass Comparison with DES-SN5YR as a Function of Redshift}\label{app:des5yr_comp}

\begin{figure}
    \centering
    \includegraphics[width=0.8\linewidth]{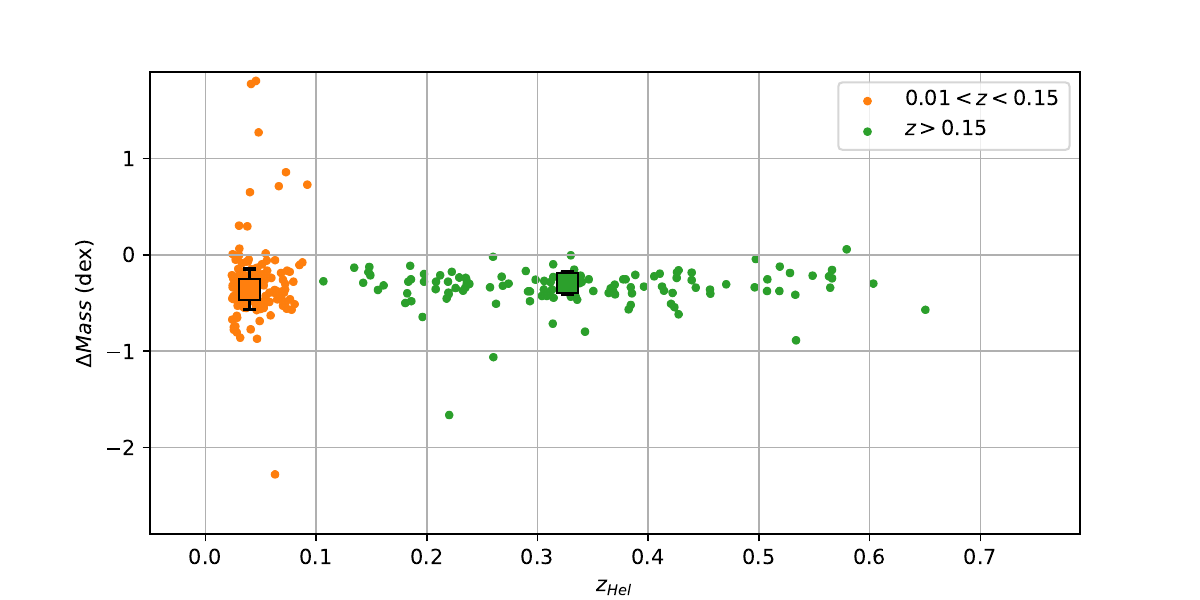}
    \caption{Comparison of LSTP masses from this study and those used in the DES-SN5YR analysis \citep{DES5YR}. There is a systematic offset between DES5YR and this study's LSTP masses, but it is consistent across the low (orange) and high (green) redshift bins, showing there is no host-mass-induced bias in their inference of cosmological parameters.}
    \label{fig:prospVSdes5yr_redshift}
\end{figure}

In \autoref{fig:prospVSdes5yr_redshift}, we show the equivalent to \autoref{fig:deltaMass_pan_union}, comparing DES-SN5YR masses to LSTP. Note this comparison is restricted to only the spectroscopically classified sample of DES SNe. Unlike the same figure shown for Pantheon+ and Union3, the DES host masses do not exhibit a redshift dependent trend against the reference LSTP mass scale, indicating their host mass determination does not introduce a bias into their cosmology inference.

\section{Correcting Pantheon+ Distance Moduli for the Host Mass Offset}
\label{app:panplus_biascorr}

\begin{figure*}
    \centering
    \includegraphics[width=0.8\linewidth]{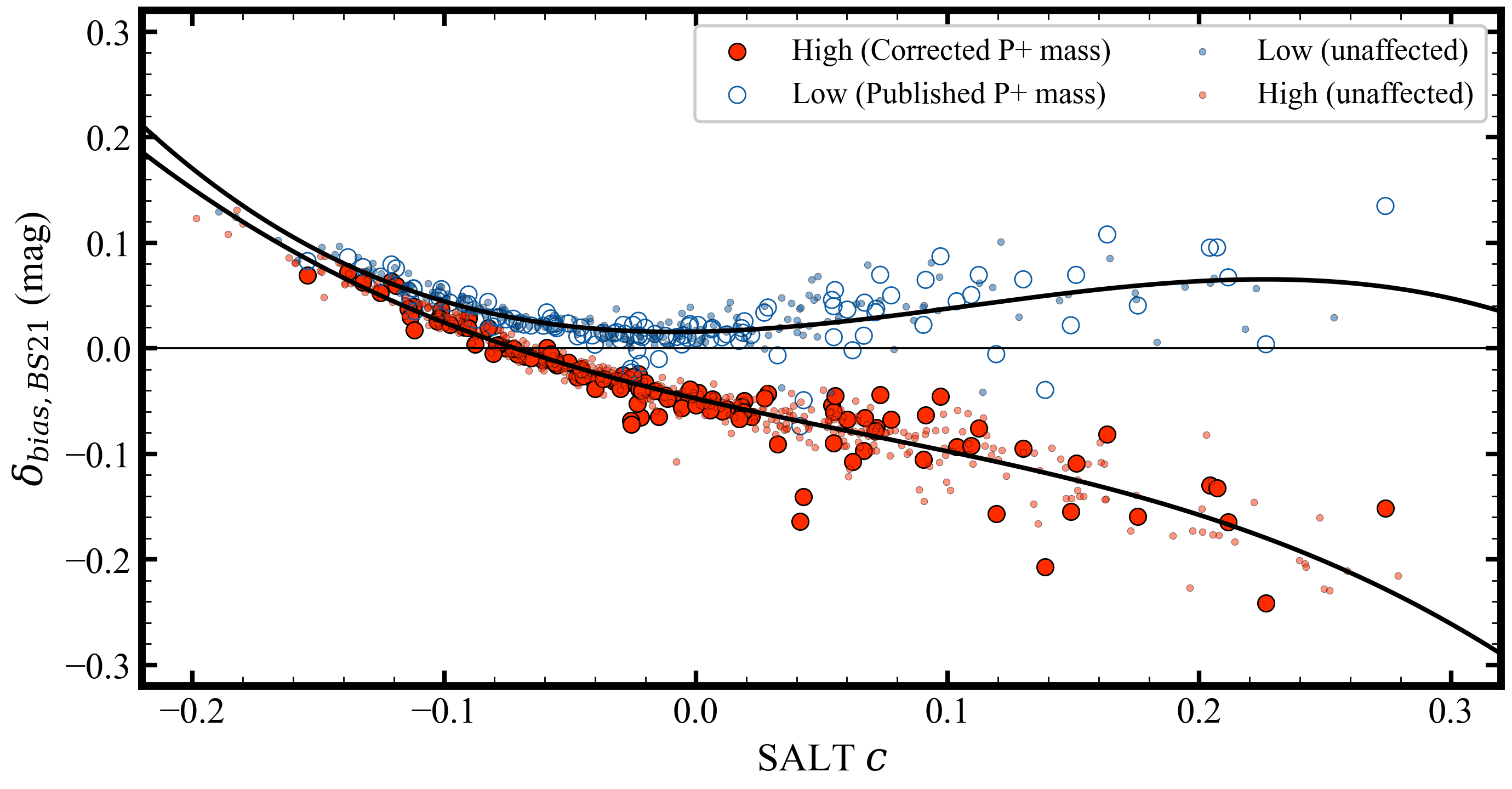}
    \caption{Updating the Pantheon+ bias corrections after correcting for the mass offset identified in their low-redshift SNe. Of the 713 SNe in the Pantheon+ low-redshift ($z<0.15$) bin for which they derived new host galaxy mass estimates, the new $0.6$~dex mass correction would shift 114 from low to high-mass, leading to significantly different bias corrections and, thus, inferred distances. Each SN of these 114 is represented by both a blue open and a red filled circle, the former representing the original bias correction value assuming a low-mass host, and the other the updated value after correction to a high-mass host. The other 599 SNe that did not change where they lie about the mass step are plotted as small dots, with blue and red representing low- or high-mass.}
    \label{fig:pplus_bias_curve}
\end{figure*}

In \autoref{fig:pplus_bias_curve}, we plot the bias corrections reported for the 713 low redshift SNe in the Pantheon+ database. We plot as large circles the 114 low-redshift ($z<0.15$) SNe in Pantheon+ that had a published host mass between 9.4 and 9.99 dex. The open blue circles represent their original bias corrections based on the systematically offset masses, while the filled red circles represent their new bias corrections based on the corrected masses. The other 599 SNe that did not change from one side of the mass step to the other are plotted as small markers. 

To estimate the updated bias corrections we computed spline representations of the two observed bias-correction vs. color relations, one for low and one for high mass, plotted here as two black curves. Then, the vertical difference between the two curves as a function of the SALT $c$ parameter was used to compute per-SN updates to the individual bias corrections. This fully preserves the way the BBC methodology employed by Pantheon+ handles galaxy host masses. The average shift in the distance modulus for the 114 SNe that flipped from low to high is $\Delta\mu=-0.073$~mag. When averaged over the entire low-redshift bin, this amounts to a distance shift of $\Delta\mu=-0.012$~mag. We add these shifts in the individual bias corrections to the $m_{b,corr}$ column in the Pantheon+ data file and use that adjusted file as input to the Cobaya cosmology fits we presented in the main text.

\section{$\Lambda$CDM Corner Plot} 

We include in this Appendix the corner plot for the main parameters inferred from the Union3.1+UNITY1.7 analysis, described in \autoref{sect:sncosmology} of the main text.

\label{app:lcdm_corner}
\begin{figure*}
    \centering
    \includegraphics[width=0.98\linewidth]{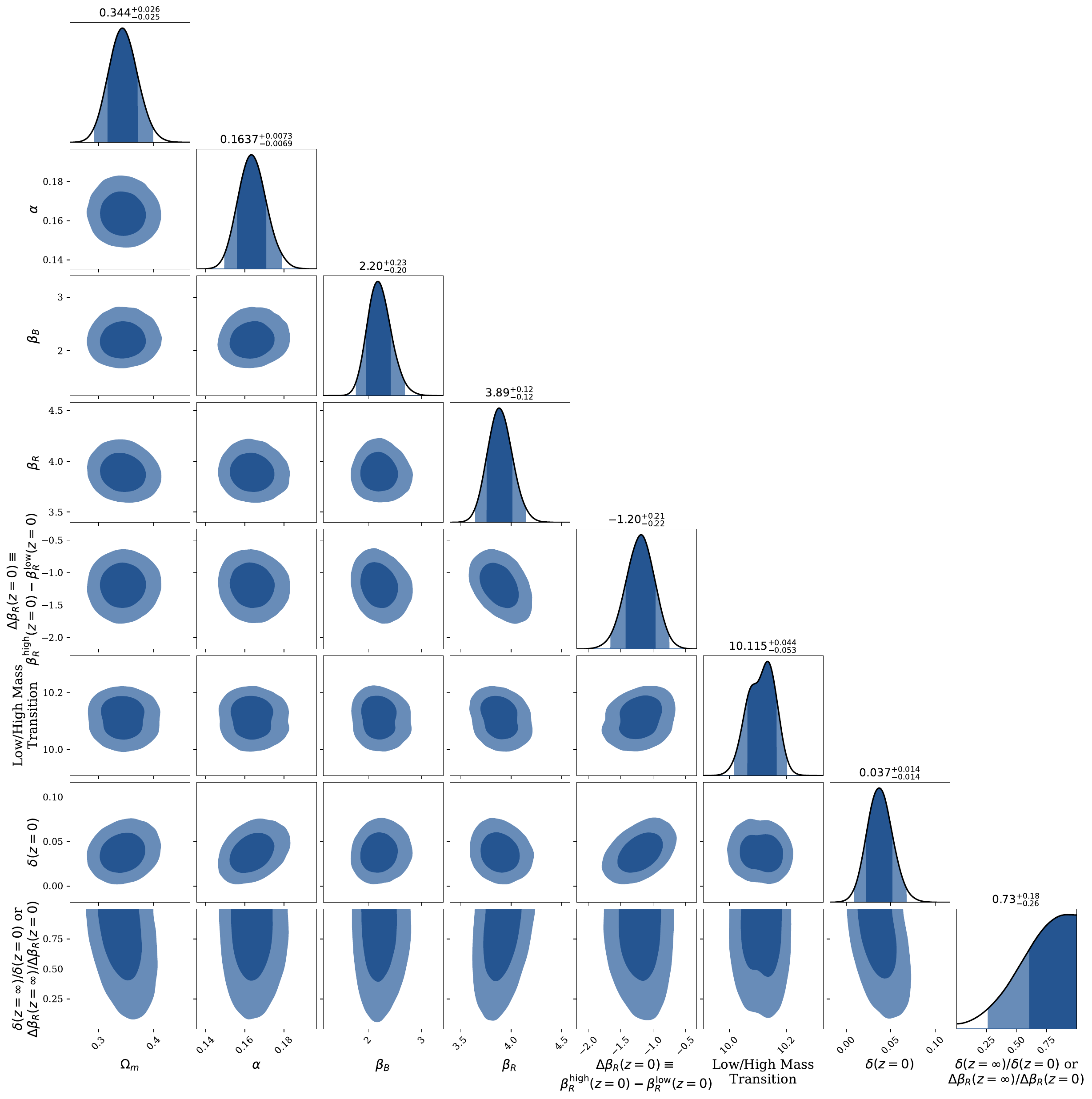}
    \caption{Corner plot of standardization parameters and $\Omega_m$ resulting from the updated Union3.1+UNITY1.7 SN-only, flat-$\Lambda$CDM cosmology analysis. The marginalized parameter constraints are printed at the top of each column and are also enumerated in \autoref{tab:fit_param_intervals} of the main text.}
    \label{fig:corner_union31}
\end{figure*}

\bibliography{bib/references}{}
\bibliographystyle{aasjournalv7}

\end{document}